\documentclass[11pt, a4paper]{article}

\usepackage{jheppub}
\usepackage[dvipsnames]{xcolor}
\usepackage[T1]{fontenc}
\usepackage[ansinew]{inputenc}
\usepackage{amsfonts}

\usepackage{physics}
\graphicspath{{Pictures/}}
\usepackage{bm}
\usepackage{float}
\usepackage{color}
\definecolor{darkblue}{cmyk}{0.9,0.9,0,0}

\usepackage{diagbox}

\usepackage{indentfirst}

\newcommand{\beq}{\begin{equation}}
\newcommand{\eeq}{\end{equation}}
\newcommand\beqa{\begin{eqnarray}}
\newcommand\eeqa{\end{eqnarray}}
\newcommand\bea{\begin{array}}
\newcommand\eea{\end{array}}

\newcommand{\Li}{\text{Li}}

\def\XXint#1#2#3{{\setbox0=\hbox{$#1{#2#3}{\int}$}
\vcenter{\hbox{$#2#3$}}\kern-.5\wd0}}

\newcommand{\p}{\partial}
\newcommand{\ii}{\mathrm{i}}
\newcommand{\ee}{\mathrm{e}}
\newcommand{\bfu}{\mathbf{u}}
\newcommand{\bfv}{\mathbf{v}}

\newcommand{\sh}{{\rm sinh}}

\def\[{\left[}
\def\]{\right]}

\def\spi{\relax{\rm \pi\kern-0.5em /}}
\def\sA{\relax{\rm A\kern-0.5em /}}
\def\sp{\relax{\rm p\kern-0.5em /}}
\def\sd{\relax{\rm \d\kern-0.5em /}}
\def\sk{\relax{\rm k\kern-0.5em /}}
\def\sn{\relax{\rm n\kern-0.5em /}}
\def\sl{\relax{\rm l\kern-0.5em /}}
\def\sP{\relax{\rm P\kern-0.7em /}}
\def\sBethe{\relax{\rm \Bethe\kern-0.5em /}}

\newcommand{\volodya}[1]{\com{V: #1}}

\usepackage{varioref}
\usepackage{makeidx}
\makeindex

\usepackage[english]{babel}

\par\vspace{1.5cm}
\vspace{2mm}

\begin{document}










    
\title{\textbf{Checkerboard CFT}}
\date{}
\author[a,b]{Mikhail Alfimov,}
\author[c]{Gwen{a}\"{e}l Ferrando,}
\author[d]{Vladimir Kazakov}
\author[e]{and Enrico Olivucci}
\affiliation[a]{HSE University, 6 Usacheva str., Moscow 119048, Russia}
\affiliation[b]{P.N. Lebedev Physical Institute of the Russian Academy of Sciences, 53 Leninskiy pr., Moscow 119991, Russia}
\affiliation[c]{School of Physics and Astronomy, Tel Aviv University,
Ramat Aviv 69978, Israel}
\affiliation[d]{Laboratoire de Physique de l'\'{E}cole Normale Sup\'{e}ri\'{e}ure, CNRS, Universit\'{e} PSL, Sorbonne Universit\'{e}, Universit\'{e} Paris Cit\'{e}, 24 rue Lhomond, 75005 Paris, France}
\affiliation[e]{Perimeter Institute for Theoretical Physics,
Waterloo, Ontario N2L 2Y5, Canada}
\emailAdd{malfimov@hse.ru}
\emailAdd{gwenael@tauex.tau.ac.il}
\emailAdd{kazakov@ens.fr}
\emailAdd{e.olivucci@gmail.com}
\abstract{The Checkerboard conformal field theory is an interesting representative of a large class of non-unitary, logarithmic Fishnet CFTs (FCFT) in arbitrary dimension which have been intensively studied in the last years. Its planar Feynman graphs have the structure 
of a regular square lattice with checkerboard colouring. Such graphs are integrable since each coloured cell of the lattice 
is equal to an R-matrix in the principal series representations of the conformal group. We compute perturbatively and numerically the anomalous dimension of the shortest single-trace operator in two reductions of the Checkerboard CFT: the first one corresponds to the Fishnet limit of the twisted ABJM theory in 3D, whereas the spectrum in the second, 2D reduction contains the energy of the BFKL Pomeron. 
We derive an analytic expression for the Checkerboard analogues of Basso--Dixon 4-point functions, as well as for the class of Diamond-type 4-point graphs with disc topology. 
The properties of the latter are studied in terms of OPE for operators with open indices. We prove that the spectrum of the theory receives corrections only at even orders in the loop expansion and we conjecture such a modification of Checkerboard CFT where quantum corrections occur only with a given periodicity in the loop order.}

\maketitle



\renewcommand{\thefootnote}{\fnsymbol{footnote}}
\setcounter{page}{1}
\setcounter{footnote}{0}
\setcounter{figure}{0}


\section{Introduction}
\label{sec:intro}

Conformal quantum field theories are ubiquitous in modern high energy physics and statistical mechanics where they describe many important physical phenomena. There exists a multitude of CFTs in spacetime dimensions $d<4$. For $d\geqslant 4$, however, the list of known CFTs is very short~(see e.g. \cite{Banks:1981nn} for the Banks--Zaks critical point), unless supersymmetry enters the game. Nevertheless, if we drop the unitarity requirement and allow for non-unitary, logarithmic CFTs \cite{Gurarie:1993xq}, an ample class of integrable theories in $d=4$ has been discovered in~\cite{Gurdogan:2015csr} as a special double-scaling limit of $\mathcal{N}=4$ super Yang--Mills theory (weak coupling combined with strong imaginary $\gamma$-deformation). The definition of these theories has later been extended to any $d$ dimensions in~\cite{Kazakov:2018qbr}. In the planar 't~Hooft limit, the perturbation theory of such CFTs is dominated by a very limited number of Feynman diagrams of a specific shape: in the simplest of these theories they are represented by the regular square lattice. 
As was shown by A.~Zamolodchikov~\cite{Zamolodchikov:1980mb}, the square lattice graphs are integrable, because they are equivalent to partition functions in a certain integrable statistical mechanical model of continuous spin variables, which are the coordinates of the vertices of the graph. These graphs are called \emph{fishnets} owing to their shape, and the CFTs that are dominated by them are dubbed Fishnet CFTs (FCFTs).

Apart from the FCFTs stemming from the double-scaling limit of $\gamma$-deformed $\mathcal{N}=4$ super Yang--Mills theory, a vast class of so-called Loom FCFTs was proposed in~\cite{Kazakov:2022dbd}. It is based on Zamolodchikov's construction of integrable Feynman graphs of a more general type, featuring arbitrary valency of the vertices, any dimension $d$, and diverse types of propagators. This construction relies on the existence, for each such diagram, of an associated Baxter lattice -- a collection of straight lines parallel to $M$ directions that we dub \emph{slopes}. The scaling powers of the propagators, which are the Boltzmann weights in the stat-mech picture, are proportional to the angles between the lattice lines (see~\cite{Zamolodchikov:1980mb,Kazakov:2022dbd} for the details).

The graphs stemming from the Loom construction are, to our knowledge, the most general integrable Feynman diagrams. Many of them present practical interest for the perturbative computations in various realistic CFTs. Consequently, they spurred the development of many computational tools such as the Yangian symmetry~\cite{Chicherin:2017cns,Chicherin:2017frs, Corcoran:2021gda, Kazakov:2023nyu}, spin chain transfer matrix~\cite{Caetano:2016ydc,Gromov:2016rrp,Gromov:2018hut}, separation of variables~\cite{Derkachov:2018lyz, Derkachov:2019tzo, Derkachov:2020zvv, Cavaglia:2021mft, Olivucci:2021cfy, Derkachov:2021ufp, Olivucci:2023tnw, Aprile:2023gnh} or quantum spectral curve~\cite{Gromov:2017cja}. Among the multitude of FCFTs originating from the Loom construction, some particular cases are more relevant than others. They include the CFTs obtained as the fishnet limits of the ABJM model~\cite{Caetano:2016ydc}, featuring Feynman graphs of the shape of regular triangular lattice, or of the $\mathcal{N}=4$ SYM theory~\cite{Gurdogan:2015csr}.

In this paper we study a class of Loom FCFTs with $M=4$ slopes that feature only quartic scalar vertices. We name this model the Checkerboard CFT because all of its planar graphs are dual to a Baxter lattice with only rectangular faces that can be bi-coloured in checkerboard style, as seen in figure~\ref{Checkerboard_lattice}. These planar Feynman graphs  also have the shape of a square lattice, with the powers of the propagators alternating along both rows and columns. The theory has two coupling constants and four complex adjoint $SU(N)$ matrix fields $Z_{j}$, $ j=1,2,3,4$, with the Lagrangian~\eqref{CheckerboardCFT} (see the detailed definition of the theory in section~\ref{sec:Checkerboard}). Unlike the bi-scalar FCFT~\cite{Gurdogan:2015csr} and its $d$-dimensional anisotropic generalisation~\cite{Kazakov:2018qbr}, each square face of a graph of Checkerboard CFT is equal to the R-matrix operator acting on unitary irreducible representations of the conformal group $SO(1,d+1)$ both in auxiliary and quantum space~\cite{Derkachov:2014gya} and depending on 3 parameters, namely the scaling dimensions that define the physical and auxiliary space representations, and the spectral parameter $u$. The integrability for such graphs is therefore explicit since they are equal to a product of conformal transfer matrices. It follows that the computation of correlation functions in Checkerboard CFT is reduced to the standard integrability methods for quantum spin-chains.\footnote{In this case, one has to deal with spins in principal series representations of the conformal group, for which the formalism is not as developed as for finite-dimensional or highest-weight representations.} In particular, the correlation functions of two single-trace, non-local operators discussed in section~\ref{sec:L_correlators} represent, in the spin-chain picture, the statistical-mechanical system of interacting conformal spins on the regular square lattice of cylindrical topology and alternating Boltzmann weights  along both rows and columns.  In section~\ref{sec:L=2} we present the all-loop computation of the simplest such correlator following the methods of~\cite{Grabner:2017pgm, Kazakov:2018qbr, Gromov:2018hut}. From this four-point function one can extract the scaling dimensions of the shortest length, \(L=2\) operators exchanged in its OPE $s$-channel, in terms of certain infinite double sums. 

The richness of the parameter space of Checkerboard CFT allows for various interesting reductions, which we consider in section~\ref{sec:reductions}. For specific choices of the scaling dimensions $\Delta_{j}$ of the fields, this theory reduces to the strongly twisted 3D ABJM FCFT~\cite{Caetano:2016ydc} or to a 2D logarithmic CFT that captures the spectrum of Lipatov's reggeized gluons~\cite{Lipa:1993pmr,Lipatov:1993yb,Kuraev:1977fs,Balitsky:1978ic}, describing the famous BFKL limit of high-energy QCD. We will apply our general results for Checkerboard CFT for the four-point functions  and the anomalous dimension  of \(L=2\) exchange operators to study these quantities in more detail (perturbatively and numerically)  for these two reductions.   

The sections~\ref{sec:BD}, \ref{sec:diamonds}, and \ref{sec:higher_periods} are devoted to the study of some Feynman diagrams that can be drawn on the disc, and completely describe the vacuum expectation value of single-trace nonlocal operators in the Checkerboard CFT, ie. single-trace multi-point correlators. We concentrate on four-point diagrams, obtained by cutting out a rectangular piece of the square lattice and then identifying some of the coordinates of the external legs. We distinguish two classes of such correlators: for the first class the edges of the rectangle are parallel to the lattice lines, whereas they are parallel to the lattice diagonals for the second class (the so-called diamond graphs). 

The first class of these four-point correlators and the related Feynman integrals are studied in section~\ref{sec:BD}. They generalise the 4D result of Basso and Dixon~\cite{Basso:2017jwq} for the case of bi-scalar FCFT, and in~\cite{Derkachov:2018ewi} for the 2D bi-scalar FCFT. In section~\ref{sec:SoV} (with the details of the computations in appendix \ref{app:SoV}) we compute, using the separation of variables method of~\cite{Derkachov:2019tzo,Derkachov:2020zvv,Derkachov:2014gya,Sklyanin:1984sb,Derkachov:2001yn,Derkachov:2018rot}, the general expression for the four-point functions of such type, emerging in the full Checkerboard CFT, for $d=2$ and $d=4$. 

The diamond correlators, which will be studied in section~\ref{sec:diamonds}, can themselves be divided into four types according to their boundary conditions. Such correlators feature drastically different properties from the ``Basso--Dixon" class. First of all, in the simplest bi-scalar FCFT these are highly divergent objects, and it is not completely clear how to give them a physical interpretation in terms of CFT correlators. Otherwise, for general parameters $\Delta_k$ such graphs are finite, and many of them evaluate either to zero or to a tree-level-like product of propagators (while being quantities defined at loop level).
This fact can be checked by simple computations based on the star-triangle identity. We will examine such correlators via their operator product expansions in section~\ref{sec:diamonds}. 

We end this section with a proof by induction that the spectrum of anomalous dimensions in Checkerboard CFTs is protected at any odd order of perturbation theory. On the other hand, when the bare dimensions of the fields satisfy additional constraints, e.g. $\Delta_1 + \Delta_3 = \Delta_2 + \Delta_4 = d/2$, the spectrum gets corrected at every order of perturbation theory. The different regimes, namely interaction at any-loop order vs protected odd-loop contributions, is also apparent in the analytic structure of poles in the SoV representation of four-point Ladder integrals from section \ref{sec:ladders}.

Moreover, the Checkerboard model can be generalised for a higher number of slopes $M>4$ to a theory of $M_1 M_2$ complex scalar fields, with $M_1+M_2=M$ as proposed in section~\ref{sec:higher_periods}. In this case, the square-lattice graphs have a general periodicity $(M_1,M_2)$ along rows and columns. We conjecture that the spectrum of anomalous dimensions gets corrections only at loop orders $k$ such that $k\,\text{mod}(M_1)=0$ or $k\,\text{mod}\,(M_2) =0$.

The concluding  section~\ref{sec:discussion} is devoted to the discussion of our results. Contextually, we will list there many unsolved problems concerning the Fishnet CFTs.

\section{Definition  of Checkerboard CFT}
\label{sec:Checkerboard}

The Checkerboard Fishnet CFT studied in this paper is a theory of four complex matrix scalars fields \(Z_j\) of $N \times N$ components in any space-time dimension $d$. The Lagrangian of the theory features in general non-local kinetic terms and two quartic interactions, 
\begin{equation}
\label{CheckerboardCFT}
    {\cal L}^{(CB)} = \! N \Tr\left[ \sum_{j=1}^{4} \bar Z_j (-\p_\mu \p^\mu)^{w_j} Z_j - \xi_1^2 \,\bar{Z}_1 \bar{Z}_2 Z_3 Z_4 - \xi_2^2\,  Z_1 Z_2 \bar{Z}_3 \bar{Z}_4 \right]\,.
\end{equation}
We impose the constraint \(w_1+w_2+w_3+w_4=d\) in order to work with dimensionless couplings $\xi_1^2, \xi_2^2$. We shall often switch from the parameters $w_j$ to another set of labels, commonly used in the $SO(1,d+1)$ spin-chain formalism \cite{Chicherin:2012yn}, namely
\begin{equation}\label{eq:Props}
    w_1=u+d-\Delta_+\,, \quad w_2=-u+\Delta_-\,, \quad w_3=u+\Delta_+\,, \quad w_4=-u-\Delta_-\,,
\end{equation}
where $\Delta_\pm$ and the spectral parameter $u$ are generic. It follows from \eqref{CheckerboardCFT} that the scaling dimensions of the fields $\bar Z_j,Z_j$ are  $\Delta_j=\frac{d}{2}-w_j$.

We are generally interested in the planar (or multi-colour) limit of the theory, $N\to \infty$. While for generic $w_j$'s the Lagrangian is UV complete and the theory is finite, there are \emph{special values} when double-trace correlators of length-$2$ operators are divergent and a corresponding counter-term must be added to \eqref{CheckerboardCFT}.
This is the case when a couple of neighbouring fields in the interaction vertices features a pair of \emph{conjugate} dimensions, i.e.
\begin{equation}
\label{special_pts}
\Delta_1 +\Delta_2 = \frac{d}{2} = \Delta_3 + \Delta_4\quad \text{or} \quad \Delta_1 + \Delta_4 = \frac{d}{2} = \Delta_2 + \Delta_3\, .
\end{equation}
 For instance, whenever the first equation is verified, the double-trace counter-terms
\begin{align}
\label{doubletraces}
{\cal L}_{\text{dt}}^{(CB)}&= \alpha(\xi_1,\xi_2) \Tr(\bar{Z}_1\bar{Z}_2) \Tr(Z_3 Z_4) + \bar\alpha(\xi_1,\xi_2) \Tr(Z_1 Z_2) \Tr(\bar{Z}_3 \bar{Z}_4)\,
\end{align} 
should be added to the Lagrangian \eqref{CheckerboardCFT}, with the couplings  $\alpha(\xi_1,\xi_2)$ adjusted to their critical values in order to preserve conformal symmetry.
The propagators of the adjoint fields \(Z_k\) are
\begin{equation}\label{propagator}
    D_{i}(x)=\langle Z_i(x) \bar Z_i(0)\rangle =\frac{\Gamma\left(\frac{d}{2}-w_i\right)}{4^{w_i}\pi^{\frac{d}{2}}\Gamma(w_i)}\frac{1}{(x^2)^{\frac{d}{2}-w_i}}\,,
\end{equation}
where the indices \(i\) and \(j\) enumerate the fields. Here we neglected the $SU(N)$ matrix indices of the fields since in the multi-colour limit we are interested in we simply have to restrict ourselves to planar graphs.



\begin{figure}
\begin{center}
\includegraphics[scale=0.5]{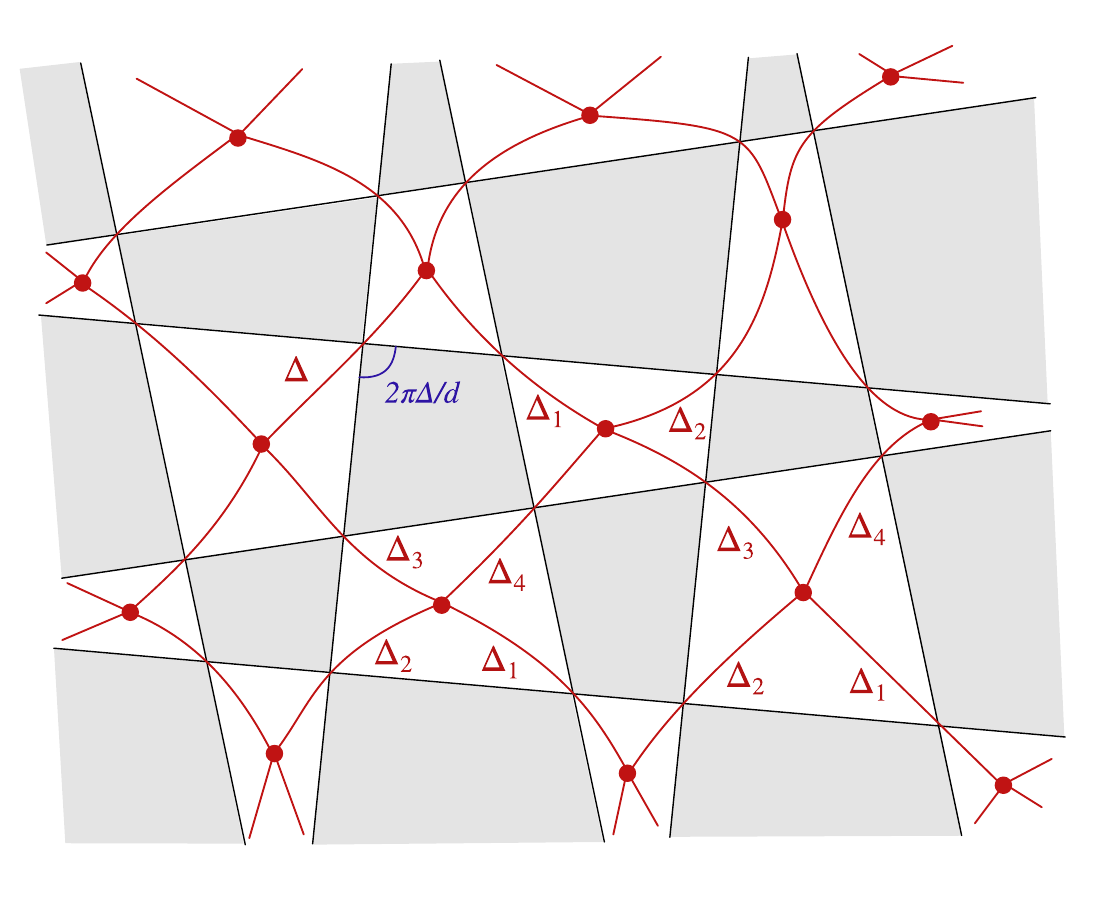}
\caption{The \(M=4\) Checkerboard Loom Baxter lattice and the associated Feynman diagram (in red). The diagram features quartic vertices emitting propagators \eqref{propagator}. Their scaling power is determined by the angle of Baxter lattice through which they pass.}
\label{Checkerboard_lattice}
\end{center}
\end{figure}
Such a simple content of Feynman diagrams of the theory is a consequence of the non-Hermiticity, i.e. the \emph{chirality}, of the interaction featured by \eqref{CheckerboardCFT}, because
\begin{equation}
    \Tr[\bar{Z}_{1} \bar{Z}_{2} Z_{3} Z_{4}]^{\dagger} =  \Tr[\bar{Z}_{4} \bar{Z}_{3} Z_{2} Z_{1}]
\end{equation}
are not allowed vertices in the theory.

Let us note that the Checkerboard CFT can be viewed as a reduction of the Loom  FCFT$^{(4)}$ with $M=4$ slopes proposed in~\cite{Kazakov:2022dbd} (see section~3.3  therein) if  we  keep only the couplings for two vertices of valence 4, namely
\beq\label{nonzero_vertices}
\Tr\left[v_1 X_3 Y_2 \bar{u}_1 \right] \quad \textrm{and} \quad \Tr\left[u_1 \bar{v}_1 \bar{X}_3 \bar{Y}_2 \right]\,,
\eeq
and set the remaining \(131-2=129\) couplings to zero. Then the fields $X_1$, $X_2$, $X_4$ and $Y_1$ and the dual fields $u_2$, $u_3$, $u_4$ and $v_1$ are effectively decoupled from the theory, while only the fields $X_3$ and $Y_2$ and the dual fields $u_1$ and $v_1$ have interactions. The latter are identified with fields in \eqref{CheckerboardCFT} as follows
\beq
u_1=Z_1\,, \quad \bar{v}_1=Z_2\,, \quad X_3=Z_3 \quad \textrm{and} \quad Y_2=Z_4\,.
\eeq
In this respect, we must point out that the number of slopes featured by the Baxter lattice in figure~\ref{Checkerboard_lattice} is reduced from four to three (or two) when either (or both) of the conditions \eqref{special_pts} is met.

In the particular case $d=4$ and $w_1=w_2=w_3=w_4=1$, the Checkerboard CFT has  the ``standard" local  Lagrangian
\begin{align}
\label{NormalCheckerboard4D}
{\cal L}^{(CB)}&= N \Tr\left[ \sum_{j=1}^{4} \p_\mu \bar Z_j \p^\mu Z_j
+
\xi_1^2 \bar{Z}_1 \bar{Z}_2 Z_3 Z_4+
\xi_2^2 Z_1 Z_2 \bar{Z}_3 \bar{Z}_4 \right] +\text{double-traces }\,.
\end{align} 
Of course it is still a non-unitary, logarithmic FCFT. This case is potentially very useful since this FCFT is dominated by standard scalar Feynman graphs which might be computable by integrability methods. Importantly, this ``isotropic" point for the checkerboard is the case when the Baxter lattice features two slopes that - in addition - are perpendicular. In this case, among the double-traces terms we must include the counter-terms of the type $\alpha'(\xi_1,\xi_2)\,\text{Tr}(Z_i \bar Z_{i+2}) \text{Tr}(\bar Z_i Z_{i+2})$; the critical values of such double-trace couplings are more involved to compute (as it is for the simplest bi-scalar Fishnet CFTs, see \cite{Gromov:2017cja,Gromov:2018hut}), though the existence is guaranteed beyond perturbation theory.

In the next sections, we will explore the properties of the Checkerboard theory and compute certain of its correlators. We will realise that its typical planar Feynman graphs, depicted in figure~\ref{Checkerboard_lattice}, have the shape of the regular square lattice, similar to the bi-scalar FCFT~\cite{Gurdogan:2015csr} with the Lagrangian~\eqref{bi-scalar}. However, in  the Checkerboard CFT  each square face of the Feynman graph in figure~\ref{Checkerboard_lattice} appears to be given by the R-matrix introduced in~\cite{Chicherin:2012yn} (see next section for explicit definition) acting in both spaces on principal series representations of conformal group and explicitly depending on spectral parameter $u$. Thus, the integrability is contained in an explicit and familiar way in the Checkerboard CFT.    

\section{Correlators of single-trace operators: spin-chain picture and \texorpdfstring{$L=2$}{L=2} case}
\label{sec:L_correlators}

Among many physical quantities, a special role is played by correlators of single-trace operators---typical quantities to study in the planar limit. Correlators of one, non-local, single-trace operator will be considered in sections~\ref{sec:BD} and \ref{sec:diamonds} for example. In this section, we work out some correlators between two, non-local, single-trace operators.

\subsection{Integrability of the Correlators}

We will consider a class of $2L$-point functions obtained by complete point-split inside the two traces. Its perturbative weak-coupling expansion can be presented in terms of Feynman diagrams with cylindrical topology and with the Checkerboard structure, one at each loop order, described in the previous section. 
Let us define a concrete instance of such correlator,
\begin{align}
\langle {\cal O}(x_1,\dots,x_L) \widetilde{\cal O}(x'_1,\dots,x_L')\rangle\,,
\label{correlator}
\end{align} where the notation for single-trace, point-split operators is 
\begin{align}
\begin{aligned}
\label{Operator_wheel_1}
&{\cal O}(x_1,\dots,x_L) = \Tr[(Z_1 Z_2)(x_1)\,(Z_1 Z_2)(x_2)\,\dots\,(Z_1 Z_2)(x_L)]\,,\\
& \widetilde{\cal O}(x'_1,\dots,x_L')=\Tr[(\bar Z_4 \bar Z_3)(x_1')\,(\bar Z_4 \bar Z_3)(x_2')\,\dots\,(\bar Z_4 \bar Z_3)(x_L')] \,.
\end{aligned}
\end{align}
Its perturbation theory can be re-summed to a finite coupling expression via the Bethe--Salpeter (BS) method. We will identify this BS kernel with the transfer matrix of an integrable non-compact spin chain with $SO(1,d+1)$ symmetry.
Quantum corrections in perturbation theory for weak couplings (and in the planar limit) consist of $L$ Feynman diagrams at each order $(\xi_2^2)^{ (n+1) L} (\xi_1^2)^{n L}$ where the loop order $n\geqslant0$ is an integer. In particular, this correlators are zero in free theory.

\begin{figure}
\begin{center}
\includegraphics[scale=0.60]{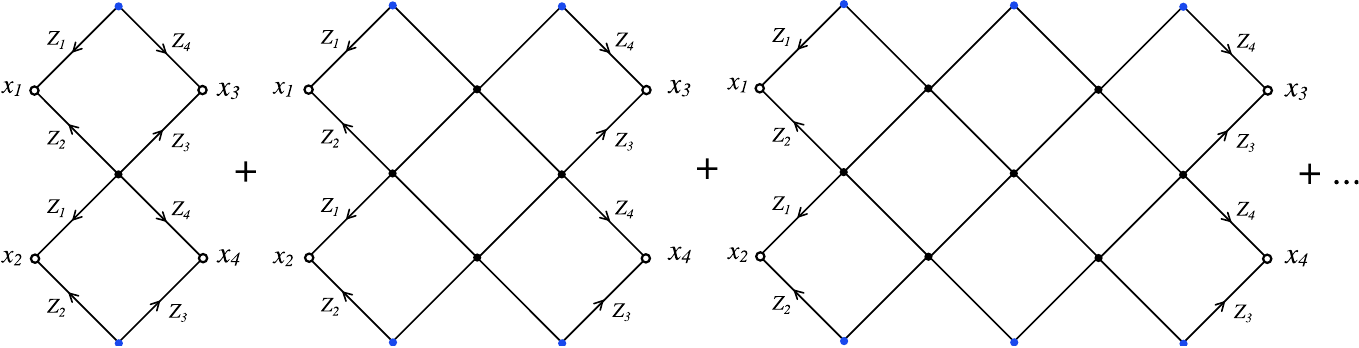}
\caption{The weak coupling expansion of $K(x_1,x_2|x_3,x_4)$ up to order $\xi_2^{12} \xi_1^{8}$ in terms of the Feynman diagrams a.k.a. integral kernels $T_n(x_1,x_2|x_3,x_4)$ for $n=1,2,3$. Arrows on top of propagators are oriented from $Z_k$ to $\bar Z_k$. Black dots are integrated vertices. Each pair of blue dots in a column denotes the same integrated vertex, since the graphs are wrapped on a cylinder.}
\label{L2_4pts}
\end{center}
\end{figure}

\begin{figure}
\begin{center}
\includegraphics[scale=1.25]{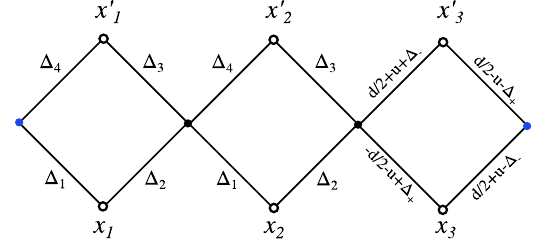}
\caption{Illustration of the formula \eqref{T-matrix_operator_form} for the case \(L=3\). Black dots are integrated vertices. The pair of blue dots in a column denotes the same integrated vertex, since the graphs are wrapped on a cylinder, which corresponds to taking trace in auxiliary space.}
\label{Checkerboard_R_matrices}
\end{center}
\end{figure}

The Feynman diagrams for a given $n$ can be expressed as a power of a certain integral ``graph-building" operator $\widehat{T}$, acting on functions of $L$ variables, say $x_1,\dots,x_L$, in $\mathbb{R}^d$. In practice, one of the $L$ diagrams is expressed as the kernel of $\widehat{T}^{\, n}$, namely
\begin{multline}
T_n( x_1,\dots,x_{L}|x'_1,\dots,x'_L)= \\
=\int \prod_{i=1}^L \dd^d y_i \, T_{n-1}( x_1,\dots,x_L|y_1,\dots,y_L) T( y_1,\dots,y_L|x_1',\dots,x_L')\,,
\end{multline}
and the others are obtained by any cyclic shift of points $x_{k}\to x_{k+j}$ with $1\leqslant j\leqslant L-1$.
Hence, the correlator at finite-coupling results from the BS resummation, namely
\begin{equation}
\langle {\cal O}(x_1,\dots,x_L) \widetilde{\cal O}(x'_1,\dots,x_L')\rangle = \sum_{j=0}^{L-1}K(x_{1+j},x_{2+j},\dots,x_{L+j}|x'_1,\dots,x'_L)\, ,
\end{equation}
with
\begin{align}
K(x_1,\dots,x_L|x'_1,\dots,x'_L)= \xi_2^{2L} \sum_{n=0}^{+\infty}(\xi_1^2 \xi_2^2)^{n L}
\,T_{n+1}( x_{1},\dots,x_{L}|x_1',\dots,x_L')\,.
\label{T-representation}
\end{align}
The operator $\widehat{T}$ is the transfer matrix of a non-compact, homogeneous spin chain with $SO(1,d+1)$ symmetry and periodic boundary conditions. Each of the $L$ sites carries the infinite-dimensional representation of a scalar field with scaling dimension $\Delta_1+\Delta_2$.
The operator $\widehat{T}$ is the trace over the auxiliary space (infinite-dimensional representation of 
dimension $\Delta_0 = \Delta_1 + \Delta_4$) of a product of $L$ solutions $\widehat R_{0k}$ of the Yang--Baxter equation~\cite{Chicherin:2012yn}, that is
\begin{equation}\label{T-matrix_operator_form}
    \widehat{T} = \Tr_0\left[\widehat{R}_{01} \widehat{R}_{02} \dots \widehat{R}_{0L}\right]\,.
\end{equation}
Each of the operators $\widehat{R}_{0k}$, for $k=1,2,\dots,L$, is an integral operator acting on functions of $x_k$ in physical space as well as on functions of $x_{0}$ in auxiliary space. Its kernel for $k=1$ reads
\beq\label{R-matrix}
R(x_1,x_0|x_{1'},x_{0'}) = \frac{c}{(x_{10}^2)^{-u-\frac{d}{2}+\Delta_+}
(x_{01'}^2)^{u+\frac{d}{2}+\Delta_-}
(x_{1'0'}^2)^{-u+\frac{d}{2}-\Delta_+}(x_{0'1}^2)^{u+\frac{d}{2}-\Delta_-}}\,,
\eeq
where $\Delta_{\pm} =(\Delta_0  \pm (\Delta_1+\Delta_2))/2$, and
\begin{multline}\label{c_coefficient}
c=\prod_{j=1}^{4}\frac{\Gamma\left(\frac{d}{2}-w_j\right)}{4^{w_j} \pi^{\frac{d}{2}} \Gamma(w_j)}= \\
=\frac{\Gamma\left(-u-\frac{d}{2}+\Delta_+\right)\Gamma\left(u+\frac{d}{2}+\Delta_-\right)\Gamma\left(-u+\frac{d}{2}-\Delta_+\right)\Gamma\left(u+\frac{d}{2}-\Delta_-\right)}{(2\pi)^{2d}\, \Gamma(u+d-\Delta_+)\, \Gamma(-u-\Delta_-)\, \Gamma(u+\Delta_+)\, \Gamma(-u+\Delta_-)}\,.
\end{multline}
The infinite-dimensional trace in \eqref{T-matrix_operator_form} is the iterative convolution of R-operators with respect to the auxiliary-space variable $x_0,x_{0}',\dots$, assuming periodic boundary $L+1\equiv 1$. This construction is depicted in figure~\ref{Checkerboard_R_matrices}. The Yang--Baxter property for $\widehat R$ guarantees that the operators $\widehat{R}$ and $\widehat{T}$ satisfy the RTT equation~\cite{Chicherin:2012yn}.
According to standard argument, this property ensures the existence of a 
complete set of commuting operators acting on $x_1,\dots, x_L$, hence the quantum integrability of the
spin chain of length~\(L\).

Notice that in the Checkerboard CFT the spectral parameter $u$ is naturally incorporated into the structure of the Feynman graphs due to the identification \eqref{eq:Props}. This feature is proper to the whole construction of the Loom FCFTs~\cite{Kazakov:2022dbd}.

The main implication of formula \eqref{T-representation} is that the calculation of the correlation functions \eqref{correlator} boils down to finding a basis of eigenfunctions and of the corresponding  eigenvalues for the integrable transfer matrix \eqref{T-matrix_operator_form}. Here, quantum integrability guarantees the existence of a non-perturbative, exact approach to the study of the Checkerboard CFT.

\subsection{Anomalous Dimensions for \texorpdfstring{$L=2$}{L=2}}
\label{sec:L=2}

In this section, we derive the exact expression for the shortest four-point correlator of the type~\eqref{correlator} in the Checkerboard CFT,
and extract the anomalous dimension of lightest single-trace operator, which dominates the OPE $s$-channel,
\begin{equation}
\label{lightest}
   \Tr[Z_1 Z_2 Z_1 Z_2](x)\,.
\end{equation}
Another candidate for the lightest operators is $\Tr[Z_3 Z_4 Z_3 Z_4]$. We will assume without loss of generality that $\Delta_1+\Delta_2 < \Delta_3+\Delta_4$, hence \eqref{lightest} is the lightest one.
We shall follow the methods of~\cite{Grabner:2017pgm,Gromov:2018hut,Kazakov:2018qbr,Derkachov:2023xqq}, i.e. perform a conformal partial wave decomposition of the s-channel and, from that, achieve an all-loop equation for the anomalous dimensions of the exchanged operators. Within the choice of shortest length $L=2$ and for operators of spin $S=0$ the solution of the spectral problem reduces to computing a two-loop, massless, Kite master integral with specific powers of the propagators. 

\begin{figure}
\begin{center}
\includegraphics[scale=1]{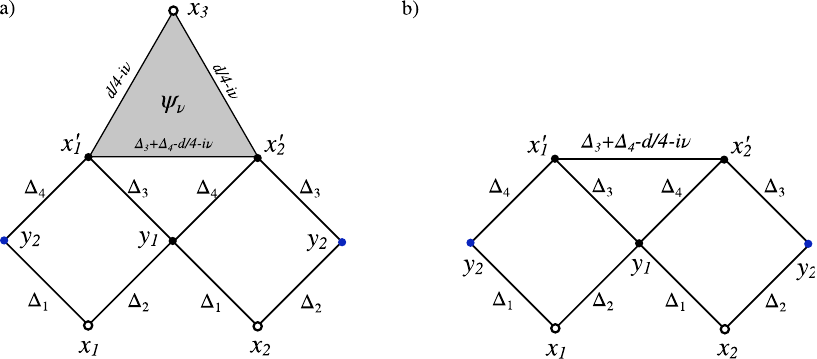}
\caption{$\widehat{T}$ operator with $L=2$ acting on the 3-point function $\Psi_{\nu}(x_1,x_2;x_3)$, in grey, before (a) and after (b) sending the point $x_3$ to $\infty$.}
\label{fig:actionT}
\end{center}
\end{figure}

We will study this problem in two concrete realisations of the Checkerboard CFT.
First, in $d=3$, for the ABJM FCFT~\cite{Caetano:2016ydc} defined in the section \ref{sec:ABJMFCFT}, we compute the anomalous dimension of \eqref{lightest} up to a few orders in the weak coupling expansion, and otherwise numerically in a large range of couplings.
Second, for the $d=2$ BFKL FCFT discussed in the section \ref{sec:BFKLlimit} we show that the anomalous dimension of the lightest operator matches the Pomeron spectrum in the Regge limit of QCD \cite{Lipatov:1993yb,Lipatov:1993qn,Faddeev:1994zg,DeVega:2001pu,Derkachov:2001yn}. Let us start within the general setting. According to what was explained above, one has\footnote{Of course, we consider a different correlator \(\langle\text{Tr}\left[(Z_1 Z_2)(x_1)(Z_1 Z_2)(x_2) \right]\text{Tr}\left[(\bar Z_2 \bar Z_1)(x_1')(\bar Z_2 \bar Z_1)(x_2')\right]\rangle\) and the similar one with the fields \(Z_3\) and \(Z_4\). The result is
\begin{multline*}
    \langle\text{Tr}\left[(Z_1 Z_2)(x_1)(Z_1 Z_2)(x_2) \right]\text{Tr}\left[(\bar Z_2 \bar Z_1)(x_1')(\bar Z_2 \bar Z_1)(x_2')\right]\rangle \\
    =\xi_2^4 \iint \dd^d x''_1 \dd^d x''_2 D_{1}(x_1-x''_1) D_{2}(x_2-x''_2)(K(x''_1,x''_2|x'_1,x'_2) + K(x''_1,x''_2|x'_2,x'_1))\,,
\end{multline*}
where \(K(x,y|x',y')\) is given by \eqref{SD_2}. Therefore, for the considered correlator the problem is again equivalent to the diagonalisation of the T-operator \eqref{eq:RR}.}
\begin{multline}\label{symmetric sum}
   \left\langle \Tr[(Z_1 Z_2)(x_1)(Z_1 Z_2)(x_2)] \Tr[(\bar Z_4 \bar Z_3)(x_1')(\bar Z_4 \bar Z_3)(x_2')] \right\rangle= \\
   =K(x_1,x_2|x'_1,x'_2) + K(x_1,x_2|x'_2,x'_1)\,,
\end{multline}
with
\begin{equation}
\label{SD_2}
    K(x_1,x_2|x'_1,x'_2)= \xi_1^{4} \sum_{n=0}^{\infty} (\xi_1^2\xi_2^2)^{2 n}\,
T_{n+1}(x_{1},x_{2}|x_1',x_2')\,.
\end{equation}
According to the formulae~\eqref{T-matrix_operator_form} and \eqref{R-matrix}, the kernel of the operator $\widehat{T}$ at $L=2$ takes the form
\begin{align}\label{eq:RR}
    T(x_1,x_2|x_1',x_2')=c^2 \iint  &\frac{\dd^d x_0 \dd^d x_{0'}}{(x_{10}^2)^{-u-\frac{d}{2}+\Delta_+} (x_{01'}^2)^{u+\frac{d}{2}+\Delta_-} (x_{1'0'}^2)^{-u+\frac{d}{2}-\Delta_+} (x_{0'1}^2)^{u+\frac{d}{2}-\Delta_-}} \notag \\
    \times &\frac{1}{(x_{20'}^2)^{-u-\frac{d}{2}+\Delta_+} (x_{0'2'}^2)^{u+\frac{d}{2}+\Delta_-} (x_{2'0}^2)^{-u+\frac{d}{2}-\Delta_+ } (x_{02}^2)^{u+\frac{d}{2}-\Delta_-}}\,,
\end{align}
where $c$ is given in equation \eqref{c_coefficient}. We can compute \eqref{symmetric sum} by simply adapting to our problem the methods of ~\cite{Grabner:2017pgm,Gromov:2018hut,Kazakov:2018qbr}. The spectral equation
reads 
\begin{equation}\label{eq:eigenvEq}
    \iint \dd^d x_{1'} \dd^d x_{2'} T(x_1,x_2|x_1',x_2')\Psi_{\nu,S}(x_{1'},x_{2'};x_3)= h(\nu,S)\,\Psi_{\nu,S}(x_1,x_2;x_3)\,,
\end{equation}
where $\nu \in \mathbb{R}$ is the continuous label of principal series and $S$ is the spin (rank of symmetric traceless tensor). Because $L=2$, the eigenfunctions of $\widehat{T}$ are entirely determined by its conformal symmetry, hence $\Psi_{\nu,S}(x_{1'},x_{2'};x_3)$ is a conformal 3-point function between two scalars of dimension $\Delta_+ - \Delta_- = \Delta_1 + \Delta_2$, at points $x_1$ and $x_2$, and one symmetric traceless tensor of spin $S$ with dimension in the principal series $\Delta=d/2 + 2 \ii \nu$, inserted at an arbitrary point $x_3$~\cite{Dobrev:1977qv,Polyakov:1970xd,Fradkin:1978pp}. For simplicity, we restrict ourselves to $S=0$. The eigenfunction $\Psi_{\nu}\equiv\Psi_{\nu,0}$ has the form
\begin{equation}\label{3pt_function}
    \Psi_{\nu}(x_1,x_2;x_3) = C(\nu)\left(x_{12} ^2 \right)^{\frac{d}{4} + \ii\nu - \Delta_1 - \Delta_2}\left(x_{13}^2 x_{23} ^2 \right)^{-\frac{d}{4}-\ii\nu}\, ,
\end{equation}
where $C(\nu)$ is a normalisation constant. Let us compute the eigenvalue. First, we take the simplifying limit $x^2_3\to\infty$ on both sides of equation~\eqref{eq:eigenvEq}. This leads to
\beq\label{eq:eigenvEq2}
    \iint \dd^d x'_{1} \dd^d x'_{2} T(x_1,x_2|x_1',x_2')\left(x_{1'2'}^2 \right)^{\frac{d}{4} + \ii\nu - \Delta_1 - \Delta_2}= h(\nu) \times \left(x_{12}^2\right)^{\frac{d}{4}+\ii\nu -\Delta_1- \Delta_2}\,,
\eeq
whose left-hand side is depicted in figure~\ref{fig:actionT} (a). From such equation we read off the eigenvalue in the form of an integral to be evaluated
\begin{equation}\label{eq:T_eigenvalue}
    h(\nu) = \left(x_{12}^2\right)^{\Delta_1+\Delta_2-\frac{d}{4}-\ii\nu}  \iint \dd^d x'_{1} \dd^d x'_{2}\, T(x_1,x_2|x_1',x_2')\left(x_{1'2'}^2\right)^{\frac{d}{4}+\ii\nu -\Delta_1- \Delta_2}\,.
\end{equation}

\begin{figure}
\includegraphics[scale=0.85]{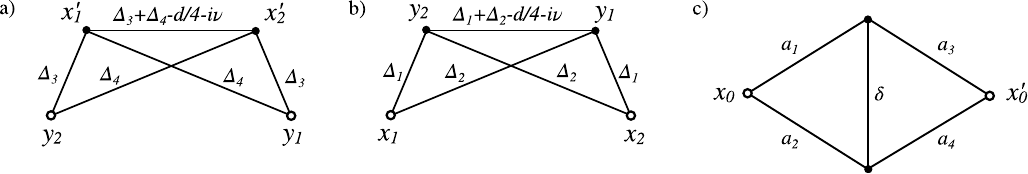}
\caption{(a), (b) integrals give respectively the factors $h_1(\nu)$ and $h_2(\nu)$ of the eigenvalue. (c) The two-loop massless master integral of \cite{Grozin:2012xi,Derkachev:2022lay}.}
\label{T_operator_eigenvalue_components}
\end{figure}

It is convenient to factor the eigenvalue into the product of two terms (see appendix \ref{app:eigenvalue} for details),
\beq\label{T_operator_eigenvalue1}
h(\nu) = h_1(\nu)h_2(\nu)\,,
\eeq
because $h_1(\nu)$ and $h_2(\nu)$ are in fact the same function
\begin{multline}\label{eq:B_Grozin}
B(a_1,a_2,\delta) = \frac{(x_{00'}^2)^{\delta+2a_1+2a_2-d}}{4^{2d-2a_1-2a_2}\pi^{2d}} \left(\frac{\Gamma(a_1)\Gamma(a_2)}{\Gamma\left(\frac{d}{2}-a_1\right)\Gamma\left(\frac{d}{2}-a_2\right)}\right)^2 \\
\times \iint\frac{\dd^d x_{1'} \dd^d x_{2'}}{(x_{1'2'}^2)^{\delta}(x_{01'}^2)^{a_1}(x_{1'0'}^2)^{a_2}(x_{0'2'}^2)^{a_1}(x_{2'0}^2)^{a_2}}\,,
\end{multline}
evaluated at different values of its parameters, i.e. 
\beq
h_1(\nu) = B\!\left(\Delta_1,\Delta_2,\Delta_3 + \Delta_4 - \frac{\Delta}{2}\right)\,, \quad h_2(\nu) = B\!\left(\Delta_3,\Delta_4,\Delta_1 + \Delta_2 - \frac{\Delta}{2}\right)\,.
\label{h12}
\eeq
The two factors of the eigenvalue of $\widehat{T}$ are schematically drawn in figure~\ref{T_operator_eigenvalue_components}. The function $B(a_1,a_2,\delta)$ can be computed by Mellin-space techniques~\cite{Derkachev:2022lay}. In fact, such function is the generalisation of a two-loop massless master integral to the case of propagators with complex dimensions, see~\cite{Grozin:2012xi}. Following \cite{Derkachev:2022lay}, one gets
\begin{equation}\label{B_integral}
B(a_1,a_2,\delta) = \frac{\Gamma\!\left(\frac{d}{2}-1\right) (\mathcal{I}_1+\mathcal{I}_2+\mathcal{I}_3)}{4^{2d-2a_1-2a_2}\pi^{d}\, \Gamma(d-2)\,  A_0(a_1)\, A_0(a_2)\, A_0(2a_1+2a_2+\delta-d)}\, ,
\end{equation}
where
\begin{equation}\label{A_0_function}
    A_0(a)=\frac{\Gamma\!\left(\frac{d}{2}-a\right)}{\Gamma(a)}
\end{equation}
and $\mathcal{I}_1$, $\mathcal{I}_2$, $\mathcal{I}_3$ are expressed as double infinite sums~\cite{Derkachev:2022lay}:

\begin{align}
     & \mathcal{I}_1=\sum\limits_{n,k=0}^{+\infty} M_n \frac{(-1)^k}{k!}\frac{\Gamma(a_2+n+k)}{\Gamma\left(\frac{d}{2}-a_2-k\right)}\frac{\Gamma\left(\frac{d}{2}+n-\delta+k\right)}{\Gamma(\delta-k)}\frac{\Gamma\left(a_1+\delta-\frac{d}{2}-k\right)}{\Gamma(d+n-a_1-\delta+k)} \notag \\
     & \times \frac{\Gamma(d+n-a_1-a_2-\delta+k)}{\Gamma\left(a_1+a_2+\delta-\frac{d}{2}-k\right)}\frac{\Gamma\left(\frac{d}{2}-a_1-a_2-k\right)}{\Gamma(n+a_1+a_2+k)}\frac{1}{\Gamma\left(\frac{d}{2}+n+k\right)}\,, \label{I1_sum} \\
     & \mathcal{I}_2=\sum\limits_{n,k=0}^{+\infty} M_n \frac{(-1)^k}{k!}\frac{\Gamma\left(n-\frac{d}{2}+a_1+a_2+\delta+k\right)}{\Gamma(d-a_1-a_2-\delta-k)}\frac{\Gamma\left(\frac{d}{2}-a_1-\delta-k\right)}{\Gamma(n+a_1+\delta+k)}\frac{\Gamma(n+a_1+k)}{\Gamma\left(\frac{d}{2}-a_1-k\right)} \notag \\
     & \times \frac{\Gamma\left(\frac{d}{2}+n-a_2+k\right)}{\Gamma(a_2-k)}\frac{\Gamma(d-2a_1-a_2-\delta-k)}{\Gamma(n-\frac{d}{2}+2a_1+a_2+\delta+k)}\frac{1}{\Gamma\left(\frac{d}{2}+n+k\right)}\,, \label{I2_sum} \\
     & \mathcal{I}_3=\sum\limits_{n,k=0}^{+\infty} M_n \frac{(-1)^k}{k!}\frac{\Gamma\left(\frac{d}{2}+n-a_1+k\right)}{\Gamma(a_1-k)}\frac{\Gamma\left(a_1+a_2-\frac{d}{2}-k\right)}{\Gamma(d+n-a_1-a_2+k)}\frac{\Gamma(d+n-a_1-a_2-\delta+k)}{\Gamma(-\frac{d}{2}+a_1+a_2+\delta-k)} \notag \\
     & \times \frac{\Gamma(2a_1+a_2+\delta-d-k)}{\Gamma\left(\frac{3d}{2}+n-2a_1-a_2-\delta+k\right)}\frac{\Gamma\left(\frac{3d}{2}+n-2a_1-2a_2-\delta+k\right)}{\Gamma(2a_1+2a_2+\delta-d-k)}\frac{1}{\Gamma\left(\frac{d}{2}+n+k\right)}\,,\label{I3_sum}
\end{align}
where
\begin{equation}
    M_n=\frac{\Gamma(n+d-2)}{n!}\left(n+\frac{d}{2}-1\right)\, .
\end{equation}
For generic parameters, one of the two sums in each $\mathcal{I}_k$ can always be computed in terms of hypergeometric functions. Further simplifications may occur for specific values of the parameters \(a_1,a_2\) and \(d\). We will study some of those in the next section.

The insertion of a resolution of the identity in \eqref{SD_2} reads
\begin{equation}
\label{SD_2_basis}
    K(x_1,x_2|x'_1,x'_2)=  \sum_{S=0}^{+\infty}\int \dd \rho(\nu,S) \int \dd^d x_3\,
{\Psi}_{\nu,S}(x_1,x_2;x_3) \bar{\Psi}_{\nu,S}(x_1',x_2';x_3) \frac{\xi_1^{4}\,h(\nu,S)}{1-\xi_1^4\xi_2^4 \,h(\nu,S)}.
\end{equation}
where $\rho(\nu,S)$ is the Fourier--Plancherel measure over principal series, see~\cite{Dobrev:1977qv,Fradkin:1990dk}. The integral over $x_3$ is a conformal partial wave (CPW). Therefore, performing the integration over $\nu$ via the residue theorem yields the CPW expansion of \eqref{symmetric sum} in the $s$-channel
(upon symmetrisation $x_1 \leftrightarrow x_2$). It can be shown that the residues of the poles coming from the measure and the CPW cancel against each other, hence only the poles in the dynamical part of the integrand contribute. For the choice $S=0$ such poles are located at the solutions of
\begin{align}\label{Dim2}
    h(\nu)=\frac{1}{\xi_1^4\xi_2^4}\,.
\end{align}
In particular we are interested in the solution of \eqref{Dim2} with tree-level value
\begin{equation}
\label{boundary}
2 i \nu\left. \right|_{\xi_1=\xi_2=0} = -d/2+2(\Delta_1+\Delta_2)\,,
\end{equation}
describing the operators that dominate the OPE limit $x_{1}\to x_{2}$, that is the lightest ones. Notice that the analogue of \eqref{Dim2} for the short operator $\text{Tr}[\bar Z_3 \bar Z_4 \bar Z_3 \bar Z_4]$ is obtained upon the replacement $\Delta_1,\Delta_2 \to \Delta_3,\Delta_4$ in \eqref{h12}, and setting the tree-level value to $-d/2+2(\Delta_3+\Delta_4)$ instead of \eqref{boundary}. This fact is coherent with the symmetry $(Z_i,\Delta_i) \to (\bar Z_{i+2}, \Delta_{i+2})$ of \eqref{CheckerboardCFT}.

In the next section we consider reductions of the Checkerboard CFT to some particular cases of Fishnet theory, and study the solutions of \eqref{Dim2} therein.

\section{Reductions of Checkerboard CFT to Familiar FCFTs}
\label{sec:reductions}

The Lagrangian \eqref{CheckerboardCFT} can be specified to particular cases of FCFTs already defined in the literature. In this respect, a first example is the ABJM Fishnet theory which is obtained as double-scaling limit of the \(\gamma\)-twisted ABJM theory in $3d$~ \cite{Caetano:2016ydc}. Another interesting instance, now in \(2d\), is what we dub ``BFKL Fishnet Theory"  which, is related, through specific physical quantities, to the Balitsky--Fadin--Kuraev--Lipatov model for the scattering of partons in the Regge limit of QCD and $\mathcal{N}=4$ SYM \cite{Kuraev:1977fs,Balitsky:1978ic,Lipatov:1985uk,Fadin:1998py,Kotikov:2000pm,Kotikov:2002ab}.

Let us note from the very beginning that the  $4d$  bi-scalar FCFT
\begin{align}\label{bi-scalar}
    {\cal L}_{d}^{(CB)} = N \Tr[\bar{X} (-\partial_{\mu}\partial^{\mu})^{-u-\Delta_-}X+\bar{Z}(-\partial_{\mu}\partial^{\mu})^{-u+\Delta_-}Z- \xi^2  X Z  \bar{X} \bar{Z}]\,, 
\end{align}
which originally was deduced as a particular case of the double scaling limit of ${\cal N}=4$  SYM theory~\cite{Gurdogan:2015csr},  is not contained in this Checkerboard theory.\footnote{ On the other hand, this bi-scalar model can be viewed as a reduction of the Loom FCFT with \(M=4\) or even with \(M=3\) slopes, defined in section~3.2 of~\cite{Kazakov:2022dbd}, to the case of $M=2$ slopes. For example,  we can take in the latter theory only one non-zero coupling -- for  the vertex  \(\text{Tr}\left(X u \bar{X} \bar u \right) \)  out of 4-valent vertices of eq.(15) in~\cite{Kazakov:2022dbd}. The fields \(X\) and \(u\) then decouple from the rest of the interactions and we obtain the \(d\)-dimensional anisotropic  generalisation of bi-scalar FCFT of~\cite{Kazakov:2018qbr} (renaming \(u\to Z\)) -- a generalisation of~\eqref{bi-scalar}.  }

\subsection{FCFT with regular triangular graphs and ABJM reduction }\label{sec:ABJMFCFT}

\begin{figure}
\begin{center}
\includegraphics[scale=0.38]{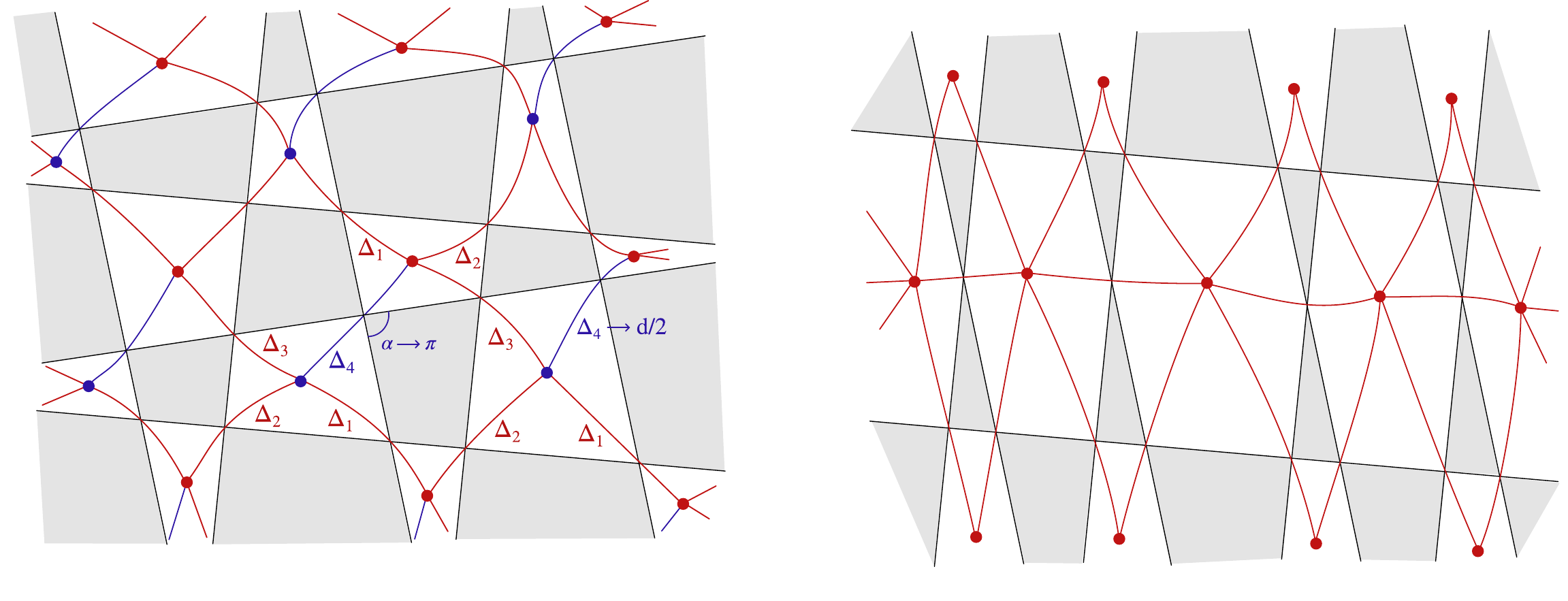}
\caption{Checkerboard CFT at $w_4=0$: The ultralocal gaussian integration over the field $Z_4$ leads to the effective shrinking of the propagator of dimension $\Delta_4=d/2$ to a point (depicted in blue). The Feynman graphs shaped as a regular square lattice of Checkerboard CFT reduce to the triangular graphs  describing a  theory of 3 complex scalars with sextic interaction, cf. \eqref{regtrig}.}
\label{fig:Integration_Z4}
\end{center}
\end{figure}

An interesting choice for the parameters of \eqref{CheckerboardCFT} is given by $w_4=-u-\Delta_-=0$.
In that case, as long as we study operators that do not depend on $Z_4$, the path integral over this field is purely Gaussian and can be performed:
\begin{equation}
    \int \mathrm{e}^{-\int \Tr[\bar{Z}_4 Z_4 - \xi_1^2 \bar{Z}_1 \bar{Z}_2 Z_3 Z_4 - \xi_2^2 Z_1 Z_2 \bar{Z}_3 \bar{Z}_4] \dd^d x} \mathcal{D}Z_4 \mathcal{D}\bar{Z}_4 \propto \mathrm{e}^{\int \xi_1^2 \xi_2^2 \Tr[\bar{Z}_3 \bar{Z}_1 \bar{Z}_2 Z_3 Z_1 Z_2] \dd^d x}
\end{equation}
where the proportionality constant is independent of the fields, hence irrelevant. For the associated Loom (Baxter) lattice this means that the angle between two types of lines (say, blue and black lines on the  figure~\ref{fig:Integration_Z4}) becomes zero. Accordingly, the Feynman diagrams change shape from a square lattice to a regular triangular lattice, as illustrated in figure~\ref{fig:Integration_Z4}.

The Checkerboard CFT then turns into the anisotropic $d$-dimensional FCFT~\cite{Kazakov:2018qbr}
\begin{multline}\label{regtrig}
    {\cal L}_{d}^{(CB)} = N \Tr\left[\bar{Z}_1 (-\partial_{\mu}\partial^{\mu})^{u+d-\Delta_+}Z_1+\bar{Z}_2(-\partial_{\mu}\partial^{\mu})^{-2u}Z_2+\bar{Z}_3(-\partial_{\mu}\partial^{\mu})^{u+\Delta_+}Z_3 \right. \\
    \left.- \xi^2 \bar{Z}_3 \bar{Z}_1 \bar{Z}_2 Z_3 Z_1 Z_2\right]\,,
\end{multline}
where $\xi^2=(\xi_1\xi_2)^2$. We regard this theory as a $d$-dimensional generalisation of the $3d$ FCFT obtained in~\cite{Caetano:2016ydc} from the ABJM theory in the double-scaling of large imaginary twist angle and weak coupling (up to the exchange of $Z_3$ with $\bar{Z}_3$). This latter theory is recovered from \eqref{regtrig} at the point $d=3$, $u=-1/2$, and $\Delta_+=3/2$ or, equivalently, $\Delta_1 = \Delta_2 = \Delta_3 = 1/2$.
Below we will study the $4$-point function \eqref{correlator} for \(L=2\) in this theory, focusing on the spectrum of lightest single-trace, local operators exchanged in its OPE $s$-channel.

\subsubsection{ABJM \texorpdfstring{$L=2$}{L=2} Fishnet}

Here we shall present the explicitly $h(\nu)$, i.e. \(h_1(\nu)\) and \(h_2(\nu)\) defined in equation \eqref{h12}, for the ABJM FCFT, namely
\beq
h_1(\nu) = B\!\left(\frac{1}{2}, \frac{1}{2}, 2 - \frac{\Delta}{2}\right)\,, \quad h_2(\nu) = B\!\left(\frac{1}{2}, \frac{3}{2},1 - \frac{\Delta}{2}\right)\, ,
\eeq
where \(\Delta=3/2 + 2\ii\nu\) for \(d=3\).
The calculation of \(h_2\) is trivial in this particular case; starting from the integral representation \eqref{eq:B_Grozin} and using the following representation of Dirac's $\delta$ function,
\begin{equation}
    \lim_{a_2 \rightarrow \frac{d}{2}} \frac{\Gamma(a_2)}{\Gamma\left(\frac{d}{2} - a_2\right) x^{2a_2}} = \pi^{\frac{d}{2}} \delta^{(d)}(x)\,,
\end{equation}
one can easily see that
\begin{equation}\label{h2_d3}
    h_2(\nu) = \frac{1}{16\pi^2}\,.
\end{equation}

We now turn our attention to $h_1$. Computing the prefactor in front of $(\mathcal{I}_{1}+\mathcal{I}_{2}+\mathcal{I}_{3})$ from equation \eqref{B_integral} we obtain
\begin{equation}\label{prefactor_h1}
h_1 = \frac{\mathcal{I}^{(1)}_1+\mathcal{I}^{(1)}_2+\mathcal{I}^{(1)}_3}{256\sqrt{\pi}\sin(\frac{\pi\Delta}{2}) \Gamma\!\left(\frac{\Delta}{2}\right) \Gamma\!\left(\frac{\Delta}{2}+\frac{1}{2}\right)}\,,
\end{equation}
where we denote by the upper index \((1)\) the sums \(\mathcal{I}_j\), \(j=1,2,3\), corresponding to \(h_1\). Combining \eqref{h2_d3} with the prefactor from \eqref{prefactor_h1} yields
\begin{equation}\label{spectralE}
h = \frac{2^{\Delta-13} \left(\mathcal{I}^{(1)}_1+\mathcal{I}^{(1)}_2+\mathcal{I}^{(1)}_3\right)}{\pi^3 \sin\!\left(\frac{\pi\Delta}{2}\right) \Gamma(\Delta)}\,.
\end{equation}
The formula \eqref{spectralE} suggests the convenient re-scaling of the sums $\mathcal{I}_{1,2,3}^{(1)}$ by a factor
\begin{equation}
    \textbf{I}_j = \frac{2^{\Delta-13}\, \mathcal{I}^{(1)}_j }{\pi^3 \sin\!\left(\frac{\pi\Delta}{2}\right) \Gamma(\Delta)}\,, \quad j=1,2,3\,.
\end{equation}
We can proceed with the calculation of the sums \(\textbf{I}_{1,2,3}\). We notice that the sums over $k$ in $\textbf{I}_1$ and $\textbf{I}_2$ are drastically simplified by $k=0$ term only. This is due to the factors \(\Gamma(d/2-a_2-k)=\Gamma(1-k)\) in the denominator of \eqref{I1_sum} and \(\Gamma(d/2-a_1-k)=\Gamma(1-k)\) in the denominator of \eqref{I2_sum}. By taking this into account, we compute the first two double sums explicitly
\begin{align}
\textbf{I}_1 & =
\frac{\cot\frac{\pi\Delta}{2}}{1024\pi^3 (\Delta-1)(\Delta-2)}\frac{1}{\Gamma\left(\frac{\Delta}{2}\right)}\sum\limits_{n=0}^{+\infty}
\frac{\Gamma\left(n+\frac{\Delta}{2}\right)}{n+\frac{\Delta}{2}-\frac{1}{2}}\frac{1}{n!} \label{I1_sum_d3} \\
& =\frac{\cos\frac{\pi\Delta}{2}\, \Gamma\left(\frac{\Delta}{2}-\frac{1}{2}\right)}{1024\pi^{\frac{5}{2}}(\Delta-1)(\Delta-2)\sin^2\frac{\pi\Delta}{2}\, \Gamma\!\left(\frac{\Delta}{2}\right)} \notag \\
\textbf{I}_2 & =\frac{1}{1024\pi^{3}(\Delta-1)(\Delta-2)\sin\frac{\pi\Delta}{2}}\frac{1}{\Gamma\left(\frac{\Delta}{2}\right)}\sum\limits_{n=0}^{+\infty}\frac{n!}{\left(\frac{3}{2}-\frac{\Delta}{2}+n\right)\Gamma\left(2-\frac{\Delta}{2}+n\right)} \label{I2_sum_d3} \\
& =\frac{1}{256\pi^{4}(\Delta-1)(\Delta-2)^2 (\Delta-3)}\,_3F_2\left(1,1,\frac{3}{2}-\frac{\Delta}{2};2-\frac{\Delta}{2},\frac{5}{2}-\frac{\Delta}{2};1\right)\,. \notag
\end{align} 
The last sum \(\mathcal{I}^{(1)}_3\) \eqref{I3_sum} is more complicated, as no truncation of infinite sums occurs there. However, after performing the sum in $k$, the summand in \(n\) can be also presented in terms of hyper-geometric functions. We obtain
\begin{align}\label{I3_sum_d3}
\textbf{I}_3 & = \frac{2^{\Delta-13}}{\pi^{4} \Gamma(\Delta)} \sum\limits_{k=0}^{+\infty}\frac{1}{k!}\frac{\Gamma\left(\frac{\Delta}{2}+k\right)}{\left(\frac{1}{2}+k\right)\left(\frac{\Delta}{2}-\frac{1}{2}+k\right)} \sum\limits_{n=0}^{+\infty}\frac{\left(\frac{1}{2}+n\right)\Gamma\left(\frac{\Delta}{2}+\frac{1}{2}+k+n\right)}{(1+k+n)\left(\frac{\Delta}{2}+k+n\right)\Gamma\left(\frac{3}{2}+k+n\right)}\\
& =\frac{2^{\Delta-12}}{\pi^{4}(\Delta-2) \Gamma (\Delta)} \sum\limits_{k=0}^{+\infty}\frac{1}{k!}\frac{\Gamma\left(\frac{\Delta}{2}+\frac{1}{2}+k\right)}{\Gamma\left(\frac{3}{2}+k\right)} \notag \\
& \times \left(\frac{1}{\left(\frac{1}{2}+k\right)\left(\frac{\Delta}{2}+k\right)}\,_3F_2\left(1,\frac{\Delta}{2}+k,\frac{\Delta}{2}+\frac{1}{2}+k;\frac{\Delta}{2}+1+k,\frac{3}{2}+k;1\right) \right. \notag \\
& \left. -\frac{1}{(1+k)\left(\frac{\Delta}{2}-\frac{1}{2}+k\right)}\,_3F_2\left(1,1+k,\frac{\Delta}{2}+\frac{1}{2}+k;2+k,\frac{3}{2}+k;1\right)\right)\,. \notag
\end{align}

The equation for the spectrum now reads
\begin{equation}\label{hxi}
h=\textbf{I}_1+\textbf{I}_2+\textbf{I}_3=\frac{1}{\zeta}\,,\qquad  \zeta=(\xi_1\xi_2)^2.
\end{equation}

In the following subsection, we will analyse these formulae in the limit of weak coupling~$\zeta$. We will also present the numerical plot $\Delta(\zeta)$ stemming from~\eqref{hxi}.

\subsubsection{Perturbation theory: \texorpdfstring{$\gamma$}{gamma}-expansion for the spectrum}
In order to solve \eqref{hxi} we need to expand the RHS of \eqref{spectralE} for small values of $\gamma$ and then invert the series. That is, one needs to expand the sums \(\textbf{I}_1\), \(\textbf{I}_2\) and \(\textbf{I}_3\) with \(\Delta=2+\gamma\). The details of the calculation are given in appendix~\ref{app:small_gamma_expansion}. We obtain
\begin{align}\label{WC_I1I2I3}
    \textbf{I}_1+\textbf{I}_2+\textbf{I}_3 &= -\frac{1}{1024\pi^{2}\gamma}+\frac{1}{1024\pi^{4}}\left(\pi^2+\pi^2 \log 2-\frac{21}{2}\zeta_3\right) \\
    & -\frac{1}{1024\pi^4}\left(\pi^2+\pi^2 \log 2-\frac{21}{2}\zeta_3+\frac{\pi^4}{40}+\frac{\log^4 2}{2}+12\Li_4\!\left(\frac{1}{2}\right)\right)\gamma+\mathcal{O}(\gamma^2)\,. \notag
\end{align}
Plugging the latter expansion into formula \eqref{hxi}, we find
\begin{align}\label{PTgamma}
    h=-\frac{a}{\gamma}+b-c\gamma+\mathcal{O}(\gamma^2)\,,
\end{align}
where
\begin{align}
    a & =\frac{1}{1024 \pi^2}\,, \\ 
    b & =\frac{1}{1024\pi^{4}}\left(\pi^2+\pi^2 \log 2-\frac{21}{2}\zeta_3\right)\,, \notag \\
    c & = \frac{1}{1024\pi^4}\left(\pi^2+\pi^2 \log 2-\frac{21}{2}\zeta_3+\frac{\pi^4}{40}+\frac{\log^4 2}{2}+12\Li_4\!\left(\frac{1}{2}\right)\right)\,.
\end{align}
We notice that this result hides features of uniform transcendentality. Indeed, if one defines a new function
\begin{equation}
\label{toexpand}
    \tilde{h}=\gamma(1+\gamma)h\,,
\end{equation}
then the coefficient of its expansion at small $\gamma$ at order $n$ has a fixed weight $n$ with respect to the basis of numbers $\{\mathfrak{L}_j=\text{Li}_j(1/2)\}$, with $j \in \mathbb{N}$.
Let us first expand \eqref{toexpand}, that is
\begin{align}
    \tilde{h} &= -a+(b-a)\gamma+(b-c)\gamma^2+\mathcal{O}(\gamma^3) \\
    &=-\frac{1}{1024\pi^4}\left(\pi^2-\left(\pi^2 \log 2-\frac{21}{2}\zeta_3\right)\gamma+\left(\frac{\pi^4}{40}+\frac{\log^4 2}{2}+12\Li_4\!\left(\frac{1}{2}\right)\right)\gamma^2+\mathcal{O}(\gamma^3)\right)\,. \notag
\end{align}
Next, we express the expansion in terms of solely \(\mathfrak{L}_j\). Order by order, the replacement reads
\begin{align}\label{constants_to_polylogarithms}
    & \log 2=\mathfrak{L}_1\,, \\
    & \pi^2 = 12\,\mathfrak{L}_2+6\,\mathfrak{L}_1^2\,, \notag \\
    & \zeta_3=\frac{8}{7}\,\mathfrak{L}_3 +\frac{8}{7}\,\mathfrak{L}_2 \,\mathfrak{L}_1+\frac{8}{21}\,\mathfrak{L}_j^3 \,. \notag
\end{align}
and finally the expansion reads
\begin{multline}
    \tilde{h}=-\frac{1}{1024\pi^4} \Bigg[12\mathfrak{L}_2+6\mathfrak{L}_1^2+\left(12\mathfrak{L}_3-2\mathfrak{L}_1^3\right)\gamma \\
    +\left(12\mathfrak{L}_4+\frac{18}{5}\mathfrak{L}_2^2+\frac{18}{5}\mathfrak{L}_2\mathfrak{L}_1^2+\frac{7}{5}\mathfrak{L}_1^4\right)\gamma^2+\mathcal{O}(\gamma^3)\Bigg]\,,
\end{multline}
from which the transcendentality at each order in \(\gamma\) is explicit. It would be interesting to check this feature at all orders of perturbation theory.
Now, the spectral equation \eqref{Dim2} reads 
\begin{equation}
     -\frac{a}{\gamma }+b\,-\,c\gamma\,+\,{\cal O}\left(\gamma^2 \right)\,=\frac{1}{\zeta}\,.
\end{equation}
and inverting the series we extract the anomalous dimension
\begin{equation}\label{gamma_weak_coupling}
    \gamma=-a\zeta-ab\zeta^2-a(b^2+ac)\zeta^3+{\cal O}\left(\zeta^4\right)
    \,,
\end{equation}
The leading term is straightforward to reproduce from a single Feynman diagram computations. After a convenient redefinition of the coupling, $\zeta = 1024 \pi^2 \eta$, the perturbative anomalous dimension reads:
    \begin{align}\label{gamma_new_coupling}
        \gamma &= -\eta-\left(1+\log 2-\frac{21\zeta_3}{2\pi^2}\right)\eta^2-\left(2+3\log 2-\frac{63\zeta_3}{2\pi^2}+\right. \\
        & \left.+\frac{\pi^2}{40}-\log^2 2+\frac{\log^4 2}{2\pi^2}-\frac{21\zeta_3 \log 2}{\pi^2}+\frac{441\zeta_3^2}{4\pi^4}+\frac{12\Li_4\!\left(\frac{1}{2}\right)}{\pi^2}\right)\eta^3+\mathcal{O}\left(\eta^4\right)\,. \notag
    \end{align}
This result \eqref{gamma_new_coupling} can be compared with the numerical calculations of the following section.

\subsubsection{Numerics and comparison with perturbation theory}\label{sec:numerics}

\begin{figure}[ht]
\begin{center}
\includegraphics[scale=0.9]{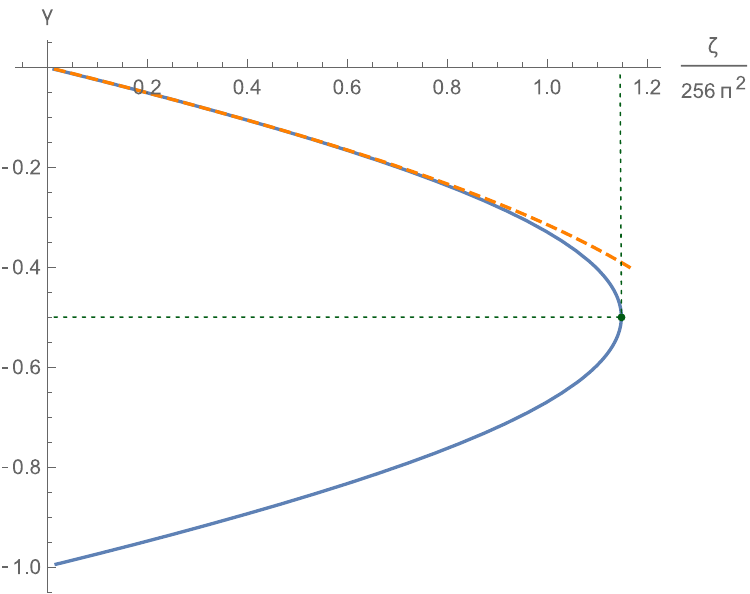}
\caption{The blue line plots the numerical anomalous dimension $\gamma=\Delta-2$ as a function of the effective coupling $\zeta=\xi_1^2\xi_2^2$. We evaluated the r.h.s. of \eqref{hxi} at the points $\gamma=-0.05k,\,\,k=1,\dots,19$ and interpolated it via $\texttt{Interpolation}$ in {\it Mathematica}. The orange dashed line is the plot of perturbation theory up to order $\zeta^3$, that is \eqref{PTgamma}. The numerical evaluation of the double sum~\eqref{I3_sum_d3} has been performed after summing over $n$ analytically, and using the $\texttt{WynnEpsilon}$ attribute option for $\texttt{NSum}$. The function $\gamma(\zeta)$ features two branches that meet at $\zeta_{crit}\simeq 2903.03$, since the curve has a maximum at $\gamma_{crit}=-0.5$, marked on the plot. The value $\zeta_{crit}$ is the branch point of $\gamma(\zeta)$, from where two complex conjugate dimensions emerge when $\zeta>\zeta_{crit}$.}
\label{Numerics_ABJM_fishnet}
\end{center}
\end{figure}

We present a numerical plot for the function $\gamma(\zeta)$ in figure~\ref{Numerics_ABJM_fishnet}, and we compare it with the plot of its perturbative expansion at order $\zeta^3$. The two curves show good agreement for $\gamma < 0.7$, and they start to significantly diverge around $\gamma\sim 1$. The numerical curve features two real branches of the same function -- hence two values of $\gamma$ for a given coupling $\zeta$. The two branches span the intervals $\gamma \in (0,-1/2)$ and $\gamma \in (-1/2,-1)$, are separated at the branch-point $\gamma_{crit}=-1/2$ and are symmetric under reflection w.r.t. $\gamma_{crit}$. This point corresponds to the coupling $\zeta_{crit}\simeq 2903.03$. For $\zeta>\zeta_{crit}$ the real part of $\gamma$ is single valued, but the function develops two branches given by complex conjugate dimensions, $\gamma(\zeta)$ and $\bar\gamma(\zeta)$, similarly to the observations of \cite{Gromov:2017cja,Grabner:2017pgm,Gromov:2018hut}.
The behavior of $\zeta(\gamma)$ for $\gamma \sim \gamma_{crit}$ is $$\zeta\simeq\zeta_{crit}-\frac{1}{C^2} (\gamma+\frac{1}{2})^2\,,$$ so that for $\zeta \sim \zeta_{crit}$ one has $\gamma\simeq -\frac{1}{2}\pm i C\sqrt{\zeta-\zeta_{crit}}$, where $C$  is a numerical constant. This  complexity is a typical feature of the non-unitary, but ``tT-invariant" Fishnet CFTs~\cite{Kazakov:2022dbd}.

\subsection{FCFT of BFKL Type From the Checkerboard}
\label{sec:BFKLlimit}

In this section we will consider $d=2$, for which the operator $\widehat{T}$ is the transfer matrix of an $SO(1,3)\sim SL(2,\mathbb{C})$ spin chain with sites in the representation of scaling dimension $\Delta_1+\Delta_2$. Our purpose is to relate Fishnet theories and the  BFKL model describing QCD amplitudes in the multi-Regge kinematics. Practically, we want to define a family of two-dimensional Checkerboard CFTs labelled by the spectral parameter $u$, such that FCFT correlators are related to certain expectation values in the BFKL model.
For that, we should identify the spin chain that describes the Checkerboard lattice with the one underlying Lipatov's Hamiltonian, that is to set $\Delta_1+\Delta_2=0$. Furthermore, without the loss of generality we choose $\Delta_0=0$, hence in eqs.~\eqref{CheckerboardCFT},~\eqref{eq:Props} one has $\Delta_{\pm}=0$, and the scaling dimensions of the fields $Z_k$ read
\begin{equation}\label{BFKLchoice}
    \Delta_1 =-1-u\,,\,\,\Delta_2 =1+u\,,\,\,\Delta_3 =1-u\,,\,\,\Delta_4 =1+u\,.
\end{equation}
The corresponding Lagrangian reads
\begin{align}\label{BFKLCFT}
{\cal L}_{d} = N \Tr\!\Big[&\bar{Z}_1 (-\bar \partial\partial)^{u+2}Z_1 + \bar{Z}_2(-\bar \partial\partial)^{-u}Z_2 + \bar{Z}_3(-\bar \partial\partial)^{u}Z_3 + \bar{Z}_4(-\bar \partial\partial)^{-u}Z_4  \notag\\
 &- \xi_1^2 \bar{Z}_1 \bar{Z}_2 Z_3 Z_4 - \xi_2^2 Z_1 Z_2 \bar{Z}_3 \bar{Z}_4\Big]\,,
\end{align} 
where we made use of holomorphic/antiholomorphic coordinates $(z,\bar z)=(x_0\pm i x_1)$.
For the choice \eqref{BFKLchoice} of scaling dimensions, the R-matrix \eqref{R-matrix} introduced in~\cite{Chicherin:2012yn}, reduces  now to the following form
\beq\label{R-matrix_BFKL_coord}   R(x_1,x_0|x'_{1},x'_{0})\,=\,\frac{c}{(x_{10}^2)^{-u-1}(x_{01'}^2)^{u+1} (x_{1'0'}^2)^{-u+1}(x_{0'1}^2)^{u+1}}\,,
\eeq
that is, in operator form
\begin{equation}\label{R-matrix_BFKL_oper}
    \widehat{R}_{10}^{BFKL} = \frac{\Gamma(-1-u)\Gamma(1-u)}{4^{2+2u}\pi^2 \Gamma(2+u)\Gamma(u)}\mathbb{P}_{01}(x^2_{10})^{u+1}(p^2_{0})^{u}(p^2_{1})^{u}(x_{10}^2)^{u-1}\, ,
\end{equation}
where $\mathbb{P}_{01}$ is the permutation operator between the two spaces. As was noticed in~\cite{Chicherin:2012yn}, the Taylor-expansion of $\widehat{R}$ around \(u=0\) delivers at linear order a differential operator
\begin{align}
\widehat{R}_{ab}^{BFKL}&=\frac{\Gamma(-1-u)\Gamma(1-u)}{4^{2+2u}\pi^2 \Gamma(2+u)\Gamma(u)}\mathbb{P}_{ab}(x_{ab}^2)^{u+1}(\hat{p}_b^2)^{u}(\hat{p}_a^2)^{u}(x_{ab}^2)^{u-1}\, \label{hbfkl} \\
&=\frac{\mathbb{P}_{ab}}{16\pi^2}\left(1+u\,\hat h_{ab}^{BFKL}+\mathcal{O}(u^2)\right)\,, \notag
\end{align}
whose explicit form is
\begin{align}\label{Hbfklab}
    \hat h_{ab}^{BFKL} & = 2\log(x^2_{ab})+x^2_{ab}\log(p^2_{a}p^2_{b})x^{-2}_{ab}-4\psi(1)-4\log 2-2 \\
    & =(p_{a}^{-2})\log(x^2_{ab})(p_{a}^2) +(p_{b}^{-2})\log(x^2_{ab})(p_{b}^2) + \log(p^2_{a}p^2_{b}) - 4\psi(1) - 4\log 2 - 2\,. \notag
\end{align}
Rewriting the formula \eqref{Hbfklab} in terms of holomorphic coordinates, we obtain
\begin{equation}\label{h_BFKL_hol_sep}
    \hat h_{ab}^{BFKL}=h^{z}_{ab}+h^{\bar{z}}_{ab}-2\,,
\end{equation}
where
\begin{equation}\label{BFKL_hol}
    h^{z}_{ab}=2\log z_{ab}+z_{ab}\log(p^z_a p^z_b)z_{ab}^{-1}-2\psi(1)
\end{equation}
and $-4\log 2$ disappeared because of the relation between ordinary and holomorphic momentum operators $p_a^2=4p^{z}_a p^{\bar{z}}_a$. The expression \eqref{BFKL_hol} coincides with the holomorphic part of the Hamiltonian density for Lipatov's \(SL(2,\mathbb{C})\) chain of reggeized gluons~\cite{Lipatov:1993yb,Lipatov:1993qn,Faddeev:1994zg,DeVega:2001pu,Derkachov:2001yn,Derkachov:2002wz,Derkachov:2002pb}. Therefore, in the case of $L=2$ we know the eigenvalue of the Hamiltonian
\begin{equation} \label{Hbfkl}
  \widehat{H}^{BFKL} = \sum_{a=1}^L\hat h_{a,a+1}^{BFKL}\,, \quad (\hat h_{L,L+1}\equiv\hat h_{L,1})\,,
\end{equation}
which is otherwise described as the logarithmic derivative of the operator \(\widehat T\) of equation \eqref{T-matrix_operator_form} (with the R-matrix \eqref{R-matrix_BFKL_oper}) at the point \(u=0\),
\begin{equation}
    \widehat{H}^{BFKL} = \widehat{T}(0)^{-1} \frac{\dd \widehat{T}}{\dd u} (0)\,.
\end{equation}


It would be tempting to formulate a BFKL Fishnet theory via a Lagrangian, as a $\log$-derivative of \eqref{BFKLCFT} around $u=0$. However, such point is too singular so we simply assume that  calculating  physical quantities we first use the Lagrangian  \eqref{BFKLCFT} at $u \neq 0$, then extract the associated quantity in the BFKL model from the $u=0$ expansion. That is, for the practical purposes in doing the calculations in this BFKL limit, such as for the correlators of the type~\eqref{correlator}, the graph-building operator formalism developed in the previous section seems to suit better. Similar $4$-point functions and related $2$-dimensional integrals were studied in \cite{Dotsenko:1984ad,Neretin:2022pbv,Derkachov:2022ytx}.

\subsubsection{Pomeron spectrum in BFKL limit of Checkerboard CFT}

Here we concentrate on the BFKL FCFT defined in the subsection~\ref{sec:BFKLlimit} and analyse in this framework the Checkerboard 4-point correlation function \eqref{symmetric sum}, as well as the dimension of the lightest exchanged operator.

The eigenvalue of the operator \(\widehat{T}\) with the weights
\begin{equation}
\label{BFKL_weights}
    \Delta_1=-1-u+\Delta_+\,, \quad \Delta_2=1+u\, \quad \Delta_3=1-u-\Delta_+\,, \quad \Delta_4=1+u\, ,
\end{equation}
is given by \eqref{T_operator_eigenvalue1} and \eqref{h12}. When $d=2$, the function $B$, defined in \eqref{eq:B_Grozin}, is now equal to
\begin{equation}\label{B_function_d=2}
    B(a_1,a_2,\delta)=\frac{(I_1+I_2+I_3)(a_1,a_2,\delta)}{4^{4-2a_1-2a_2}\pi^2 A^2_0(a_1) A^2_0(a_2)} \, ,
\end{equation}
where \(I_{1,2,3}(a_1,a_2,\delta)\) are expressed in terms of generalised hypergeometric functions \cite{Derkachev:2022lay}, see appendix \ref{BFKLdoublesums}, and we recall that \(A_0(a)\) is defined by \eqref{A_0_function}. Notice that we have momentarily kept $\Delta_+\neq 0$ because some of the functions $I_i$ are singular in the limit $\Delta_+\to 0$. The sum $I_1+I_2+I_3$ is however finite, as is shown in appendix \ref{BFKLdoublesums}. We do not have an explicit formula for it for arbitrary $u$ but we can expand it around $u=0$.
The eigenvalue is then
\begin{align}
    h(\Delta) &= \lim_{\Delta_+\to 0} B\left(-1-u+\Delta_+,1+u,2-\Delta_+ - \frac{\Delta}{2}\right) B\left(1-u-\Delta_+,1+u,\Delta_+-\frac{\Delta}{2}\right)\\
    &= \frac{u^4}{256\pi^8(1+u)^4} \lim_{\Delta_+\to 0} (I_1+I_2+I_3)\left(-1-u+\Delta_+,1+u,2-\Delta_+ - \frac{\Delta}{2}\right)\\
    &\ \,\qquad\qquad\qquad\times \lim_{\Delta_+\to 0}(I_1+I_2+I_3)\left(1-u-\Delta_+,1+u,\Delta_+-\frac{\Delta}{2}\right)\\
    &= \frac{1}{256\pi^4}\left[1+4u\left(-1-2\psi(1)+\psi\left(\frac{\Delta}{2}\right)+\psi\left(1-\frac{\Delta}{2}\right)\right)+\mathcal{O}\left(u^2\right)\right]\,. \label{hBFKL}
\end{align}
Writing \(\Delta=1+2i\nu\), the latter expression coincides with the energy of the Pomeron state, obtained in the Regge limit of QCD 
or in \({\cal N}=4\) SYM theory by~\cite{Kuraev:1977fs,Balitsky:1978ic} up to a constant coming from the equation \eqref{h_BFKL_hol_sep}
\begin{equation}
\omega(\nu)=4\left(2\psi(1)-\psi\left(\frac{1}{2}+i\nu\right)-\psi\left(\frac{1}{2}-i\nu\right)\right)\,.
\end{equation}
This establishes the direct link between the Checkerboard theory and the BFKL limit of QCD. Various physical quantities for both models, such as the graph-building operator appearing in the work~\cite{Kazakov:2018qbr}~(eq.(7) at \(d=2,\xi\to 0\)),  or Lipatov's Hamiltonian \eqref{hbfkl} with nearest-neighbour spin interactions, are different commuting charges of the same integrable $SL(2,\mathbb{C})$-symmetric spin chain.

Now we can compute the anomalous dimension of the short operators \eqref{lightest}.
Let us introduce an effective coupling $\eta$ through
\begin{equation}
\xi_1\xi_2 = 4\pi(1 - u \eta)\, ,
\end{equation}
which we keep finite in the limit $u\to 0,\,\,\xi_1\xi_2\to 4\pi$.
We obtain from~\eqref{Dim2} and \eqref{hBFKL} in the limit \(u\to 0\) the equation for the spectrum of conformal dimensions $\Delta(\eta)$ of exchange operators in BFKL FCFT (at \(L=2\))
\begin{align}
\eta=\psi\left(\frac{\Delta}{2}\right)+\psi\left(1-\frac{\Delta}{2}\right)-2\psi(1)-1+\mathcal{O}(u)\,.
\end{align}

\section{Single-Trace Correlators}
\label{sec:BD}

\begin{figure}
    \centering
    \includegraphics{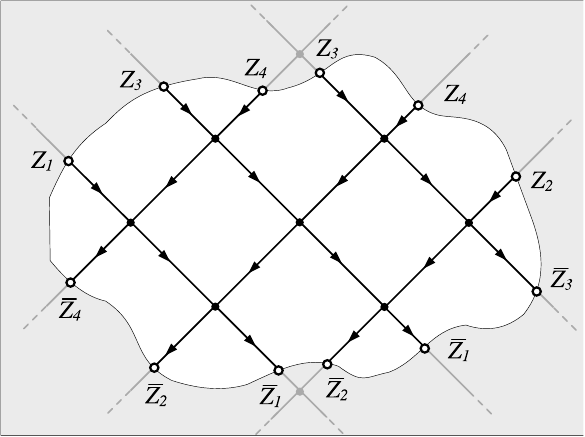}
    \caption{A closed region of disk topology carved out of a  square lattice describes a Checkerboard Feynman diagram. Crossing between the contour and the propagators define the fields inside a single-trace correlator $\left\langle \Tr[Z_1(x_1)\bar Z_4(x_2) \bar Z_2(x_3)\cdots Z_4(x_{12})Z_3(x_{13})] \right\rangle$. Other Feynman diagrams that would contribute to such correlator are subleading in the planar limit.}
    \label{fig:carving}
\end{figure}

In this section we shall compute a few classes of correlators obtained by point-split of the VEV of a single-trace local operator. We consider single-trace correlators featuring a number $m_1+m_2+\dots+m_n$ of external fields grouped into $n$ coinciding positions, which have the general form:
\begin{equation}\label{npt_ST}
    \frac{1}{N} \left\langle \Tr[(\Phi_{1,1}\cdots \Phi_{1,m_1})(x_1)(\Phi_{2,1}\cdots \Phi_{2,m_2})(x_2)\cdots (\Phi_{n,1}\cdots \Phi_{n,m_n}) (x_n)] \right\rangle  \,.
\end{equation}
The fields $\Phi_{n,m}$ are chosen among $Z_k,\bar Z_k$ and each pair of brackets $(\dots )$ delimits a product of fields located at the same point and with open $SU(N)$ indices, e.g.
\begin{equation}\label{open_indices_op}
    (Z_1 Z_2 \bar Z_3 Z_1 Z_1\cdots \bar Z_2)_{ij}(x) =\sum_{a_1\dots a_L} (Z_1)_{i a_1}(x) (Z_2)_{a_1 a_2}(x)\cdots(\bar Z_2)_{a_L j}(x) \,.
\end{equation}
These type of correlators are fundamental objects in the planar limit of Fishnet theories, since they usually get quantum corrections by at most one Feynman integral \cite{Chicherin:2017cns,Chicherin:2017frs}. In practice one can define a single-trace planar correlator with a simple procedure, depicted in figure~\ref{fig:carving}: let us draw a closed contour on a large Checkerboard square lattice. The propagators which are crossed by the contour define the fields inside the trace and are initially located at different spacetime points. Now, one can check that no other Feynman diagram can contribute to the perturbation theory of this single-trace correlator in the planar limit $N\to \infty$.

The OPE data of local operators of type~\eqref{open_indices_op} can be extracted from the OPE of single-trace correlators of type~\eqref{npt_ST}. The spectrum of such \emph{open-index} operators in planar theory is easier to study wrt that of single-trace operators: the correlator of two single-trace operators is affected by finite-volume effects (wrappings, spiraling, \cite{Caetano:2016ydc}), whereas a two-point single-trace correlator receives quantum corrections by at most one Feynman diagram. In the language of vertex models \cite{Zamolodchikov:1980mb}, the latter can be is the partition function of a Checkerboard square-lattice with boundary conditions fixed by the coordinates of fields under trace (similar to arbitrary fixed configuration of spins on the boundary of a $2d$ lattice spin system).

In this section and in the next one, we will analyse two types of single-trace 4-point correlators. One type is the rectangular correlators, i.e. the generalisation of the \emph{Basso--Dixon} four-point Fishnets \cite{Basso:2017jwq} to the Checkerboard square lattice. The second type is dubbed \emph{diamond} correlators and have not yet been studied even in the bi-scalar FCFT.

\subsection{Rectangular Fishnets}

\begin{figure}
    \centering
    \includegraphics[scale=0.55]{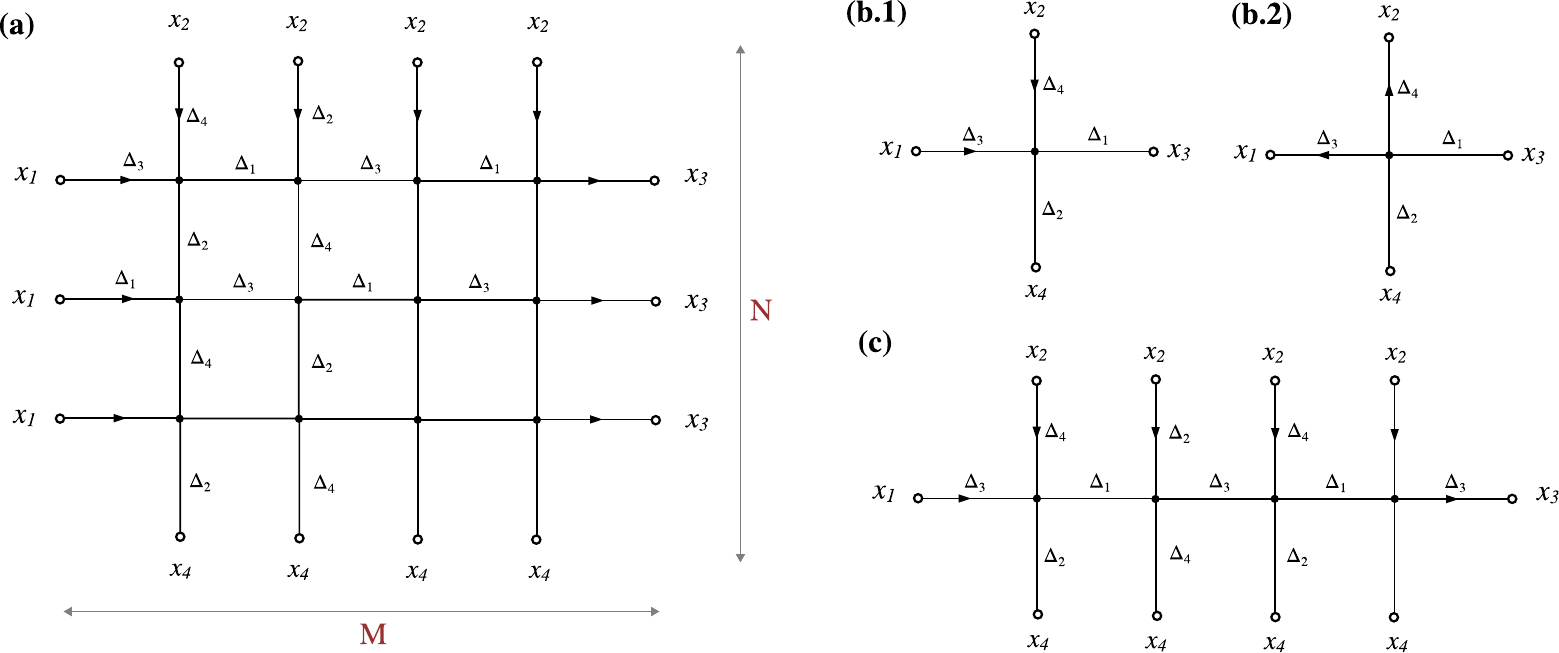}
    \caption{\textbf{(a)} A rectangular Fishnet integral corresponding to the four-point correlator $I_{3,4}$ of the Checkerboard CFT. The dimension of propagators alternates along rows/columns of the square lattice. \textbf{(b.1),(b.2)} are the two cross-integrals corresponding to the two vertices of the theory. The diagram \textbf{(c)} is an example of Checkerboard Ladder of length $M=4$.}
    \label{fig:checker_BD}
\end{figure}

In this section we consider four-point rectangular Fishnet diagrams in the Checkerboard theory, see figure~\ref{fig:checker_BD}. We shall compute them using the method of separation of variables~\cite{Derkachov:2018rot,Derkachov:2019tzo,Derkachov:2020zvv,Derkachov:2021ufp}. This approach, inspired by the early work~\cite{Derkachov:2001yn} for two dimensions,  was applied in general dimension $d$ to the rectangular Fishnets of the bi-scalar theory in \cite{Derkachov:2021ufp}. We will present explicit computations in $d=2,4$ where the SoV representation is particularly well established.

For a rectangle of size $2n \times 2m$ the corresponding correlator reads
\begin{equation}\label{checker_BD_even}
   I_{2n,2m} = \frac{1}{N}\left\langle \Tr[(Z_1 Z_3)^{n}(x_1) (\bar Z_2 \bar Z_4)^{m}(x_4) (\bar Z_3 \bar Z_1)^{n}(x_3) (Z_4 Z_2)^{m}(x_2)]\right\rangle\, ,
\end{equation}
whereas for more generic size the expression is slightly cumbersome due to the alternating nature of fields along the rectangle edges. For example, with reference to the case in figure~\ref{fig:checker_BD},
\begin{equation}\label{checker_BD_oddeven}
   I_{2n+1,2m} = \frac{1}{N}\left\langle \Tr[(Z_3 (Z_1 Z_3)^{n})(x_1) (\bar Z_2 \bar Z_4)^{m}(x_4) ((\bar Z_3 \bar Z_1)^{n} \bar Z_3)(x_3) (Z_2 Z_4)^{m}(x_2)] \right\rangle\, ,
\end{equation}
and the other cases are similar. In particular, we could define Ladder integrals of two types, corresponding to $N=1$ and any $M$, or vice-versa. The first case is depicted in figure~\ref{fig:checker_BD} (c) for $M=2m=4$,
\begin{equation}\label{Ladder_I}
    I_{1,2m} = \frac{1}{N}\left\langle \Tr[Z_3(x_1) (\bar Z_2 \bar Z_4)^{m}(x_4) \bar Z_3(x_3) (Z_2 Z_4)^{m}(x_2)]\right\rangle\, ,
\end{equation}
whereas for odd $M=2m+1$, the Ladder corresponds to the correlator:
\begin{equation}\label{Ladder_II}
    I_{1,2m+1} = \frac{1}{N}\left\langle \Tr[Z_3(x_1) ((\bar Z_2 \bar Z_4)^{m}\bar Z_2)(x_4) \bar Z_1(x_3) (Z_4(Z_2 Z_4)^{m})(x_2)]\right\rangle\, .
\end{equation}
The main difference between this class of Feynman integrals and the one of the bi-scalar FCFT~\eqref{bi-scalar} is, evidently, the alternating nature of fields/propagators along rows/columns of the Fishnet square-lattice.

\subsubsection{SoV Representation}
\label{sec:SoV}

We present here the SoV representations of the rectangular diagrams in the $2D$ and $4D$ theories. Details regarding the derivation in two dimensions are given in appendix \ref{app:SoV}. This generalises (some of) the results of \cite{Derkachov:2001yn,Derkachov:2018rot} (in 2D) and \cite{Basso:2017jwq,Derkachov:2019tzo,Derkachov:2020zvv} (in 4D).

The first thing to notice is that the diagrams can be built from the repeated application of some graph-building operators. Due to the alternating nature of the lattice, two such operators $\Lambda_{N}$ and $\Lambda'_{N}$ are needed, where $N$ is the height of the diagram. They are related through the exchange of the weights $\Delta_i$ of the propagators: 
\begin{equation}
  \Lambda'_{N}=  \Lambda_{N}\big|_{(\Delta_1,\Delta_2)\leftrightarrow (\Delta_3,\Delta_4)}\, .
\end{equation}
The kernel of $\Lambda_{N}$ is
\begin{multline}
    \bra{y_1,\dots,y_{N}} \Lambda_{N}\ket{z_1,\dots,z_{N}}= \\
    =\prod_{k=0}^{\left\lfloor\frac{N-1}{2}\right\rfloor} (y_{2k+1} - z_{2k+1})^{-2\Delta_1} y_{2k,2k+1}^{-2\Delta_2} \prod_{k=1}^{\left\lfloor\frac{N}{2}\right\rfloor} (y_{2k} - z_{2k})^{-2\Delta_3} y_{2k-1,2k}^{-2\Delta_4}\, ,
\end{multline}
where $y_0 = 0$ and we use the standard notation $\bra{y}\ket{z} = \pi^{\frac{d}{2}} \delta^{(d)}(y-z)$. The rectangular diagrams we want to compute are then
\begin{equation}\label{IN2L-1}
    I_{N,2L-1}(y,z) = \bra{y,\dots,y} \prod_{k=0}^{\left\lfloor\frac{N-1}{2}\right\rfloor} y_{2k,2k+1}^{2\Delta_4} \prod_{k=1}^{\left\lfloor\frac{N}{2}\right\rfloor} y_{2k-1,2k}^{2\Delta_2} (\Lambda'_{N}\Lambda_{N})^L \ket{z,\dots,z}\,,
\end{equation}
and
\begin{equation}\label{IN2L}
    I_{N,2L}(y,z) = \bra{y,\dots,y} \prod_{k=0}^{\left\lfloor\frac{N-1}{2}\right\rfloor} y_{2k,2k+1}^{2\Delta_4} \prod_{k=1}^{\left\lfloor\frac{N}{2}\right\rfloor} y_{2k-1,2k}^{2\Delta_2} (\Lambda'_{N}\Lambda_{N})^L \Lambda'_{N} \ket{z,\dots,z}\, ,
\end{equation}
where we have chosen, for brevity, to ignore the constant coefficient in the propagators \eqref{propagator}, so that they are simply $D_i(x) = x^{-2\Delta_i}$.

We show in appendix~\ref{app:SoV2D} how to construct orthogonal bases of left and right eigenvectors of $\Lambda'_{N}\Lambda_{N}$. Assuming that the bases are complete, one can then compute the diagrams. For an even number of graph-building operators, one has\footnote{A necessary condition for the SoV representation to be valid is that each semi-infinite series of poles in the integrand, coming from the various Gamma functions, is either strictly above or below the real axis. This means that $\Re(\Delta_1)>|\!\Re(\Delta_2 - \Delta_4)|/2$ and $\Re(\Delta_3)>|\!\Re(\Delta_2 - \Delta_4)|/2$.}
\begin{equation}\label{rectangular BD even}
    \begin{aligned}
    I_{N,2L-1}(y,z) & = A_0(\Delta_3)^{LN+\left\lfloor\frac{N-1}{2}\right\rfloor \left\lfloor\frac{N+1}{2}\right\rfloor} A_0(\Delta_1)^{LN+\left\lfloor\frac{N-2}{2}\right\rfloor \left\lfloor\frac{N}{2}\right\rfloor}\\
    & \times A_0(\tilde{\Delta}_2)^{\left\lfloor\frac{N}{2}\right\rfloor \left\lfloor\frac{N+2}{2}\right\rfloor} A_0(\tilde{\Delta}_4)^{\left\lfloor\frac{N-1}{2}\right\rfloor \left\lfloor\frac{N+1}{2}\right\rfloor} \frac{y^{2Y_N} z^{2Z_N}}{N!}\\
    & \times\!\!\! \sum_{m_1,\dots,m_N = -\infty}^{+\infty} \int_{\mathbb{R}^N}  \prod_{k=1}^N P^{(d)}_{m_k}(\theta) f_{N,2L-1}(r;\bfu_k)\, \mu^{(d)}_N(\bfu_1,\dots,\bfu_N) \dd u_1 \dots \dd u_N\, ,
    \end{aligned}
\end{equation}
where $\bfu_k = (m_k,u_k)\in\mathbb{N}\times\mathbb{R}$ and the measure for the $2d$ case  is
\begin{equation}
    \mu^{(2)}_N(\bfu_1,\dots,\bfu_N) = \frac{1}{(2\pi)^{N}} \prod_{1\leqslant i < j \leqslant N} \left[(u_i-u_j)^2 + \frac{(m_i - m_j)^2}{4}\right]
\end{equation}
and for the $4d$ case
\begin{multline}
    \mu^{(4)}_N(\bfu_1,\dots,\bfu_N) = \frac{\prod_{k=1}^{N} (m_k+1)}{(2\pi)^{N}} \prod_{1\leqslant i < j \leqslant N} \left[(u_i-u_j)^2 + \frac{(m_i - m_j)^2}{4}\right]\\
    \times \left[(u_i-u_j)^2 + \frac{(2+m_i + m_j)^2}{4}\right]\, .
\end{multline}
The summand also contains
\begin{equation}
    P^{(2)}_m(\theta) = \ee^{\ii m\theta} \quad \text{or} \quad P^{(4)}_m(\theta) = \frac{\ee^{\ii (m+1)\theta}}{\ee^{\ii \theta} - \ee^{-\ii \theta}}\, ,
\end{equation}
and the functions
\begin{equation}
    f_{N,2L-1}(r;\bfu) = r^{2\ii u}
    \left(A_m(\tilde{a}(u)) A_m(\tilde{b}(u))\right)^{L+\left\lfloor\frac{N}{2}\right\rfloor} \left(A_m(\tilde{a}'(u)) A_{m}(\tilde{b}'(u))\right)^{L+\left\lfloor\frac{N-1}{2}\right\rfloor}\, ,
\end{equation}
where $r$ and $\theta$ parameterise the cross-ratio through
\begin{equation}
r = \frac{|y|}{|z|}\, ,\qquad \cos\theta = \frac{y\cdot z}{|y| |z|}\, .
\end{equation}
We used the notation $\tilde\alpha = d/2 - \alpha$ and the functions $A_m(\alpha) = \Gamma(\tilde\alpha+m/2)/\Gamma(\alpha+m/2)$, as well as
\begin{align}
    a(u) &= \frac{\tilde{\Delta}_2}{2} + \frac{\Delta_1-\Delta_3}{4} - \ii u = \frac{\Delta_1}{2} + \frac{\Delta_4 - \Delta_2}{4} - \ii u\, ,\\
    b(u) &= \frac{\tilde{\Delta}_2}{2} + \frac{\Delta_3-\Delta_1}{4} + \ii u = \frac{\Delta_3}{2} + \frac{\Delta_4 - \Delta_2}{4} + \ii u\, .
\end{align}
Finally, we have also introduced the short notations
\begin{equation}
  a'=  a\big|_{(\Delta_1,\Delta_2)\leftrightarrow (\Delta_3,\Delta_4)}\, ,\quad b'=  b\big|_{(\Delta_1,\Delta_2)\leftrightarrow (\Delta_3,\Delta_4)}\, ,
\end{equation}
and
\begin{equation}\label{power y}
    Y_N = - \left\lfloor\frac{N+1}{2}\right\rfloor \left(\frac{\Delta_3}{2} + \frac{\Delta_2 - \Delta_4}{4}\right) - \left\lfloor\frac{N}{2}\right\rfloor \left(\frac{\Delta_1}{2} + \frac{\Delta_4 - \Delta_2}{4}\right)\, ,
\end{equation}
\begin{equation}\label{power z}
    Z_N = - \left\lfloor\frac{N+1}{2}\right\rfloor \left(\frac{\Delta_1}{2} + \frac{\Delta_2 - \Delta_4}{4}\right) - \left\lfloor\frac{N}{2}\right\rfloor \left(\frac{\Delta_3}{2} + \frac{\Delta_4 - \Delta_2}{4}\right)\, .
\end{equation}
The diagrams realised by an odd number of graph-building operators are given by
\begin{multline}\label{rectangular BD odd}
    I_{N,2L}(y,z) = A(\Delta_3)^{LN+\left\lfloor\frac{N^2+3}{4}\right\rfloor} A(\Delta_1)^{LN+\left\lfloor\frac{N^2}{4}\right\rfloor} \left(A(\tilde{\Delta}_2) A(\tilde{\Delta}_4)\right)^{\left\lfloor\frac{N^2}{4}\right\rfloor} \frac{y^{2Y_N} z^{2Z'_N}}{N!} \times \\
    \times \prod_{k=1}^N \sum_{m_k = -\infty}^{+\infty} P^{(d)}_{m_k}(\theta) \int_{\mathbb{R}^N} F_{N,2L}(r;\bfu_1,\dots,\bfu_N) \mu^{(d)}_N(\bfu_1,\dots,\bfu_N) \dd u_1 \dots \dd u_N\, ,
\end{multline}
where
\begin{equation}
    F_{N,2L}(r;\bfu_1,\dots,\bfu_N) = \frac{\prod_{k=1}^N f_{N,2L-1}(r;\bfu_k) \times \begin{cases}
    A_{m_k}(\tilde{a}(u_k)) & \text{if $N$ is even}\\
    A_{m_k}(\tilde{a}'(u_k)) & \text{if $N$ is odd}
    \end{cases}}{A_{m_N}(b(u_N)) A_{m_{N-1}}(b'(u_{N-1})) A_{m_{N-2}}(b(u_{N-2})) \cdots}\, .
\end{equation}


\subsubsection{Ladder diagrams}
\label{sec:ladders}

In the particular case of the ladder diagrams, $N=1$, the previous formulae read
\begin{equation}
\begin{aligned}
\label{Ladder}
    I_{1,M}(y,z) & = \frac{A_0(\Delta_1)^{\left\lfloor\frac{M+1}{2}\right\rfloor} A_0(\Delta_3)^{\left\lfloor\frac{M+2}{2}\right\rfloor}}{|y|^{\Delta_3+\frac{\Delta_{24}}{2}}} z^{2 Z_1(M)}\sum_{m = 0}^{+\infty} P^{(d)}_m(\theta)\\
    & \times\int_{-\infty}^{+\infty} r^{2\ii u} \left(\frac{\Gamma\left(\frac{\Delta_1+m}{2}+\frac{\Delta_{24}}{4}+\ii u\right) \Gamma\left(\frac{\Delta_1+m}{2}+\frac{\Delta_{42}}{4}-\ii u\right)}{\Gamma\left(\frac{\Delta_4+m}{2}+\frac{\Delta_{31}+d}{4}-\ii u\right) \Gamma\left(\frac{\Delta_2+m}{2}+\frac{\Delta_{31}+d}{4}+\ii u\right)}\right)^{\left\lfloor\frac{M+1}{2}\right\rfloor} \\
    & \times\left(\frac{\Gamma\left(\frac{\Delta_3+m}{2}+\frac{\Delta_{24}}{4}-\ii u\right) \Gamma\left(\frac{\Delta_3+m}{2}+\frac{\Delta_{42}}{4}+\ii u\right)}{\Gamma\left(\frac{\Delta_4+m}{2}+\frac{\Delta_{13}+d}{4}+\ii u\right) \Gamma\left(\frac{\Delta_2+m}{2}+\frac{\Delta_{13}+d}{4}-\ii u\right)}\right)^{\left\lfloor\frac{M+2}{2}\right\rfloor} \mu_1^{(d)}(\bfu) \, \dd u\, ,
    \end{aligned}
\end{equation}
where $\Delta_{ij} = \Delta_i - \Delta_j$, the measure is simply
\begin{equation}
    \mu_1^{(2)}(\bfu) = \frac{1}{2\pi}\quad \text{or}\quad \mu_1^{(4)}(\bfu) = \frac{m+1}{2\pi}\, ,
\end{equation}
and the exponent $Z_1(M)$ depends only on the parity of $M$:
\begin{equation}
    2Z_1(M) = \begin{cases}
    2Z_1 = - \Delta_1 + \frac{\Delta_4 - \Delta_2}{2}& \text{if $M$ is odd}\\
    2Z'_1 = - \Delta_3 + \frac{\Delta_2 - \Delta_4}{2} & \text{if $M$ is even}
    \end{cases}\, .
\end{equation}
From the SoV analysis of Ladders one can extract information about the dilation operator of the theory. In the light-cone OPE limit $x_{12}^2,x_{34}^2\to 0$, which in terms of \eqref{Ladder} becomes
\begin{equation}
\label{LC_OPE}
    r \to 0\,, \, \theta \to -\ii \infty\,,\,\, 0<t =r e^{\ii \theta} <\infty\,,
\end{equation}
the Ladder diagrams are dominated by a leading UV behaviour
\begin{equation}
\sim \log^n (r e^{-\ii \theta}) \times F_{M}(t)\,.
\end{equation}
For a Ladder of length $M$ in the general Checkerboard CFT we argue that $n= \lfloor{M}/{2}\rfloor$. The OPE logs are obtained from \eqref{Ladder} by taking residues at higher-order poles wrt the integration variable $u$. In the limit \eqref{LC_OPE}, the integration is performed by closing the contour in the lower half of the complex plane, i.e. encircling poles with $\text{Re}(\ii u) >0$. Hence, the SoV integration features poles of order $\lfloor M/2\rfloor +1$
\begin{equation}
 \ii u = \frac{\Delta_1+m}{2}+\frac{\Delta_4-\Delta_2}{4}+k\,,\,\, k \in \mathbb{N}\,,
\end{equation}
which produce UV logarithms with $n=\lfloor M/2\rfloor$. It features also poles of order $\lfloor (M+1)/2\rfloor$,
\begin{equation}
 \ii u = \frac{\Delta_3+m}{2}+\frac{\Delta_2-\Delta_4}{4}+k\,,\,\, k \in \mathbb{N}\,,
\end{equation}
which give either poles of the same order $n$ or of lower order $n-1$, respectively for $M$ odd or even. On the other hand, the log-behaviour of Ladder integrals in the bi-scalar Fishnet CFT~\eqref{bi-scalar} is that of a stronger divergence $\sim \log^M(re^{\ii \theta})$~\cite{Usyukina:1993ch}. More generally, whenever a certain condition of type \eqref{special_pts} is met, the two series of poles merge into one of order $M+1$, enhancing the $\log$-divergence to order $n=M$. Indeed, setting $\Delta_1+\Delta_4=\Delta_2+\Delta_3$, there is a series of $(M+1)$-order poles
\begin{equation}
 \ii u = \frac{m}{2}+\frac{\Delta_1+\Delta_3}{4}+k\,,\,\, k \in \mathbb{N}\,.
\end{equation}
The fact that leading-logs of ladder integrals have ``halved" power wrt the length $M$ hints that operators in the Checkerboard theory are protected at one-loop and quantum corrections start at two-loops---unless the parameters are tuned to \eqref{special_pts}.

\section{Diamond Correlators}
\label{sec:diamonds}

\begin{figure}
    \centering
    \includegraphics[scale=0.40]{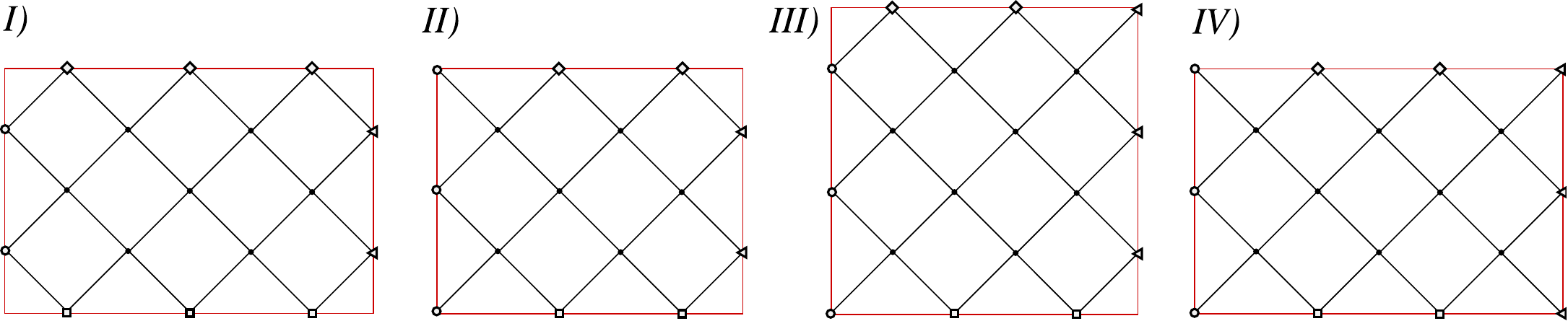}
    \caption{From left to right, one instance of each of the four possible ``types" I-IV of Diamond Fishnet diagrams  delimited by a red rectangle. They represent four different four-point correlators. External points that lie on one same edge of the rectangle and have the same shape ($\circ, \triangle, \square, \diamond$) are merged, i.e. they have the same coordinates. Their different boundaries reflect in the number and position of legs connected to the vertices of the rectangle.}
    \label{fig:diamonds_bc}
\end{figure}

In this section we will consider a new class of four-point single-trace correlators which we dub ``Diamonds". We define them on the Checkerboard square lattice for generic lengths $m,n$ by drawing a rectangle of size $m\times n$ with edges that cut the lattice cells along their diagonals, crossing the lattice vertices. We depict four instances in figure \ref{fig:diamonds_bc}, which cover the four possible choices of boundary for the square lattice, modulo reflection symmetries. In the following we will compute some Diamond four-point correlators, defined by merging the external points that lie on one same edge of the rectangle in figure \ref{fig:diamonds_bc}. In particular, we will concentrate on type I and type II boundaries; a similar analysis can be carried out for Diamonds of type III and IV.\footnote{In fact, a simple proof shows that diagrams of type IV are equivalent to type I, based on a systematic iteration of star-triangle identities.} In general, we call $m,n$ the number of \emph{entire} squares appearing in a column or in a row inside the rectangle. With reference to the figure, this is $(m,n)=(2,3)$ for type I, $(m,n)=(2,2)$ for types II, III, and IV.

\subsection{Diamond Correlators of Type I}
\label{sec:diamonds_I}

In this section we will focus on the single-trace four-point functions described by Diamond diagrams of type I. We realise the four-point correlator by labelling with $x_i$ the position of external fields in clockwise order,
\begin{equation}
\label{checker_4pt_I}
   G^{(I)}_{m,n} = \frac{1}{N}\left\langle \Tr[(\bar Z_4 Z_1)^{m}(x_1) (\bar Z_1 \bar Z_2)^{n}(x_4) (Z_2 \bar Z_3)^{m}(x_3) (Z_3 Z_4)^{n}(x_2)]\right\rangle\,,
\end{equation} 
and it is given by a single
Feynman diagram at loop order $\xi_1^{2n(m-1)}\xi_2^{2m(n-1)}$. We can regard this diagram as the convolution of $m\times n$ copies of the R-matrix \eqref{R-matrix} along $m$ rows and $n$ columns.

\begin{figure}
\begin{center}
\includegraphics[scale=0.7]{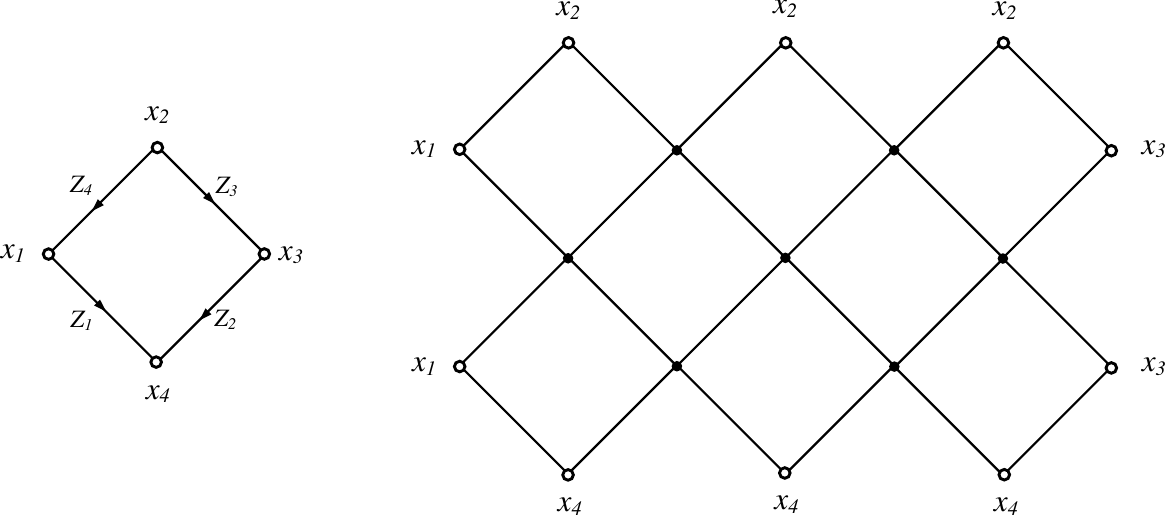}
\caption{The Diamond integral defining the correlator~\eqref{checker_4pt_I} for $(m,n)=(1,1)$ (left) and for $(m,n)=(2,3)$ (right). The first one contributes at zero coupling, while the second contributes at order $(\xi_1^{2})^3(\xi_2^{2})^4$, i.e. seven loops. We marked each propagators with the corresponding field $Z_k$, and the orientation of arrows goes as $Z \to \bar Z$.}
\label{class_I_diag}
\end{center}
\end{figure}

Notice that \eqref{checker_4pt_I} is one of the possible realisation of Diamond correlators of type I. The other ones are related by cyclic permutation of the external fields, hence we carry on the analysis with little loss of generality.
The case $m=n=1$ is purely tree-level (figure~ \ref{class_I_diag}, left), while for larger $m,n$ the correlator is equal to the product of a connected component at loop level times the $m=n=1$ term \begin{align}
\begin{aligned}
\label{checker_4pt_I_prime}
    &G^{(I)}_{m,n} = G^{(I)}_{1,1} \times \widetilde G^{(I)}_{m,n}\,,\\ &\widetilde G^{(I)}_{m,n} = \frac{1}{N} \left\langle \Tr[(Z_1 \bar Z_4 )^{m-1}(x_1) (\bar Z_2 \bar Z_1)^{n-1}(x_4) (\bar Z_3 Z_2)^{m-1}(x_3) (Z_4 Z_3)^{n-1}(x_2)]\right\rangle\,,
\end{aligned}
\end{align}
where
$$G^{(I)}_{1,1}=D_4(x_{12})D_3(x_{23})D_2(x_{34})D_1(x_{41})\,.$$

Let us focus on the connected integrals. First of all, these are not always well defined in integer dimension: they are finite for general\footnote{Here we mean general under the constraint $\Delta_1+\Delta_2+\Delta_3+\Delta_4=d$ which is required for dimensionless couplings $\xi_1,\xi_2$.} scaling dimensions $\{\Delta_1,\Delta_2,\Delta_3,\Delta_4\}$ whereas they develop UV divergence at any of the special points 
\begin{equation}
\label{special}
    \Delta_i+\Delta_{(i+1)\, \text{mod}\, 4}=\frac{d}{2}\,.
\end{equation}
For instance, for $\Delta_4+\Delta_1=d/2= \Delta_2+\Delta_3$ a couple of the open-index operators inside the trace in \eqref{checker_4pt_I_prime}, namely $(Z_1 \bar Z_4)^{m-1}$ and $(\bar Z_3 Z_2)^{m-1}$ get quantum corrections to their two-point functions and need to be renormalised. 

\subsubsection{Digression: one-loop spectrum}
Let us look at the divergence of a Diamond correlator at points \eqref{special} for the one-loop $\xi_2^2$ case $(m,n)=(2,1)$. In fact, this case reduces to the two-point function
 \begin{align}
\begin{aligned}
  \widetilde G^{(I)}_{2,1} =\frac{1}{N} \left\langle \Tr[(Z_1 \bar Z_4)(x_1) (\bar Z_3 Z_2)(x_3)]\right\rangle\,,
\end{aligned}
\end{align}
 responsible for the mixing of length-$2$ open-index operators
\begin{equation}
\label{primer_log_2}
  O_1= Z_1 \bar Z_4\,,\,  O_2= \bar Z_3 Z_2 \,.
\end{equation}
The matrix of Feynman integrals that mix this multiplet read $\left\langle \Tr[O_i(x) O_j^{\dagger}(0)]\right\rangle=$
\begin{center}
    \includegraphics[scale=0.5]{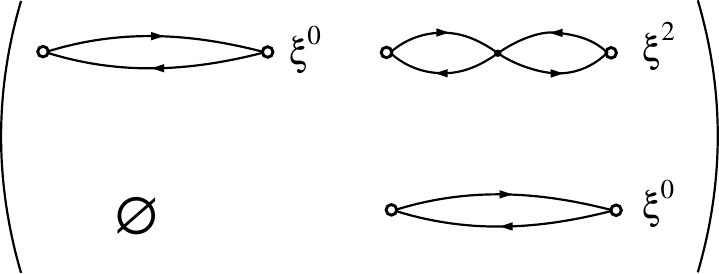}
\end{center}
with no other contribution (recall that we deal with the planar limit of single-trace correlators).
The renormalised operators are given via the mixing matrix
\begin{equation}
\label{wf_renorm}
    O^{(r)}_{i}=\sqrt{Z_{i1}}\, O_1 + \sqrt{Z_{i2}} \,O_2\,;\,\,\, \sqrt{Z_{ij}} = \begin{pmatrix}
        1 && -\xi_2^2 \frac{\gamma_{1}}{\epsilon}\\
        0 && 1
    \end{pmatrix}\,.
\end{equation}
Notice that \eqref{wf_renorm} is not diagonalisable -- a signature of non-unitarity of the theory which results in the existence of Jordan blocks in the spectrum \cite{Gurarie:1993xq}. In short, Hermitian conjugation and renormalisation do not commute: $$O^{\dagger}_{i,r}=\sqrt{Z_{1i}} O_1^{\dagger} + \sqrt{Z_{2i}} O_2^{\dagger} \neq (O_{i,r})^{\dagger}.$$
The matrix of renormalised $2$-pts functions reads
\begin{equation}
\frac{1}{N}    \left\langle \Tr[O_{i,r}(x) O_{j,r}^{\dagger}(0)] \right\rangle = \begin{pmatrix}
        \frac{1}{(x^2)^{2}} && -\xi^2 {\gamma_1} \frac{\log(x^2 \mu^2)}{(x^{2})^2}\\
        0 &&  \frac{1}{(x^2)^{2}}
    \end{pmatrix}\,,
\end{equation}
for $\gamma_1$ the coefficient of the pole $1/\epsilon$ in the one-loop diagram. 

When none of \eqref{special} is satisfied, the $1$-loop anomalous dimension $\gamma_1$ is zero, coherently with the mismatch $[O_1]\neq [O_2]$ of bare dimensions. We can check it by direct computation: the one-loop Feynman diagram generating $\gamma_1$ is proportional to 
\begin{equation}
\label{zero_2pt}
 \int \frac{\dd^{d}y}{(x-y)^{2(\Delta_1+\Delta_4)} y^{2(\Delta_2+\Delta_3)}}= \pi^{d} \frac{\Gamma\left(\frac{d}{2}-\Delta_2-\Delta_3\right)\Gamma\left(\frac{d}{2}-\Delta_1-\Delta_4\right)}{\Gamma\left(\Delta_1+\Delta_4\right)\Gamma\left(\Delta_2+\Delta_3\right)} \,\delta^{(d)}(x)\,,
\end{equation}
which evaluates to zero since $x \neq 0$.
The same arguments hold after insertion of derivatives on the fields, that is for spinning $2$-pt functions. Indeed, take 
\begin{equation}
\label{zero_2pt_der}
\frac{1}{N}    \left\langle\Tr[(\partial^{J_1} Z_1 \partial^{J_4} \bar Z_4) (x) (\partial^{J_3} \bar Z_3 \partial^{J_2}  Z_2)(0)]\right\rangle\,,
\end{equation}
where $\partial \equiv n \cdot \partial$ with generic $n^{\mu}$. The evaluation of the one-loop (and only) Feynman diagram is proportional to
\begin{equation}
(n \cdot \partial)^{J_1+J_2+J_3+J_4} \delta^{(d)}(x)\,.
\end{equation}
All together, we conclude that the dilation operator of the planar Checkerboard theory
\begin{equation}
    \mathbb{D} = \mathbb{D}_0 + \mathbb{D}_{1,0} \,\xi_1^2   +  \mathbb{D}_{0,1}\,\xi_2^2 +  \mathbb{D}_{2,0}\, \xi_1^4+ \mathbb{D}_{1,1}\, \xi_1^2 \xi_2^2 +\ldots
\end{equation}
has zero one-loop components $\mathbb{D}_{1,0},\, \mathbb{D}_{0,1}$ whenever the theory is defined on a Loom with four arbitrary slopes. Otherwise, at any special point \eqref{special}, the one-loop spectrum becomes non-trivial. The action of dilations at one-loop can be described as an operator acting on nearest-neighbour spin-chain vectors that represent consecutive fields inside the $SU(N)$ trace, say $$|Z_i , Z_j\rangle \equiv (Z_i Z_j)(0).$$ In this language our statements read
\begin{align}
\begin{aligned}
\mathbb{D}_{1,0} | Z_1 , Z_2\rangle= \delta_{2,w_1+w_2}\times \gamma_1| \bar Z_4 , \bar Z_3\rangle \,,\,\,\,\,\,\,
\mathbb{D}_{0,1} | Z_2 , \bar Z_3 \rangle=  \delta_{2,w_1+w_4}\times \gamma_1  |  Z_1 , \bar Z_4\rangle \,,
\end{aligned}
\end{align}
and similarly for cyclic permutations of fields $Z_1,Z_2,\bar Z_3,\bar Z_4$.

\subsubsection{General formula}

\begin{figure}
    \centering
\includegraphics[scale=0.38]{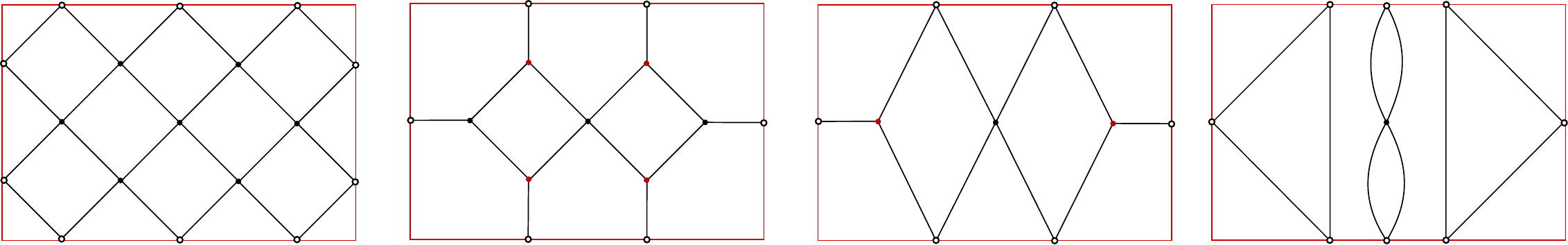}
    \caption{Star-Triangle computation of $G_{2,3}^{(I)}$. From left to right: from (1) to (2): merge points with the same coordinates on each of four boundaries and drop disconnected factor; from (2) to (3): integrate  red dot vertices by star-triangle; (4) the result given by two triangles of disconnected propagators times a two-point one-loop integral of the type \eqref{zero_2pt}.}
    \label{fig:diamond_STR_23}
\end{figure}

\begin{figure}
    \centering
\includegraphics[scale=0.38]{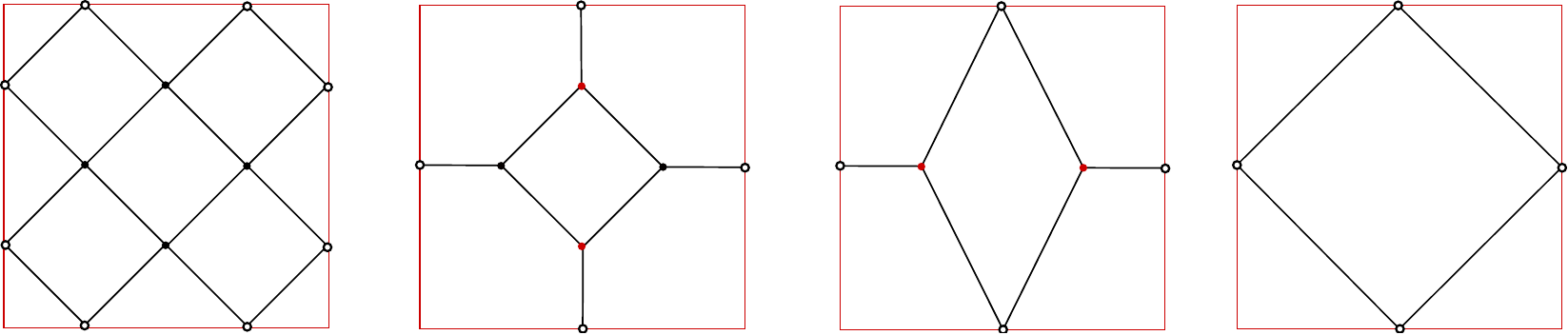}
    \caption{Star-Triangle computation of $G_{2,2}^{(I)}$. The same passages as for $m=2,n=3$ lead to a product of propagators.}
    \label{fig:diamond_STR_22}
\end{figure}

With the assumption to avoid \eqref{special} we move on to the computation of Diamonds of any size. The procedure is exemplified in figure \ref{fig:diamond_STR_23} and \ref{fig:diamond_STR_22} for $m=2$ and $n=3,2$ respectively.

The computation is based on the iterative use of star-triangle identities, and result for general $m,n$ reads
\begin{equation}
\label{mn_class_I}
   \left(D_{1}(x_{14})D_{2}(x_{34})D_{3}(x_{23})D_{4}(x_{12})\right)^2 \times  \left(\delta^{(d)}(x_{24})\right)^{\theta(m-n)}\left(\delta^{(d)}(x_{13})\right)^{\theta(n-m)}\,,
\end{equation}
where $\theta(a)=a$ if $a>0$ and zero otherwise. Dealing with the correlator of fields inserted at four different points in Euclidean space-time, the formula \eqref{mn_class_I} evaluates either to zero or, when $m=n$, to a product of propagators.

This result is atypical hence intriguing, and requires further explanations. First, notice that the last formula has the typical ``tree-level" form  but it is a contribution to the correlator ${G}_{m,m}^{(I)}$ at loop order $(\xi_1^2 \xi_2^2)^{ m(m-1)}$. The usual CFT four-point function at loop level is described by a transcendental functions of cross-ratios of the distances $x_{ij}^2$, hence it develops logarithmic divergences in the limit of light-cone distances. This is the case, for instance, of the Ladders in section \ref{sec:ladders}. The absence of logs in \eqref{mn_class_I} signals that a light-cone OPE decomposition of ${G}_{m,m}^{(I)}$ features only the exchange of operators with zero anomalous dimension.
\begin{figure}[t]
\begin{center}
\includegraphics[scale=0.48]{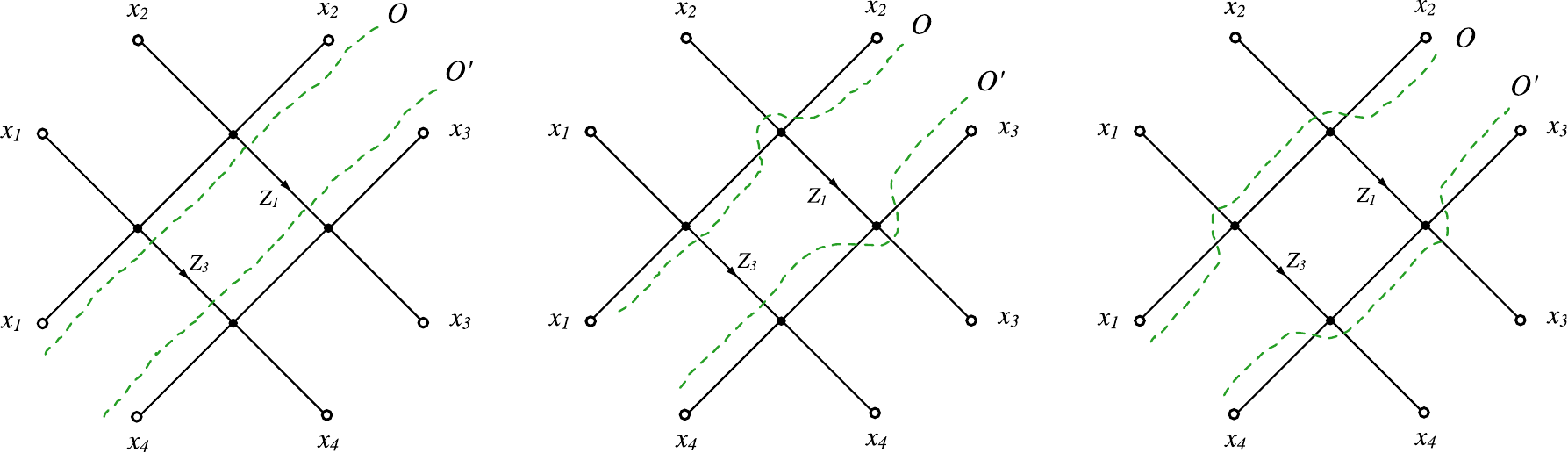}
\caption{Three instances of bare exchange operators flowing in the $s$-channel of $\widetilde{G}_{2,2}^{(I)}$. Green dashed lines cut the Diamond between points $x_1x_2$ and $x_3x_4$, defining two open-index operators $O$ and $O'$.}
\label{class_I_OPE_fig_1}
\end{center}
\end{figure}

Let us analyse the OPE of \eqref{mn_class_I} in the s-channel $x_1\to x_2,\, x_3 \to x_4$, and focus on the exchange of two-point functions $\langle O(x) O^{'\dagger}(0)\rangle$ of spinless operators. All such possibilities can be detected at glance by slicing the Diamond diagram between $x_1x_2$ and $x_3x_4$ as done in figure~\ref{class_I_OPE_fig_1} for $m=2$. The picture is general enough as other cases are either related by re-labelling fields or zero due to $[O]\neq [O']$.

\begin{figure}[t]
\begin{center}
\includegraphics[scale=0.65]{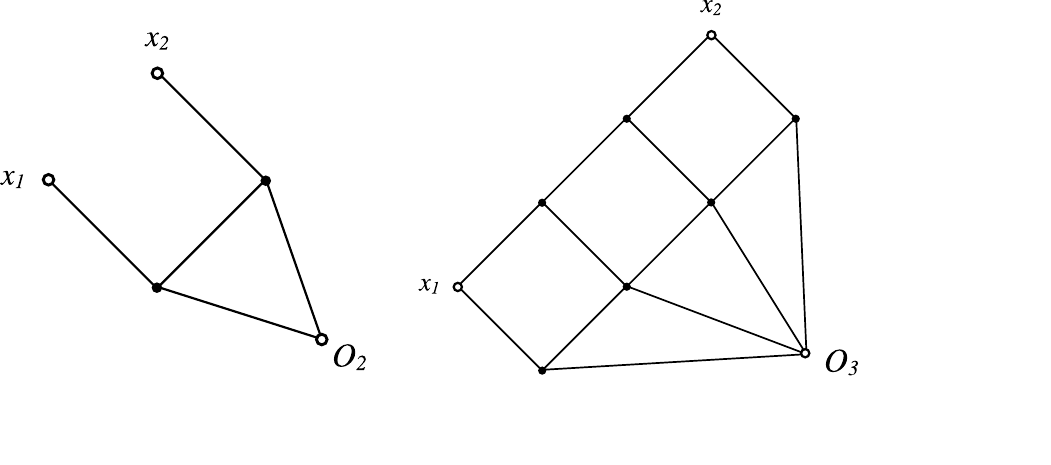}
\caption{The single-trace three-point functions $\langle\text{Tr}\,(Z_4 Z_3)(x_2)(Z_1 \bar Z_4)(x_1) (\bar Z_3\bar Z_1)(x_0)\rangle$ and $\langle\text{Tr}\,(Z_4 Z_3)^2(x_2)(Z_1 \bar Z_4)^2(x_1) (\bar Z_3\bar Z_1)^2(x_0)\rangle$ receive loop-level contributions by the Feynman diagrams depicted on the left and right respectively. Their evaluation by star-triangle identities delivers a nonzero result.}
\label{class_I_OPE_3pt}
\end{center}
\end{figure}
The shortest exchanged operator for any $m$ is
\begin{equation}
\label{shortest_m}
\frac{1}{N} \left\langle \Tr[(Z_1 Z_3)^{(m-1)}(x) (\bar Z_3 \bar Z_1)^{(m-1)}(0)]\right\rangle\,.
\end{equation}
It has non zero tree-level but it is protected from loop corrections in the planar theory. 
At the same time, these operators form single-trace three-point functions at loop order $(\xi_1^2 \xi_2^2)^{\frac{m(m-1)}{2}}$ with the external states, depicted in figure~\ref{class_I_OPE_3pt} for $m=2,3$. These three-point functions for general $m=L+1$ evaluate to
\begin{align}
&\frac{1}{N}\left\langle\Tr[(Z_4 Z_3)^L(x_2)(Z_1 \bar Z_4)^L(x_1) (\bar Z_3\bar Z_1)^L(x_0)]\right\rangle= \\
& \qquad\qquad\qquad\qquad\qquad=\frac{\left(\xi_1^{2} \xi_2^{2} \, A(\Delta_1+\Delta_4,\Delta_2,\Delta_3+\Delta_4,2-\Delta_4) \right)^{\frac{L(L+1)}{2}}}{(x_{10}^2)^{L \Delta_1}(x_{20}^2)^{L \Delta_3}(x_{12}^2)^{L \Delta_4}}\,, \notag \\
&\frac{1}{N}\left\langle\Tr[(Z_1 Z_3)^L(x_0) (\bar Z_2 \bar Z_1)^L(x_4) (\bar Z_3 Z_2)^L(x_3)]\right\rangle= \notag \\
& \qquad\qquad\qquad\qquad\qquad =\frac{\left( \xi_1^{2} \xi_2^{2} \, A(\Delta_2+\Delta_3,\Delta_4,\Delta_1+\Delta_2,2-\Delta_2) \right)^{\frac{L(L+1)}{2}}}{(x_{30}^2)^{L\Delta_3}(x_{40}^2)^{L\Delta_1}(x_{34}^2)^{L \Delta_2}}\,, \notag
\end{align}
where, as in equations \eqref{IN2L-1} and \eqref{IN2L}, we have taken the propagators to be simply $D_i(x) = x^{-2\Delta_i}$ so as to make the expressions shorter. We will do the same in all the other diagram computations of this section.

Notice that the OPE considerations made so far hold essentially for the exchange of any spinning operator $(\partial^{J_1} Z_3 \partial^{J_1'} Z_1)\cdots (\partial^{J_{m-1}} Z_3 \partial^{J_{m-1}'} Z_1)$, which have the same, minimal, twist $\Delta-J = (m-1)(\Delta_1+\Delta_3)$ and thus dominate the light-cone OPE limit $x_{12}^2,\,x_{34}^2\to 0$.

\begin{figure}
\begin{center}
\includegraphics[scale=0.7]{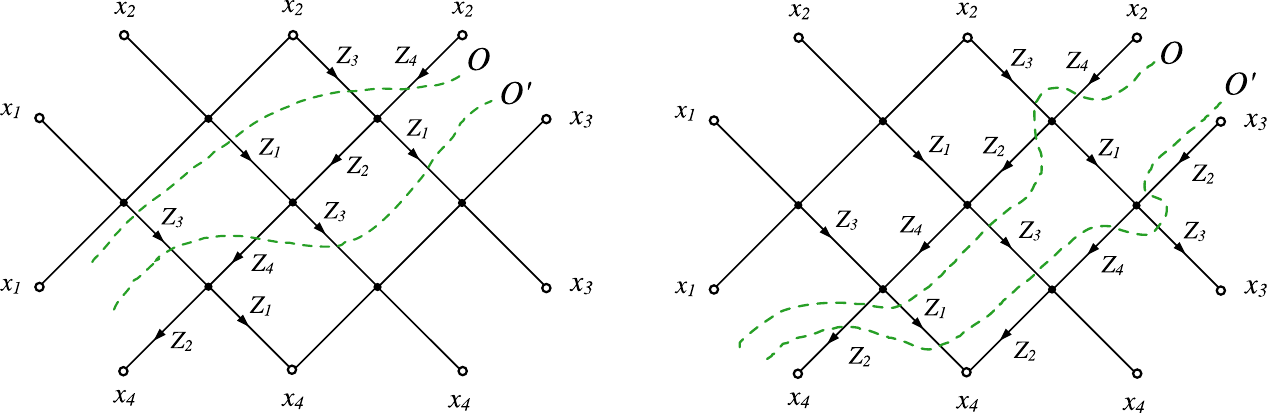}
\caption{The Feynman integral contributing to $\widetilde{G}_{2,3}^{(I)}$ at loop order $(\xi_1^2)^4(\xi_2^2)^3$.On the left, two choices $O,O'$ for a shortest exchanged operator (length $4$). On the right the exchange of two operator with nonzero two-point function though protected scaling dimension, of the type \eqref{only_normalization}.}
\label{class_I_OPE_cut_rec}
\end{center}
\end{figure}

Let us complete the analysis for $m=2$ with the other two exchanges in figure~\ref{class_I_OPE_fig_1}. One case (right picture) involves tree-level three-point function and four-loop two-point which evaluates to zero whenever $\Delta_2\neq \Delta_4$. The central picture is rather curious: the two-point function for $O=Z_4Z_3\bar Z_2$ and $O{'}=\bar Z_2 Z_3 Z_4$ is non-zero but it is not divergent either:
\begin{equation}
\label{only_normalization}
\frac{1}{N}  \left\langle \Tr[O(x) O^{'\dagger}(0)]\right\rangle 
  = \xi_1^2 \xi_2^2  \frac{A(\Delta_1,4-\Delta_1)}{(x^2)^{\Delta_2+\Delta_3+\Delta_4}}\,.
\end{equation}
At the same time, it is simple to verify that the three-point functions are zero. Hence the correlators for $m=2$ can be decomposed over the exchange of $(\partial^{J} Z_3 \partial^{J'} Z_1)$ only, plus descendants. For clarity, let's point out that for general $m$ the result  \eqref{mn_class_I} does not contradict the exchange of operators which receive loop corrections in the normalisation without getting anomalous dimension. The absence of these in the OPE expansion should be checked by computing three-point functions, as we did for $m=2$.

The analysis can be repeated for $m\neq n$, for which the Diamond integrals evaluate to zero. Take for instance the connected diagrams for $m=2,n=3$ in figure~\ref{class_I_OPE_cut_rec}. We focus on the shortest operators in the s-channel. For $m \neq n$ it is not possible to cut the diagram along a row of lattice, and the cut must bend horizontally or vertically for some cells and crossing propagators of type $Z_2$ or $Z_4$ as well. The way of bending is not unique, and there are more options of shortest operator. For example, with reference to Fig.~\ref{class_I_OPE_cut_rec} (left), the two operators of length-$4$ are
\begin{equation}
    O = Z_4 Z_3 Z_1 Z_3 \,,\,\,\,O' = Z_1 Z_3  Z_4 Z_3 \,.
\end{equation}
These operators mix the diagram of Ladder type $L_2$  enclosed between dashed cuts in Fig.~\ref{class_I_OPE_cut_rec}, which is $1/\epsilon$-divergent as a two-point function, as argued in section \eqref{sec:ladders}. Therefore $\langle \Tr[OO'^{\dagger}]\rangle$ gets anomalous dimension. Nevertheless, the three-point function of $O$, $O'$ with the external open-index states evaluate to zero. One can consider also longer operators as those depicted on the right of Fig.~\ref{class_I_OPE_cut_rec},
\begin{equation}
    O = Z_4 Z_3 \bar Z_2 Z_3 Z_1\,,\,\, O' = \bar Z_2 Z_3 Z_4 Z_3 Z_1\,.
\end{equation}
In this case the two-point function is of the type \eqref{only_normalization}, hence non zero. Once gain, the three-point functions vanish, in agreement with the result \eqref{mn_class_I}.

All together, since the OPE of Diamonds with $m \neq n$ must reproduce a zero result, one always encounters either operators with nonzero two-point functions $\langle \Tr[OO^{'\dagger}]\rangle$ which multiply vanishing three-point functions, or whenever the three-point functions are non-zero (e.g. tree level) the exchanged operators have $[O]\neq [O']$ for general $\Delta_k$, so they cannot mix.

\subsection{Diamond Correlators of Type II}
\label{sec:diamonds_II}

\begin{figure}
\centering
\includegraphics[scale=0.7]{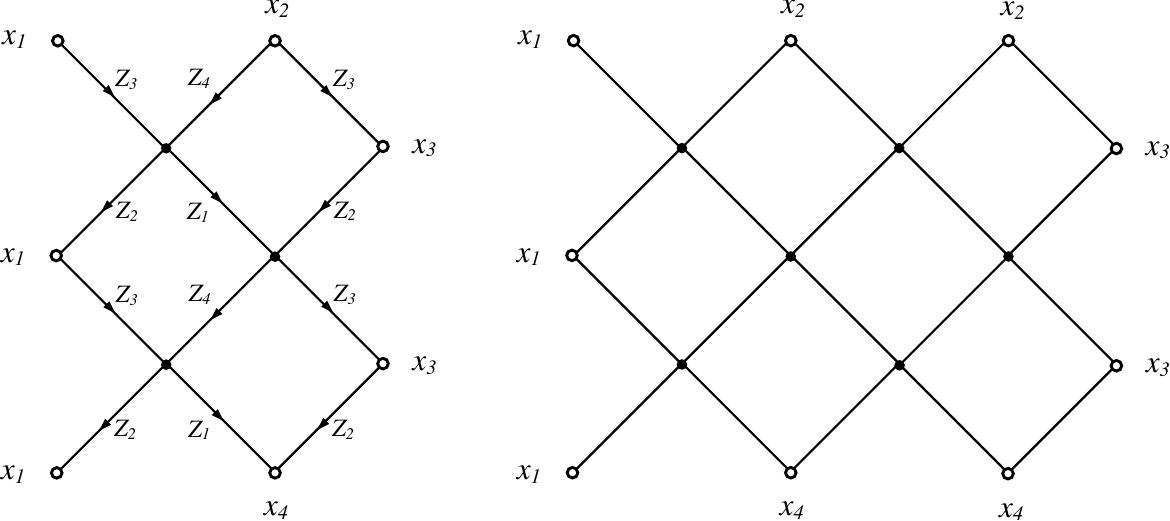}
\caption{The Diamond Feynman integrals of type II for $m=2,n=1,2$.We mark with letters $Z_k$ the propagator of Checkerboard fields. The orientation of arrows is $Z \to \bar Z$.}
\label{class_II_diag}
\end{figure}

\begin{figure}
\begin{center}
\includegraphics[scale=0.58]{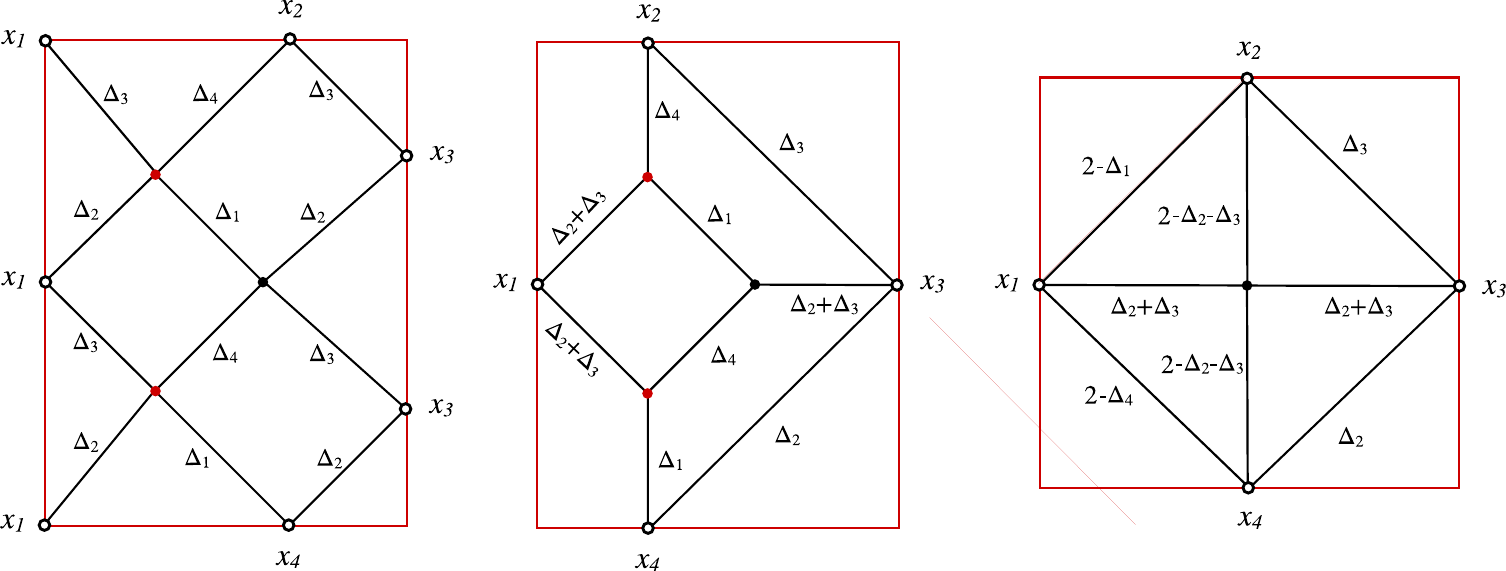}
\caption{\textbf{Left:} The Diamond $G^{(II)}_{m,n}$ four-point integral $(m,n)=(2,1)$. \textbf{Centre:} The same integral with merged points. Red cubic vertices are integrated by star-triangle identity. \textbf{Right:} the result is a conformal cross-integral with powers $\Delta_2+\Delta_3$ and $2-\Delta_2-\Delta_3$, times disconnected terms of type $(x^2_{i,i+1})^{\beta}$.}
\label{class_II_str}
\end{center}
\end{figure}

In this section we analyse the Diamond correlators of type II, and repeat the OPE analysis of the previous section. The correlators are defined for any $n$ and $m$ as
\begin{equation}
\label{checker_4pt_II_odd_m}
  G^{(II)}_{m,n}= \frac{1}{N} \left\langle \Tr[(Z_3 \bar Z_2)^{m}(x_1) (\bar  Z_1 \bar Z_2)^{n}(x_4) (Z_2 \bar Z_3)^{m}(x_3) (Z_3 Z_4)^{n}(x_2)]\right\rangle\,.
\end{equation}
These correlators are zero in the free theory, and receive loop corrections at order \newline
$\xi_1^{2n(m-1)} \xi_2^{2mn}$ due to one Fishnet diagram of the type in figure~\ref{class_II_diag}.
For generic parameters $\Delta_k$, the open-index operators that enter the trace in \eqref{checker_4pt_II_odd_m} are protected and the four-point integral is finite. Using only the star-triangle identity, one can compute the following:
\begin{itemize}
    \item $m<n$: the integral evaluates to zero, $G_{m<n}^{(II)}=0$.
    \item $m=n$: the integral with $2m^2-m$ loops evaluates to a product of powers of the distances $x_{ij}^2$, similarly to Diamonds of type I:
    \begin{equation}
        G_{m,m}^{(II)} = \xi_1^{2m(m-1)} \xi_2^{2m^2} \frac{\pi^{4m^2-2m} A(\Delta_1,\Delta_4,\Delta_2+\Delta_3)^{m^2} A(\Delta_1+\Delta_2,\Delta_3+\Delta_4)^{\frac{m(m-1)}{2}}}{(x_{12}^2)^{m(2-\Delta_1)}(x_{14}^2)^{m(2-\Delta_4)}(x_{24}^2)^{m(2-\Delta_2-\Delta_3)}(x_{23}^2)^{m\Delta_3}(x_{34}^2)^{m\Delta_2}}\,.
    \end{equation}
    The computation technique is exemplified in figure~\ref{fig:STR_II_22} for $m=n=2$.
    \begin{figure}
        \centering
        \includegraphics[scale=0.50]{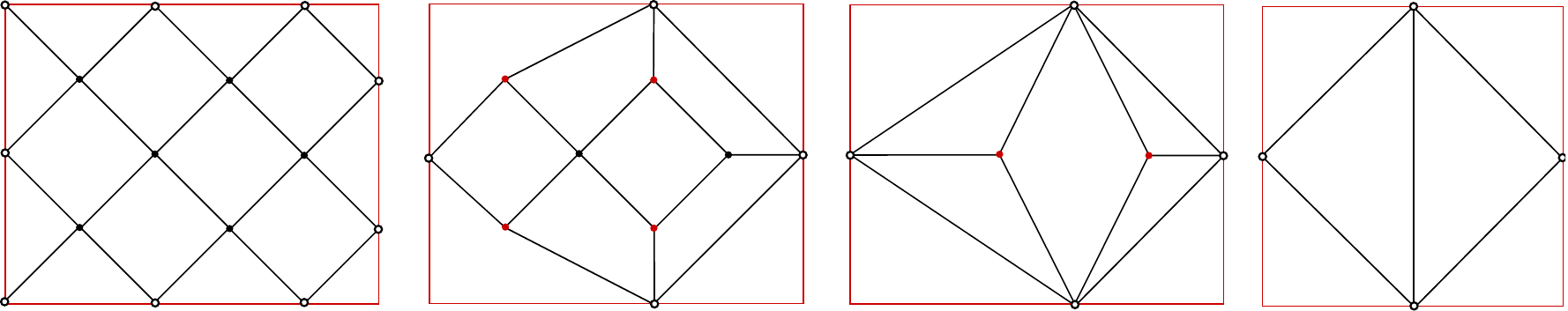}
        \caption{The star-triangle computation of $G_{2,2}^{(II)}$. From left to right: merging of external points, followed by star-triangle integration of points marked with red dots.}
        \label{fig:STR_II_22}
    \end{figure}
    \item $m>n$: the integral can be brought into the form of a Rectangular Basso--Dixon Fishnet of size $(m-n)\times n$. The latter Fishnet has horizontal/vertical propagators with the scaling dimensions $\omega=\Delta_2+\Delta_3$ and $\tilde\omega = \tfrac{d}{2}-\omega$ respectively.
    \begin{align}
    \begin{aligned}
    \label{nontrivial_Diamond}
        G_{m>n}^{(II)} &= \pi^{2 mn+2n(n-1)} A(\Delta_1,\Delta_4,\Delta_2+\Delta_3)^{m n} A(\Delta_1+\Delta_2,\Delta_3+\Delta_4)^{\frac{n(n-1)}{2}} \times \\ &\times \frac{\xi_1^{2n(m-1)} \xi_2^{2mn}}{(x_{12}^2)^{n(2-\Delta_1)}(x_{23}^2)^{n\Delta_3}(x_{34}^2)^{n\Delta_2}(x_{14}^2)^{n(2-\Delta_4)}} \times I^{(BD)}_{m-n,n}(\omega; x_1,x_2,x_3,x_4) \,.
    \end{aligned}
    \end{align}
    The transformation from Diamond to BD diagrams is explained in figure~\ref{class_II_str} for the simplest case $(m,n)=(2,1)$ and it is otherwise more cumbersome (we checked it only up to $m=4, n<m$).
\end{itemize}

These diagrams can also be computed using SoV techniques, as explained in appendix~\ref{app:diamond SoV}. This is much more direct than the SoV computations of section \ref{sec:BD} since we do not need to construct new resolutions of the identity. However, these SoV expressions make it harder to observe the simple behaviour of $G_{m,n}^{(II)}$ when $m\leqslant n$. As a consistency check, we spell out in details the SoV expressions for $(m,n)\in\{(1,1),(1,2),(2,1)\}$ in appendix~\ref{app:diamond SoV check}.

Whenever $m\leqslant n$, Diamonds of type II show similar behaviour as Diamonds of type I. Thus, an OPE analysis would lead to analogue considerations about protected anomalous dimensions and/or single-trace three-point functions. 

\subsubsection{From diamonds to rectangular fishnets}

In the following we will focus on the case $m>n$. The Feynman diagrams of figure~\ref{class_II_diag} always have a trivial term $G^{(II)}_{1,0}=D_2(x_{34})D_3(x_{23})$. We will concentrate on its nontrivial connected component $\widetilde G^{(II)}_{m,n}$,
\begin{align}
\begin{aligned}
\label{checker_4pt_II_conn}
    &G^{(II)}_{m,n} =\frac{1}{N} G^{(II)}_{1,0} \times \widetilde G^{(II)}_{m,n}\,,\\ &\widetilde G^{(II)}_{m,n} = \left\langle \Tr[(\bar Z_2 Z_3)^{n}(x_1) ((Z_4 Z_3)^{m-1}Z_4)(x_2) (\bar Z_3 Z_2)^{n}(x_3) ((\bar Z_1 \bar Z_2)^{m-1}\bar Z_1)(x_4)]\right\rangle\,.
\end{aligned}
\end{align}

The case at hand is the first instance of Diamond integral that evaluates to a non-trivial function of the conformal cross-ratios
\begin{equation}
    I^{(BD)}_{m-n,n}(\omega; x_1,x_2,x_3,x_4) = \frac{ \mathcal{I}^{(BD)}(\omega;z,\bar z)}{(x_{12}^2 x_{34}^2)^{2m -n}},\,\,\, z \bar z = \frac{x_{12}^2x_{34}^2}{x_{14}^2x_{23}^2}\,,\,\,(1-z)(1-\bar z) = \frac{x_{13}^2x_{24}^2}{x_{14}^2x_{23}^2}\,,
\end{equation}
where $\mathcal{I}_{m-n,m}^{(BD)}(\omega;z,\bar z)$ was solved via separation of variables in \cite{Derkachov:2019tzo}. Most importantly, in the limit $z \to 0$, it  develops a leading UV divergence \begin{equation}
\label{div_BD}
    \mathcal{I}^{(BD)}(\omega;z,\bar z) \sim (\log z)^{(m-n)n} \times \Phi_{m-n,n}(\omega;\bar z)\,,
\end{equation}
where $\Phi_{a,b}(\omega;\bar z)$ is an analytic function around $\bar z = 0$. The log-behaviour implies that the $s$-channel of $G^{(II)}_{m>n>0}$ must include the exchange of operators with non-zero anomalous dimension.

Let us proceed with a qualitative OPE analysis.  We have argued that the spectrum of scaling dimensions is protected at one-loop unless $\Delta_k$ satisfy \eqref{special}. In addition, by conformal symmetry, 
the maximal light-cone $s$-channel divergence of a correlator at $n$-loops is $\gamma_{1,O}^n \log^n z $, where $\gamma_{m,O}$ is the $m$-loop coefficient of the anomalous dimension of an exchanged operator $O$. This does not contradict \eqref{nontrivial_Diamond}; indeed, at loop order $2 m n -m$, namely $\xi_1^{2n(m-1)} \xi_2^{2mn}$, the maximal divergence of a Diamond \eqref{div_BD} is weaker:
\begin{equation}
    (m-n)n < 2mn -n\,,\,\,\, m,n>0\,.
\end{equation}
More in detail, let us consider the three-loop integral for $m=2,n=1$ in figure~\ref{class_II_diag}.
It lacks both the leading-logarithm $\sim\log^3 z$, and the first sub-leading one $\sim\log^2 z$, which could multiply, respectively, $\gamma_{1,O}^3$ or a combination of $\gamma_{1,O} \gamma_{2,O}$ and $\gamma_{1,O}^2$. Let us denote by $C_O^{(k)}$ the $k$-th loop order correction to structure constants in the OPE. Coherently with our result, the maximal log that appears (also) in the  terms that do not include $\gamma_{1,O}$, such as $C_O^{(1)} \gamma_{2,O}$ or $C_O^{(0)} \gamma_{3,O}$.

\begin{figure}
\begin{center}
\includegraphics[scale=0.57]{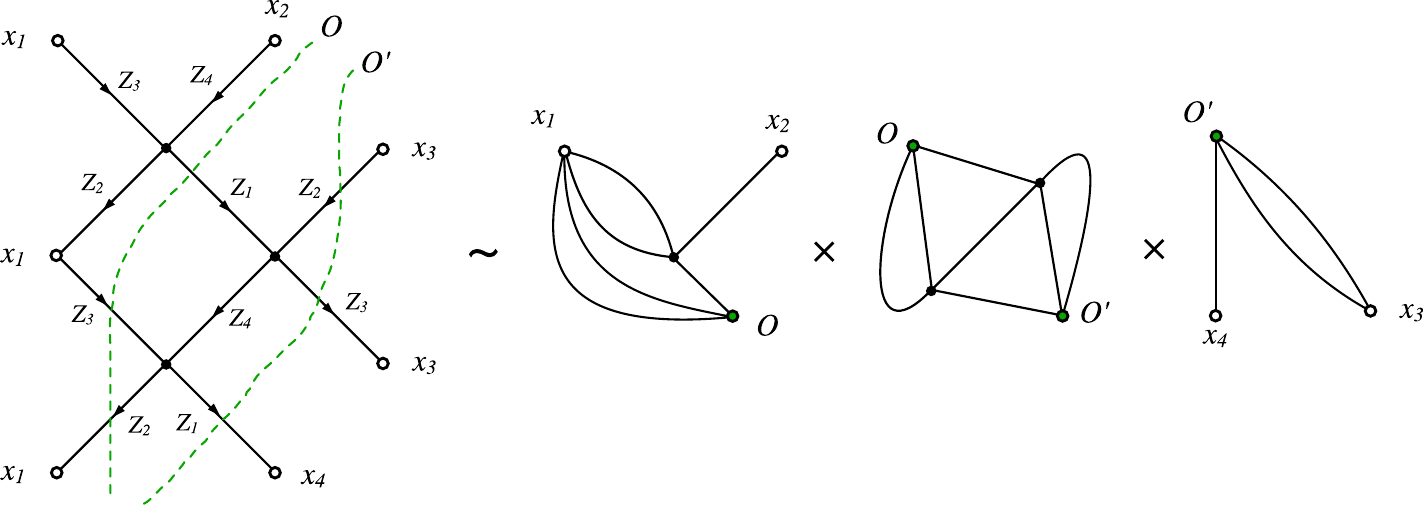}
\caption{Projection of $\widetilde G^{(II)}_{2,1}$ over the exchange of $O,O'$ in the $s$-channel. The two-point function $\left\langle \Tr[O O'^{\dagger}]\right\rangle = \left\langle \Tr[(Z_1 Z_3 \bar Z_2) (Z_2 \bar  Z_3 \bar Z_1)]\right\rangle$ is $1/\epsilon$-divergent OPE limit of a Ladder diagram $L_2$. It multiplies finite three-point functions at order $\xi_1^{2}$ (left) and tree-level (right) computed in \eqref{3pt_example}.}
\label{class_II_ope_Gamma}

\vspace*{4 mm}

\includegraphics[scale=0.57]{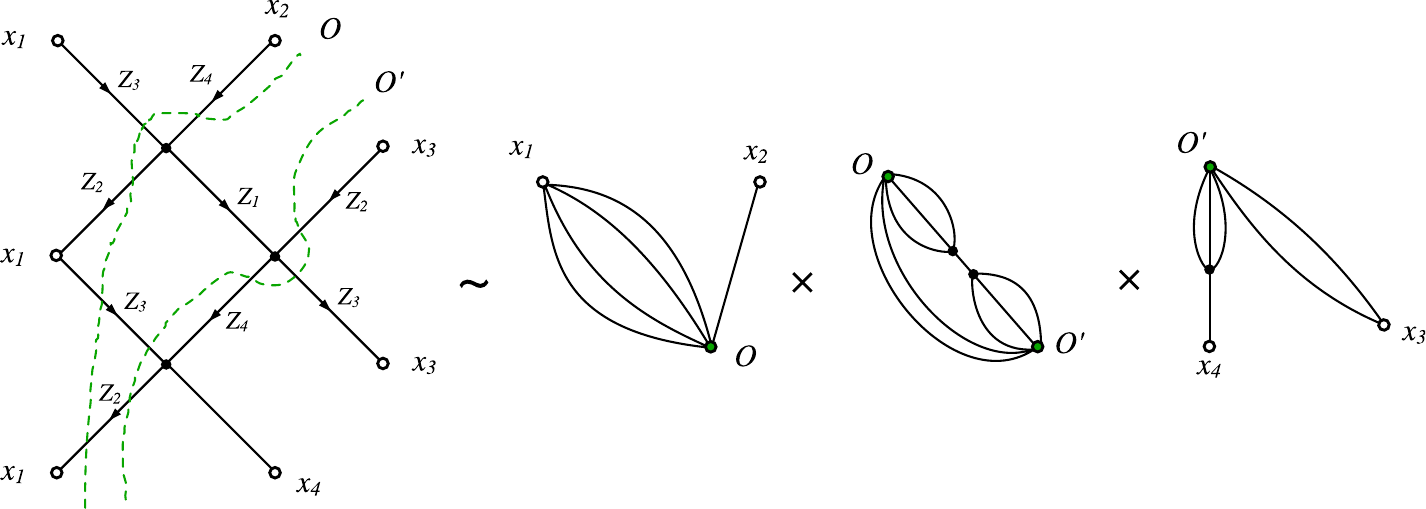}
\caption{The two-point function $\left\langle \Tr[O O'^{\dagger}]\right\rangle = \left\langle \Tr[(Z_4 Z_3 \bar Z_2 Z_3\bar Z_2) (Z_2 \bar Z_3 \bar Z_4 \bar Z_3 Z_2)]\right\rangle$ is given by a  finite integral at loop level $\xi_1^2 \xi_2^2$, of the type \eqref{only_normalization}. Though, it multiplies a vanishing three-point function at order $\xi_1^{2}$ (on the right), hence it cannot appear in the OPE.}
\label{class_II_ope_NGamma}
\end{center}
\end{figure}

The analysis can be pushed a little further. The rigid square-lattice structure of planar Checkerboard diagrams allows to read at glance all possible operators that flow in the $s$-channel.
The shortest scalar two-point function exchanged in the $s$-channel, according to the illustration in figure~\ref{class_II_ope_Gamma}, reads
\begin{equation}
    \label{twopt_2loop}
  \frac{1}{N}  \left\langle \Tr[(Z_1 Z_3 \bar Z_2)(x)(Z_2 \bar  Z_3 \bar Z_1)(0)]\right\rangle = \xi_1^2 \xi_2^2  \lim_{x_1\to x_2,\,x_3\to x_4} I_{1,2}(x_1,x_2,x_3,x_4)\,,
    \end{equation}
where we refer to the notation for Ladders \eqref{Ladder_I}. This diagram is $1/\epsilon$-divergent in dimensional regularisation, hence it generates anomalous dimension at two-loops $\gamma_{2,O}\neq 0$. One can check that this (plus spinning counterpart and descendants) is the only non-protected two-point function flowing in the $s$-channel of \eqref{checker_4pt_II_conn} with $n=2,m=1$, and it accounts for the UV behaviour $\sim \log z$. The three-point functions involved in the OPE for the spinless operator, also depicted in figure~\ref{class_II_ope_Gamma}, are given by:
\begin{align} \label{3pt_example}
    &\frac{1}{N}\left\langle \Tr[Z_4(x_2) (Z_3 \bar Z_2 Z_3 \bar Z_2)(x_1) (Z_2 \bar Z_3 \bar Z_1)(x_0)]\right\rangle= \\
    & \qquad\qquad\qquad\qquad\qquad =\frac{\xi_1^2\, A(\Delta_2+\Delta_3,\Delta_1,\Delta_4)}{(x_{10}^2)^{2-\Delta_4+\Delta_2 +\Delta_3}(x_{20}^2)^{2-\Delta_2-\Delta_3}(x_{12}^2)^{2-\Delta_1}}\,, \notag \\
    &\frac{1}{N}\left\langle \Tr[(\bar Z_2 Z_3 Z_1)(x_0) \bar Z_1 (x_4) (\bar Z_3 \bar Z_2) (x_3)]\right\rangle = \frac{1}{(x_{30}^2)^{\Delta_2+\Delta_3}(x_{40}^2)^{\Delta_1}}\,. \notag
\end{align}

\subsubsection{Spectrum and logarithmic multiplets}

The lightest two-point function \eqref{twopt_2loop} pairs two different operators $O$ and $O'$, hence it is just a component of a larger multiplet. There are three open-index operators that can be mixed by \eqref{CheckerboardCFT} in the planar limit, that is
\begin{align}
\begin{aligned}
\label{full_multi}
O_h \in \{Z_1 Z_3 \bar Z_2 \, ,\, Z_1 \bar Z_4 Z_1 \,, \, \bar Z_2 Z_3 Z_1 \}\,.
\end{aligned}
\end{align}
We draw the explicit Feynman diagrams which mix these operators to form a multiplet:

\begin{center}
\includegraphics[scale=0.58]{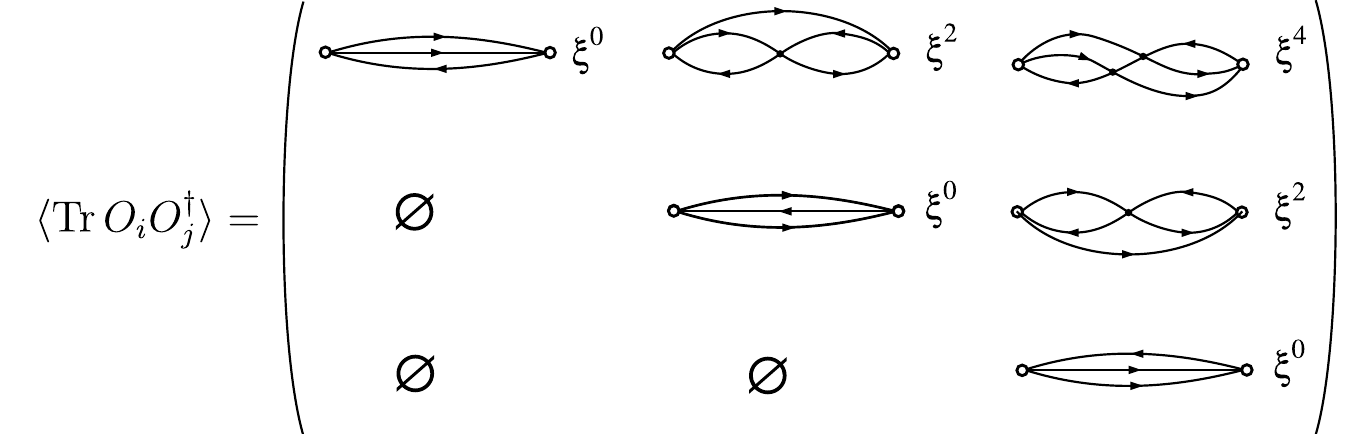}
\end{center}

The matrix is not diagonalisable, and the multiplet is logarithmic. The diagrams of order $\xi_i^2$ actually vanish for generic $\Delta_k$, so the mixing matrix $\sqrt{Z_{ij}}$ reads
\begin{equation}
\label{Z_mix_II}
    \sqrt{Z_{ij}} = \begin{pmatrix}
        1 && 0 && -\frac{\gamma_2}{2\epsilon} \\
        0 && 1 && 0 \\
        0 && 0&& 1
    \end{pmatrix}\,,
\end{equation}
where $\gamma_2$ is the residue at pole $1/\epsilon$ of the aforementioned Ladder integral at $\xi_1^2\xi_2^2$. Thus, we get a $\log$-multiplet of rank two which mixes $\bar Z_2 Z_3 Z_1$ and $Z_1 Z_3 \bar Z_2$, whereas $Z_1 \bar Z_4 Z_1$ stays protected. The two-point functions of renormalised operators follows from \eqref{Z_mix_II}. If we denote $\Delta=\Delta_1+\Delta_2+\Delta_3$ they read
\begin{align}
    \begin{aligned}
       &\frac{1}{N} \left\langle \Tr[ O^{(r)}_{1} O^{(r)\dagger}_{1}] \right\rangle = \frac{1}{(x^2)^{\Delta}}\,, &\frac{1}{N} \left\langle \Tr[ O^{(r)}_{1} O^{(r)\dagger}_{2}] \right\rangle = \xi_1^2 \xi_2^2 \,\frac{\log(x^2 \mu^2)}{(x^2)^{\Delta}}\,,\\
        &\frac{1}{N} \left\langle \Tr[ O^{(r)}_{2} O^{(r)\dagger}_{1}] \right\rangle = 0\,, &\frac{1}{N} \left\langle \Tr[ O^{(r)}_{2} O^{(r)\dagger}_{2}] \right\rangle = \frac{1}{(x^2)^{\Delta}}\,.
    \end{aligned}
\end{align}
Let us conclude the section with a look at the multiplet  \eqref{full_multi} at the special point $\Delta_2+\Delta_3=\Delta_1+\Delta_4=2$, when the one-loop spectrum of Checkerboard CFT becomes non-trivial. The Feynman integrals at order $\xi_i^2$ develop $1/\epsilon$-divergence, whereas the integral at order $\xi_1^2 \xi_2^2$ gets a stronger leading term $1/\epsilon^2$. The mixing matrix evaluates to
\begin{equation}
\label{Z_mix_II_special}
    \sqrt{Z_{ij}} = \begin{pmatrix}
        1 && -\frac{\gamma_1}{2\epsilon} && -\frac{\gamma_2}{2\epsilon}-\frac{\gamma_1^2}{4\epsilon^2} \\
        0 && 1 && -\frac{\gamma_1}{2\epsilon} \\
        0 && 0&& 1
    \end{pmatrix}\,,
\end{equation}
 and consequently, the two-point functions of renormalised operators form a $\log$-multiplet of rank three: 
\begin{equation}
\frac{1}{N}   \left\langle \Tr[O^{(r)}_{i}(x) O^{(r)\dagger}_{j}(0)] \right\rangle = \frac{1}{(x^2)^{\Delta_1+\Delta_2+\Delta_3}} \times \begin{pmatrix}
        1 && \xi_1^2 \log(x^2 \mu^2) && \xi_1^2\xi_2^2 \log^2 (x^2 \mu^2) \\
        0 && 1 && \xi_2^2 \log(x^2 \mu^2) \\
        0 && 0&& 1
    \end{pmatrix}\,.
\end{equation}
Notice that special points of this type include all theories where $\Delta_1=\Delta_3$ and $\Delta_2=\Delta_4$, when the quartic vertices feature two orthogonal pairs of propagators with dimensions $\Delta,\Delta'=2-\Delta$ as in anisotropic bi-scalar FCFT \cite{Kazakov:2018qbr}. 



\subsubsection{Parity of the all-loop spectrum}

The OPE analysis of this section hints that the spectrum of anomalous dimensions in a Checkerboard CFT with four slopes is zero at odd orders in the loop expansion. Hence, for any local operator $O$ in the theory its anomalous dimension expands as
\begin{equation}
\label{theconjecture}
{\gamma}_O(\xi_1,\xi_2)=\sum_{n=1}^{\infty}\sum_{r=0}^{n} \xi_1^{2r}\xi_2^{4n-2r} \gamma_O^{(r,2n-r)}\,.
\end{equation}
The last statement can be proved by induction.
Since the discussion of Ladders in section \ref{sec:ladders} we found that $\gamma_O^{(1)}=0$ by looking at the analytic structure of the cross-integral $L_1$. More trivially, this statement follows from conformal symmetry: the only one-loop diagram is a quartic Checkerboard vertex between two couples of fields with mismatching bare dimensions.

\begin{figure}
        \centering
        \includegraphics[scale=0.58]{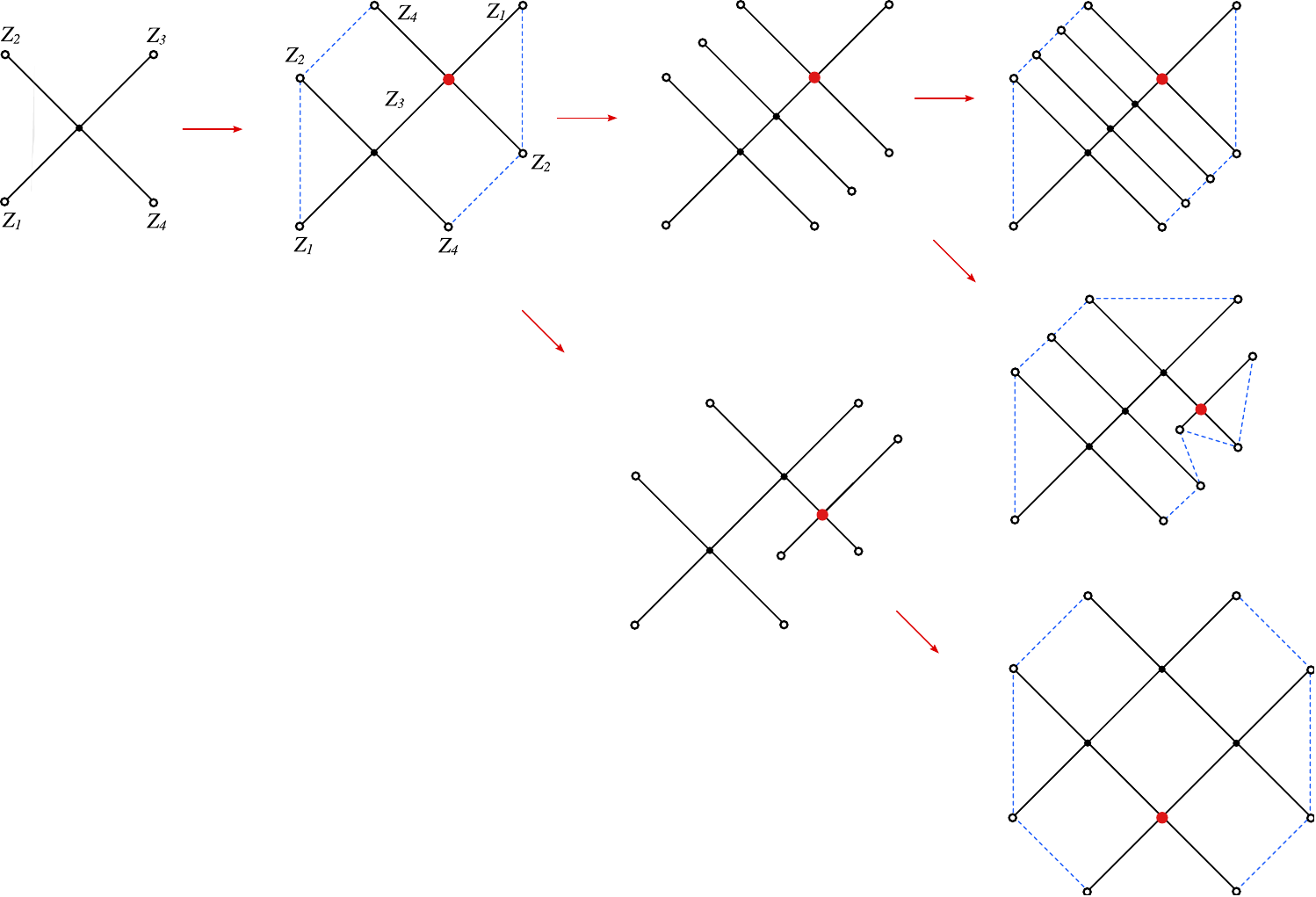}
        \caption{Starting from the vertex of the theory, i.e. a cross-integral at one-loop, we generate higher-loop Checkerboard Fishnets by adding a vertex in various ways. We connect with a blue dashed lines group of points that, once merged, form a two-point functions of open-index operators with the same bare dimension. In the text we argue that this can happen only at even loop order.}
        \label{fig:enter-label}
\end{figure}

Next, the Ladder integral $L_2$ is $\log$-divergent in the OPE limit for general $\Delta_k$. This is enough to guarantee that $\gamma_O^{(2)}\neq 0$.
Coherently, the two-point function obtained in this limit involves two operators with the same bare dimension.


\begin{figure}
\begin{center}
\includegraphics[scale=0.9]{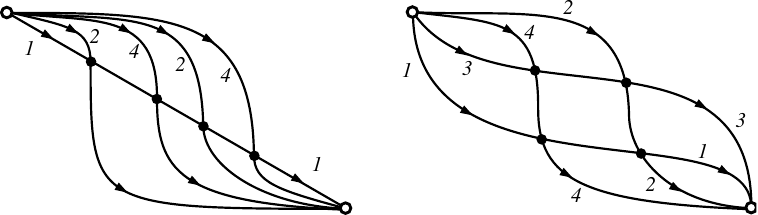}
\caption{There are only two four-loop two-point diagrams (modulo cyclic permutations of fields) that correct the spectrum of the theory. Both appear at order $\xi_1^4 \xi_2^4$. The first (left) is obtained identifying a pair of points in the Ladder $L_4$, the second is the result of the same identification in the Rectangular Fishnet $I_{2,2}$. Both diagrams have a leading divergence $\sim 1/\epsilon^2$ in dimensional regularisation.}
\end{center}
\label{ladder_spectr_4L}
\end{figure}
Let us formulate the induction. Suppose that for a given $m$-loop Checkerboard Fishnet it is possible to group external legs into two operators $O_1,O_2$ such that $[O_1]=[O_2]$. Then, adding one vertex to the diagram increases the total scaling dimension of external legs by
 $n_1 \Delta_1 +n_2 \Delta_2 +n_3 \Delta_3 +n_4 \Delta_4$ with $n_k = \pm 1$. We distinguish three cases:
 \begin{itemize}
     \item Add three legs on the external point of a leg of type $Z_j$ or $\bar Z_j$ and make it a vertex: $n_j=-1,\,n_{k\neq j}=1$;
     \item Merge two legs in one external point and add two legs on top, making the point into a vertex: $n_j=n_{j+1}=-1,\,n_{j+2}=n_{j+3}=1$;
    \item Merge three legs in one external point and add one leg on top: $n_j=1 ,\,n_{k\neq j}=-1$;
 \end{itemize}
 In order to see if the external fields of the Checkerboard at $m+1$ loops can be grouped in two new operators $O_1',O_2'$ with $[O_1']=[O_2']$ we can also re-distribute $p_k$ fields of type $Z_k$ or $\bar Z_k$ among $O_1, O_2$. All together we can write
\begin{equation}
    [O_1'] = [O_1]+\sum_{k} n_k \Delta_k - \sum_{k} p_k \Delta_k\,,\,\,\, [O_2'] = [O_2]+ \sum_{k} p_k \Delta_k\,,\,\,\, p_k \in \mathbb{Z}
\end{equation}
Imposing that bare dimensions are equal and the constraint $\sum_k \Delta_k =4$, one gets the equation
\begin{equation}
\label{theeq_gamma}
    8 p_4 + 2\sum_{k=1}^3 (p_k-p_4) \Delta_k = 4 n_4 +\sum_{k=1}^3 (n_k-n_4) \Delta_k\,.
\end{equation}
We want a solution valid for any choice of $\Delta_{k=1,2,3}$ hence we can differentiate \eqref{theeq_gamma} getting three equations \begin{equation}
\label{theeq_final}
    2 (p_k-p_4) = n_k-n_4\,.
\end{equation}
Imposing \eqref{theeq_final} into \eqref{theeq_gamma} one gets $n_4 = 2 p_4$, which contradicts $n_h=\pm 1$. On the other hand, any way of adding two vertices, ie. two loops, to a Checkerboard diagram requires $n_h=0,\pm 2$. Therefore, there is always a solution $p_h = n_h/2 \in \mathbb{Z}$ that describes a way of grouping fields into operators of matching classical dimensions at $m+2$ loops.
   
All together, if there is a non-zero two-point function in Checkerboard theory at $m$ loops, there cannot be one at $m+1$ loops, but there certainly is one at $m+2$ loops. This inductive step together with $\gamma_O^{(1)}=0$ and $\gamma_O^{(2)}\neq 0$ is the proof of \eqref{theconjecture}. Of course, equation \eqref{theeq_gamma} admits solutions even at $m+1$ loops with respect to $\Delta_k$, such as \eqref{special}. For example, setting $p_1=p_2=p_4=0,p_3=1$, $n_4=-1$ and $n_k=1$, the solution is \begin{equation}
    \Delta_3 +\Delta_4 =\Delta_1 +\Delta_2\,.
\end{equation}
Two remarks are due. First, our derivation holds for two-point functions with disk topology (ie. single-trace) with \emph{any} structure of $SU(N)$ indices in the open-index operators. Indeed, we never assumed that $O_k$ or $O_k'$ are made with consecutive fields inside the trace. Second, our conclusion holds true also for multi-trace two-point functions (for instance, the two-point function of single trace operators). Indeed, a multi-trace correlator can be cut and opened onto the disk. If a cut crosses $\ell$ propagators, $2 \ell$ new external legs are generated, adding a total scaling dimension $2 \sum_{h=1}^{\ell} \Delta_{k(h)}=2 \sum_{k=1}^{4} q_k \Delta_{k}$. Equation \eqref{theeq_final} holds after the redefinition $p_k \to p_k-q_k$, hence the cutting has no ``net effect" on our derivation.

We conclude by saying that since $\mathbb{D}_{1,0}=\mathbb{D}_{0,1}=0$, the spectral problem is described at weak-coupling (two-loops) by a next-to-nearest-neighbour spin-chain Hamiltonian.

\section{Checkerboard Lattice with Higher Periods}
\label{sec:higher_periods}

In this section we are going to fully exploit the Loom construction \cite{Kazakov:2022dbd} in order to define a more general Checkerboard Lagrangian. Recall that the Checkerboard theory \eqref{CheckerboardCFT} is defined for $M=4$ slopes in the Loom, which allows to select two quartic vertices featuring four different fields 
\begin{equation}
  \text{Tr}\left[\bar{Z}_1 \bar{Z}_2 Z_3 Z_4\right]\,,\,\, \text{Tr}\left[ Z_1 Z_2 \bar{Z}_3 \bar{Z}_4\right].
\end{equation}
The planar Checkerboard diagrams have the shape of a Fishnet square-lattice where these two vertices alternate along rows/columns. We shall say that it is a Fishnet lattice with periods $(2,2)$ as opposed to the $(1,1)$-periodicity square lattice of the bi-scalar FCFT~\eqref{bi-scalar}. 

\begin{figure}
    \centering
    \includegraphics[scale=0.35]{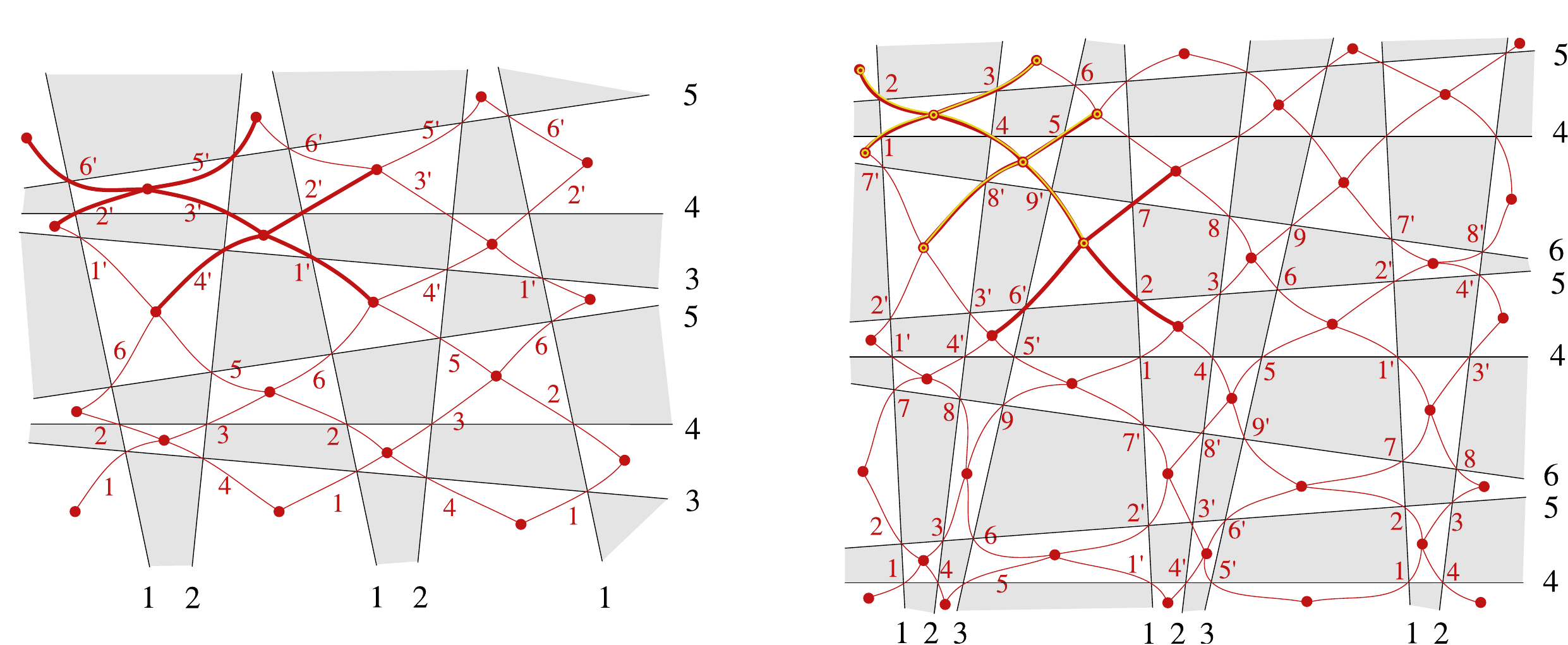}
    \caption{A Baxter lattice made of a number $M_1=2$ (left) or $M_1=3$ (right) alternating \emph{vertical} slopes and $M_2=3$ alternating \emph{horizontal} slopes. The dual Fishnet diagrams are drawn in red (specific diagrams are further highlighted in yellow) and they form a square lattice where the propagators alternate with periodicity $M_1$ and $M_2$ along rows/columns. The case $M_1=M_2=2$ would belong to the Checkerboard CFTs.}
    \label{fig:higher_checker}
\end{figure}

The observations of previous sections about anomalous dimensions being protected at odd-loop order can be regarded as a signature of the Fishnet periodicity. Accordingly, the dilation operator in the weak coupling approximation should be described by a next-to-nearest-neighbour spin-chain Hamiltonian.
Notice that also for $M=4$ slopes there are other Checkerboard theories than \eqref{CheckerboardCFT}, obtained with a different choice of non-vanishing couplings in the full Loom FCFT$^{(4)}$.
\begin{align}
    \begin{aligned}
   &(1,1)\,\,\,\, \text{Tr}\left[Z_1 Z_2 \bar{Z}_1 \bar{Z}_2 \right]\\
    &(1,2)\,\,\,\,  \text{Tr}\left[\bar{Z}_1 \bar{Z}_2 Z_1 Z_4\right]\,,\,\, \text{Tr}\left[ Z_1 Z_2 \bar{Z}_1 \bar{Z}_4\right]\\
    &(2,1)\,\,\,\,  \text{Tr}\left[\bar{Z}_1 \bar{Z}_2 Z_3 Z_2\right]\,,\,\, \text{Tr}\left[ Z_1 Z_2 \bar{Z}_3 \bar{Z}_2\right] \\
     &(2,2)\,\,\,\,  \text{Tr}\left[\bar{Z}_1 \bar{Z}_2 Z_3 Z_4\right]\,,\,\, \text{Tr}\left[ Z_1 Z_2 \bar{Z}_3 \bar{Z}_4\right]\\
    \end{aligned}
\end{align}
Clearly, the bi-scalar theory $(1,1)$ is also obtained from a Loom with $2$ or $3$ slopes, and the $(1,2)$ and $(2,1)$ theories both require only $3$ slopes to be defined from a Loom. Logically, if one starts from a Loom with $M$ slopes, it is possible to define a Checkerboard CFT with periodicity $(M_1,M_2)$ whenever $M\geqslant M_1+M_2$.

Let us consider the two examples in figure~\ref{fig:higher_checker} in order to learn about the general features of these theories. In the first case we increase the number of slopes of the original Checkerboard theory by one slope, $(M_1,M_2)=(2,3)$ and obtain a Fishnet square lattice with periods $(2,3)$ formed by $2\times 3$ chiral vertices built with twelve matrix fields
\begin{equation}
    Z_1,Z_2,\dots,Z_{6}, Z_1',Z_2',\dots Z_6'\,,
\end{equation}
together with their hermitian conjugates. Here fields $Z_k$ and $Z_k'$ have scaling dimensions related by $\Delta_k = d/2-\Delta_k'$, i.e. their propagators cross the same vertices of the Baxter lattice, but across supplementary angles. The Lagrangian for the Checkerboard CFT$^{(M_1,M_2)}$ is a special reduction of the Loom FCFT$^{(M_1+M_2)}$ theory. We can read the interaction vertices for $(M_1,M_2)=(2,3)$ from the diagram in figure~\ref{fig:higher_checker},
\begin{align}
\begin{aligned}
\label{vertex_32}
 & \text{Tr}\left[Z_1 Z_2 \bar Z_3 \bar Z_4 \right]\,,\,\, \text{Tr}\left[Z_6 Z_1' \bar Z_4' \bar Z_5 \right]\,,\,\, \text{Tr}\left[Z_2' Z_6' \bar Z_5' \bar Z_3' \right]\,,\\
  &\text{Tr}\left[Z_3 Z_5 \bar Z_6 \bar Z_2 \right]\,,\,\,\text{Tr}\left[Z_4' Z_3'\bar Z_2' \bar Z_1' \right]\,,\,\,\text{Tr}\left[Z_5' Z_4\bar Z_1 \bar Z_6' \right]\,.
\end{aligned}
\end{align}
As for the spectrum of anomalous dimensions, it is protected at one-loop (and none of the interactions \eqref{vertex_32} generates double-trace counter-terms in the action). On the other hand, the anomalous dimensions are, in general, non-trivial at two-loops. Take the length-$2$ ladder, highlighted with thick red propagators in the left picture of figure~\ref{fig:higher_checker}, and which describes the four-point correlator $$\langle\text{Tr}\left[Z_6'(x_1)(\bar Z_5'\bar Z_2')(x_2)\bar Z_1'(x_3)(\bar Z_4' Z_2')(x_4)\right]\rangle\,.$$
Its OPE channel $x_1\to x_2,\,x_3\to x_4$, features the exchange of operators with non-trivial anomalous dimension. Take the lightest one $$\langle\text{Tr}\left[(Z_6'\bar Z_5'\bar Z_2')(x_2)(\bar Z_1' Z_4' Z_2')(x_4)\right]\rangle\,.$$
First, the two operators have equal bare dimensions by virtue of  $\Delta_5'+\Delta_6' = \Delta_1'+\Delta_4'$, as it follows from the scale-invariance of the second vertex in \eqref{vertex_32}. Indeed, this two-point function has a pole $1/\epsilon$ in dimensional regularisation.  

We also depicted in figure~\ref{fig:higher_checker} the Checkerboard theory corresponding to Fishnet integrals with the shape of a square lattice with periods $(3,3)$, hence formed by $3\times 3$ chiral vertices, built with sixteen matrix fields
\begin{equation}
    Z_1,Z_2,\dots,Z_{8}, Z_1',Z_2',\dots Z_8'\,,
\end{equation}
together with their hermitian conjugates. The Lagrangian for this theory is a reduction, with all but $9$ couplings switched to zero, of the Loom FCFTs$^{(6)}$ theory. The interaction vertices of the theory can be read from the diagram in figure~\ref{fig:higher_checker},
\begin{align}
\begin{aligned}
 & \text{Tr}\left[Z_1 Z_2 \bar Z_3 \bar Z_4 \right]\,,\,\, \text{Tr}\left[Z_5 Z_6 \bar Z_2' \bar Z_1' \right]\,,\,\, \text{Tr}\left[Z_4' Z_3'\bar Z_6' \bar Z_5' \right]\,,\\
  &\text{Tr}\left[Z_7 Z_1'\bar Z_4' \bar Z_8 \right]\,,\,\,\text{Tr}\left[Z_9 Z_5'\bar Z_1 \bar Z_7' \right]\,,\,\,\text{Tr}\left[Z_8' Z_4 \bar Z_5 \bar Z_9' \right]\,,\\
  &\text{Tr}\left[Z_2' Z_7'\bar Z_8' \bar Z_3' \right]\,,\,\,\text{Tr}\left[Z_6' Z_9'\bar Z_7 \bar Z_2 \right]\,,\,\,\text{Tr}\left[Z_3 Z_8 \bar Z_9 \bar Z_6 \right]\,.
\end{aligned}
\end{align}
It should be clear at this point that  one-loop anomalous dimensions are zero in this theory.
Take the length-$2$ and length-$3$ ladders from the portion of diagram with thicker (red, yellow) propagators in the picture on the right of  figure~\ref{fig:higher_checker}. Take respectively the two-point functions featuring the lightest operators exchanged in the OPE limit $x_1\to x_2,\,x_3\to x_4$, for the two- and three-loops ladders:
$$\langle\text{Tr}\left[(Z_2\bar Z_3\bar Z_5)(x_2)(\bar Z_9' Z_8' Z_1)(x_4)\right]\rangle\,,$$
and
$$\langle\text{Tr}\left[(Z_2\bar Z_3\bar Z_5\bar Z_7)(x_2)(\bar Z_2 Z_6' Z_8' Z_1)(x_4)\right]\rangle\,.$$
The first involves operators with different bare dimensions, sine the condition $\Delta_2+\Delta_3+\Delta_5 = \Delta_1 +\Delta_8' +\Delta_9'$ is never satisfied for general slopes. Hence, this Feynman integral is actually zero (i.e. evaluates to $\propto \delta(x_{24}^2)$). In the second correlator, on the contrary, the two operators share the same bare dimension by virtue of
$$\Delta_3+\Delta_5+\Delta_7 = \Delta_1 +\Delta_8' +\Delta_9'.$$ The two-point integral is divergent as $1/\varepsilon$ in dim. reg. thus the spectrum of anomalous dimensions is non-trivial starting at three-loops in the theory. 

The argument made here can be extended into the following general statement: the spectrum of anomalous dimensions in the Checkerboard theory with periods $(M_1,M_2$ is protected until order $\ell = \min (M_1,M_2)$ in the weak coupling expansion. The dilation operator at order $\ell$ can be realised as a spin-chain Hamiltonian acting on a state of $L \geqslant \ell$ sites as a (next-to)$^{\ell-1}$-nearest-neighbour Hamiltonian.

\section{Discussion}
 \label{sec:discussion}
 
The Checkerboard Fishnet CFT~\eqref{CheckerboardCFT}  introduced and studied in this paper is one representative of a huge family of generalised, Loom FCFTs~\cite{Kazakov:2022dbd} of arbitrary dimension, inspired by the observations of~\cite{Zamolodchikov:1980mb}. A remarkable property of the Checkerboard CFT is that the typical planar graphs  are represented (at least in the bulk, the boundary conditions depend on the particular physical quantities one is looking at)  by an integrable statistical mechanical system of conformal spins  with nearest-neighbour interactions on the square lattice. Each row of the lattice is a spectral-parameter-dependent transfer matrix (or monodromy matrix, if the boundary is open) with generic conformal representations in both quantum and auxiliary spaces, i.e. it  automatically encodes all the conservation laws. This integrability opens great opportunities in doing non-perturbative analysis of physical quantities of this FCFT in the planar limit, as well as to push the analytic computations of the underlying planar graphs to high (and sometimes arbitrary) loop orders.

The action of the Checkerboard CFT \eqref{CheckerboardCFT} has a few interesting reductions. In 4D and with all $w_i=1$, it has a local action of four complex scalars with chiral quartic, interactions.  In 3D at  $w_1=w_2=w_3=1,\,\,w_4=0$ it also has a local action with one sextic chiral interaction of three complex fields (a fishnet reduction of ABJM model).  For a particular reduction in 2D the   graph-building operator for planar Feynman graphs is equal to Lipatov's Hamiltonian of interacting reggeized gluons (BFKL limit). 

We presented a few analytic calculations of non-trivial physical quantities based on integrability and conformality of the Checkerboard CFT.  Using the methods of the papers~\cite{Grabner:2017pgm,Gromov:2018hut} we computed  certain four-point correlation functions and extracted from them the anomalous dimension of the shortest operator exchanged in their OPE. It would be good to compute the dimensions of other operators, in particular those with non-zero spin, as well as to extract from the 4-point function the related structure constants. Our results are based on the kite integral~\eqref{eq:B_Grozin} computed in terms of certain, rather complicated double sums~\cite{Grozin:2012xi,Derkachev:2022lay}. The computations are more involved than for the cases considered in~\cite{Gromov:2018hut,Kazakov:2018gcy}, so our results are not as explicit. Nevertheless, we managed to obtain for the dimension of this operator a few orders of perturbation theory in the 3D ABJM reduction of Checkerboard CFT. 
We observed certain transcendentality properties of the perturbative results, not directly for the anomalous dimension but rather after a certain functional transformation. It would be interesting to understand these properties in depth.

We also showed that the graph-building operator in 2D, at a certain limit of the spectral parameter, reduces to Lipatov's Hamiltonian for reggeized gluons. In the particular case of the aforementioned shortest operator, the dimension coincides with the BFKL pomeron eigenvalue~\cite{Kuraev:1977fs,Balitsky:1978ic}. Hence this limit of Checkerboard CFT may be very useful for the understanding of BFKL physics and for the computations of related physical quantities. An interesting question here is related to the question of positiveness of spectrum in this BFKL. As was noticed in~\cite{Kazakov:2022dbd}, the FCFTs have certain properties of positivity of spectrum of anomalous dimensions: they can be only real or appear complex conjugate pairs. Do these complex conjugate pairs appear in the BFKL limit Checkerboard CFT, as we observed for the ABJM reduction in sec.\eqref{sec:numerics}? What does it mean for the BFKL physics?   

There are still many interesting questions related to the Checkerboard CFT: 

\begin{itemize}
    \item What is the generalisation of fishchain picture of~\cite{Gromov:2019aku,Gromov:2019bsj,Gromov:2019jfh} to the Checkerboard case? 

    \item  Can we compute the wheel-type graphs  and the related conformal dimensions in the Checkerboard CFT? 
    Using the techniques of quantum integrability one could try to compute these correlation functions exactly at any \(L\), at all orders. In practice, this task is already complicated for the usual bi-scalar fishnet, even at small values of \(L\) (see the results of such a calculation for \(L=3\) in~\cite{Gromov:2017cja}). 
    More generally, one would need a systematic quantum integrability formalism for computations of spectra and of eigenfunctions of non-compact spin chains in principal series representations. So far, only  the \(SL(2,\mathbb C)\) case, related to the BFKL model, is relatively well studied~\cite{Derkachov:2001yn,DeVega:2001pu}.

    \item   It would be interesting to obtain the sigma-model representation for the Checkerboard fishnets with cyllindric topology, analogously to~\cite{Basso:2018agi,Basso:2019xay}, and to establish the related TBA equations. 

   \item  Could one find a useful application of Yangian symmetry~\cite{Chicherin:2017cns,Chicherin:2017frs,Kazakov:2023nyu} for Checkerboard graphs?

   \item Can we find among 2D Checkerboard graphs those which generalise the Calabi-Yau invariants of~\cite{Duhr:2022pch,Duhr:2023eld}? 

   \item Can we generalise the results of \cite{Basso:2021omx,Kostov:2022vup} regarding the thermodynamic limit of the Basso--Dixon correlators to their Checkerboard analogues computed in section \ref{sec:BD}?

   \item Can we understand the integrability structure behind the SoV approach developed in section \ref{sec:BD} and appendix \ref{app:SoV}? Because of the alternating pattern in the Checkerboard diagrams, we have had to introduce two non-commuting graph-building operators. It would be interesting to understand these operators in terms of (open) non-compact spin chains. It also seems natural to try to apply this SoV formalism to the calculation of the wheel diagrams.
    

    \item  It is possible that Checkerboard graphs feature the ``arctic circle" phenomenon~\cite{Jockusch1998random}. For the traditional spin models, like the 6-vertex model, the arctic circle means in fact an oval-shaped area inside the finite rectangular lattice outside of which the spins are almost frozen. The methods of \cite{di2023arctic} could in principle allow to study this phenomenon in the Checkerboard CFT (if it is present there). 

\end{itemize}

A few questions can be posed also in the context of stydy of diamond fishnet diagrams of the sections \ref{sec:diamonds},\ref{sec:higher_periods}:

\begin{itemize}
    \item To complete the classification of diamond graphs and of their properties for all types of boundary shapes.

    \item  To try to compute the cylindrical diamond configurations and the related anomalous dimensions of the Checkerboard CFT.
\end{itemize}

\section*{Acknowledgements}

We thank B.~Basso, F.~Levkovich-Maslyuk and I.~Kostov for useful discussions. V.K. thanks Perimeter Institute where a part of this work was done, for hospitality and support.
This research was supported in part by Perimeter Institute for Theoretical Physics. Research at Perimeter Institute is supported by the Government of Canada through the Department of Innovation, Science, and Economic Development, and by the Province of Ontario through the Ministry of Colleges and Universities. G.F. is grateful to the Azrieli Foundation for the award of an Azrieli Fellowship. The work of M.A. was supported in part by the TUBITAK 2221 Programme.

\newpage
\section*{Appendices}

\appendix


\section{Factorisation of the Eigenvalue of \texorpdfstring{$\widehat{T}$}{T}}
\label{app:eigenvalue}

In this section we are to transform the eigenvalue of the $\hat{T}$-operator \eqref{eq:T_eigenvalue}. Let us rearrange the order of integration in that formula:
\begin{multline}\label{eq:lambda}
    h(\Delta) = (x_{12}^2)^{\Delta_1+\Delta_2-\frac{\Delta}{2}}\int\frac{\dd^d x_{0} \dd^d x_{0'}}{(x_{10}^2)^{\Delta_1} (x_{0'1}^2)^{\Delta_2} (x_{20'}^2)^{\Delta_1} (x_{02}^2)^{\Delta_2}} \\
    \times \int \frac{\dd^d x_{1'} \dd^d x_{2'}}{(x_{1'2'}^2)^{\Delta_1+\Delta_2-\frac{\Delta}{2}}(x_{01'}^2)^{\Delta_4}(x_{1'0'}^2)^{\Delta_3}(x_{0'2'}^2)^{\Delta_4}(x_{2'0}^2)^{\Delta_3}}\, .
\end{multline}
The second integral in the formula above is exactly a Kite integral:
\begin{equation}\label{eq:B_constant}
    \int \frac{\dd^d x_{1'} \dd^d x_{2'}}{(x_{1'2'}^2)^{\Delta_1+\Delta_2-\frac{\Delta}{2}}(x_{01'}^2)^{\Delta_4}(x_{1'0'}^2)^{\Delta_3}(x_{0'2'}^2)^{\Delta_4}(x_{2'0}^2)^{\Delta_3}} = \frac{B\left(\Delta_3,\Delta_4,\Delta_1+\Delta_2-\frac{\Delta}{2}\right)}{(x_{00'}^2)^{\Delta_3+\Delta_4-\frac{\Delta}{2}}}\, .
\end{equation}
Substituting \eqref{eq:B_constant} into \eqref{eq:lambda}, and noticing that the remaining integral is of the same form, we obtain
\begin{align}
    h(\Delta) &= B\left(\Delta_3,\Delta_4,\Delta_1+\Delta_2-\frac{\Delta}{2}\right) (x_{12}^2)^{\Delta_1+\Delta_2-\frac{\Delta}{2}}\nonumber\\
    &\qquad\qquad\qquad\qquad\times \int\frac{\dd^d x_{0} \dd^d x_{0'}}{(x_{00'}^2)^{\Delta_3+\Delta_4-\frac{\Delta}{2}} (x_{10}^2)^{\Delta_1} (x_{0'1}^2)^{\Delta_2} (x_{20'}^2)^{\Delta_1} (x_{02}^2)^{\Delta_2}}\\
    &= B\left(\Delta_3,\Delta_4,\Delta_1+\Delta_2-\frac{\Delta}{2}\right) B\left(\Delta_1,\Delta_2,\Delta_3+\Delta_4-\frac{\Delta}{2}\right)\, .
\end{align}

\section{Small \texorpdfstring{$\gamma$}{gamma} Expansion of the Eigenvalue of \texorpdfstring{$\widehat{T}$}{T}}
\label{app:small_gamma_expansion}

Let us analyse the small \(\gamma\) of the sum \(\mathbf{I}_1+\mathbf{I}_2+\mathbf{I}_3\) given by \eqref{I1_sum_d3}, \eqref{I2_sum_d3} and \eqref{I3_sum_d3}. We will use it in the form of double series
\begin{align}\label{I_sums}
\mathbf{I}_1+\mathbf{I}_2&+\mathbf{I}_3= \sum\limits_{n=0}^{+\infty}\sum\limits_{k=0}^{+\infty}\left(\frac{\cot\frac{\pi\Delta}{2}}{1024\pi^3 (\Delta-1)(\Delta-2)}\frac{1}{\Gamma\left(\frac{\Delta}{2}\right)}
\frac{\Gamma\left(n+\frac{\Delta}{2}\right)}{n+\frac{\Delta}{2}-\frac{1}{2}}\frac{1}{n!}\delta_{k,0}+\right. \\
&+\frac{1}{1024\pi^{3}(\Delta-1)(\Delta-2)\sin\frac{\pi\Delta}{2}}\frac{1}{\Gamma\left(\frac{\Delta}{2}\right)}\frac{n!}{\left(\frac{3}{2}-\frac{\Delta}{2}+n\right)\Gamma\left(2-\frac{\Delta}{2}+n\right)}\delta_{k,0}+ \notag \\
&\left.+\frac{2^{\Delta-13}}{\pi^{4}}\frac{1}{\Gamma\left(\Delta\right)}\frac{1}{k!}\frac{\Gamma\left(\frac{\Delta}{2}+k\right)}{\left(\frac{1}{2}+k\right)\left(\frac{\Delta}{2}-\frac{1}{2}+k\right)}\frac{\left(\frac{1}{2}+n\right)\Gamma\left(\frac{\Delta}{2}+\frac{1}{2}+k+n\right)}{(1+k+n)\left(\frac{\Delta}{2}+k+n\right)\Gamma\left(\frac{3}{2}+k+n\right)}\right)\,. \notag
\end{align}
Expanding in the first two orders around the point \(\Delta=2+\gamma\), we obtain the following series at the orders \(\gamma^{-1}\) and \(\gamma^0\)
\begin{align}\label{I_sums_LO_NLO1}
    & \mathbf{I}_1+\mathbf{I}_2+\mathbf{I}_3=\sum\limits_{n=0}^{+\infty}\sum\limits_{k=0}^{+\infty}\left(-\frac{1}{128\pi^4(2n+1)^2 \gamma}\delta_{k,0}+\left(\frac{2-\gamma_E}{256\pi^4(2n+1)^2}- \right.\right. \\
    & \left.\left.-\frac{\psi(n+1)}{256\pi^4(2n+1)^2}+\frac{\psi'(n+1)}{1024\pi^4(2n+1)}\right)\delta_{k,0}+\frac{2n+1}{1024\pi^4(2k+1)^2(n+k+1)^2}+\mathcal{O}(\gamma)\right)\,. \notag
\end{align}
After taking the summation over \(k\) in \eqref{I_sums_LO_NLO1} we get the following result
\begin{align}\label{I_sums_LO_NLO2}
    & \mathbf{I}_1+\mathbf{I}_2+\mathbf{I}_3=\sum\limits_{n=0}^{+\infty}\left(-\frac{1}{128\pi^4(2n+1)^2 \gamma}+\frac{1-\gamma_E-\log 2}{128\pi^4(2n+1)^2}-\frac{\psi(n+1)}{128\pi^4(2n+1)^2}+\right. \\
    & \left.-\frac{\psi'(n+1)}{512\pi^4(2n+1)}+\mathcal{O}(\gamma)\right)\,. \notag
\end{align}
The last two sums from \eqref{I_sums_LO_NLO2} can be found in \cite{milgram2017sums}, which leads us to the following formula
\begin{equation}
    \mathbf{I}_1+\mathbf{I}_2+\mathbf{I}_3=-\frac{1}{1024\pi^{2}\gamma}+\frac{1}{1024\pi^{4}}\left(\pi^2+\pi^2 \log 2-\frac{21}{2}\zeta_3\right)+\mathcal{O}(\gamma)\,.
\end{equation}

Now we turn to the calculation of the term of the order \(\gamma^1\). To do this let us write down the corresponding term of the small \(\gamma\) expansion of \eqref{I_sums}, but before that we make some change in our summation variables. Namely, we replace the variable \(k\) by \(m=n+k\). After this the summation in \eqref{I_sums} is changed to
\begin{equation}
\sum\limits_{n=0}^{+\infty}\sum\limits_{k=0}^{+\infty} \rightarrow \sum\limits_{n=0}^{+\infty}\sum\limits_{m=n}^{+\infty}\,.
\end{equation}
It appears that the sum over \(m\) can be taken, which leads to the following sum
\begin{align}\label{3rd_order_sums}
    & \frac{1}{2048\pi^4}\sum\limits_{n=0}^{+\infty}\left(\frac{1}{\left(n+\frac{1}{2}\right)^3}+\frac{24-4\gamma_E+\pi^2-8\log 2}{4(n+1)\left(n+\frac{1}{2}\right)}-\frac{1}{\left(n+\frac{1}{2}\right)^4}+\frac{-2+\gamma_E+2\log 2}{(n+1)^2}+\right. \\
    & +\frac{-48+24\gamma_E-12\gamma_E^2+\pi^2-24\gamma_E \log 2-12\log^2 2+24\log 2}{6\left(n+\frac{1}{2}\right)^2}-\frac{\psi^2\left(n+\frac{1}{2}\right)}{2(n+1)^2}- \notag \\
    & -\frac{\psi^2(n+1)}{\left(n+\frac{1}{2}\right)^2}-\frac{\psi'\left(n+\frac{1}{2}\right)}{2(n+1)\left(n+\frac{1}{2}\right)}+\left(\frac{1}{\left(n+\frac{1}{2}\right)^2}-\frac{3}{(n+1)\left(n+\frac{1}{2}\right)}+\frac{(4-\gamma_E-2\log 2)}{2(n+1)^2}\right) \times \notag \\
    & \times \psi\left(n+\frac{1}{2}\right)+\frac{1}{4}\left(\frac{\psi\left(n+\frac{1}{2}\right)}{n+1}-\frac{\psi(n+1)}{n+\frac{1}{2}}\right)+\frac{(-4+3\gamma_E+2\log 2)\psi'(n+1)}{2\left(n+\frac{1}{2}\right)}+ \notag \\
    & \left.+\frac{(4-3\gamma_E-2\log 2)\psi(n+1)}{\left(n+\frac{1}{2}\right)^2}-\frac{\psi\left(n+\frac{1}{2}\right)\psi'\left(n+\frac{1}{2}\right)}{2(n+1)}+\frac{\psi(n+1)\psi'(n+1)}{n+\frac{1}{2}}\right)\,. \notag
\end{align}
All the sums in \eqref{3rd_order_sums} can be found in the literature (see \cite{milgram2017sums} and the references therein) except for the last two ones. In what follows we are going to find these sums using the formula (C.52) derived in \cite{milgram2017sums}
\begin{align}\label{psi_double_sum1}
& \sum\limits_{n=1}^{+\infty}\left(\frac{1}{\left(n+\frac{1}{2}\right)^2}+\frac{1}{(n+1)^2}\right)\sum\limits_{m=1}^{n}\frac{\psi\left(m+\frac{1}{2}\right)}{m+\frac{1}{2}}=\left(7\psi\left(\frac{1}{2}\right)+7\log 2\right)\zeta_3+ \\
& +\left(\left(\log 2-\frac{4}{4}\right)\psi\left(\frac{1}{2}\right)-\frac{8}{3}\log 2+\frac{1}{3}\log^2 2\right)\pi^2-\frac{1}{3}\log^4 2+\frac{79}{360}\pi^4-8\Li_4\!\left(\frac{1}{2}\right)\,. \notag
\end{align}
Then, by changing the summation order on the left hand side
\begin{align}
    & \sum\limits_{n=1}^{+\infty}\left(\frac{1}{\left(n+\frac{1}{2}\right)^2}+\frac{1}{(n+1)^2}\right)\sum\limits_{m=1}^{n}\frac{\psi(m+1)}{m+\frac{1}{2}}=\sum\limits_{m=1}^{+\infty}\frac{\psi(m+1)}{m+\frac{1}{2}}\sum\limits_{n=m}^{+\infty}\left(\frac{1}{\left(n+\frac{1}{2}\right)^2}+\right. \\
    & \left.+\frac{1}{(n+1)^2}\right)=\sum\limits_{m=1}^{+\infty}\frac{\psi(m+1)}{m+\frac{1}{2}}\left(\psi'\left(m+\frac{1}{2}\right)+\psi'(m+1)\right)= \notag \\
    & =\sum\limits_{l=1}^{+\infty}\left(\frac{\psi(l+1)\psi'\left(l+\frac{1}{2}\right)}{l+\frac{1}{2}}+\frac{\psi(l+1)\psi'(l+1)}{l+\frac{1}{2}}\right) \notag
\end{align}
and adding one term to the sum there, we obtain
\begin{multline}\label{sums_eq1}
\sum\limits_{l=0}^{+\infty}\left(\frac{\psi(l+1)\psi'\left(l+\frac{1}{2}\right)}{l+\frac{1}{2}}+\frac{\psi(l+1)\psi'(l+1)}{l+\frac{1}{2}}\right)=\frac{79}{360}\pi^4-\frac{1}{3}\log^4 2- \\
-\frac{1}{3}\pi^2 \log 2 \left(3\gamma_E+5\log 2\right)-8\Li_4\!\left(\frac{1}{2}\right)-7(\gamma_E+\log 2)\zeta_3\,.
\end{multline}
One can see that the left hand side of \eqref{sums_eq1} already contains one of the sums to be found. Let us transform the first sum on the left hand side of \eqref{sums_eq1}
\begin{equation}
    \sum\limits_{l=0}^{+\infty}\frac{\psi(l+1)\psi'\left(l+\frac{1}{2}\right)}{l+\frac{1}{2}}=\sum\limits_{n=0}^{+\infty}\frac{1}{\left(n+\frac{1}{2}\right)^2}\sum\limits_{m=0}^{n}\frac{\psi(m+1)}{m+\frac{1}{2}}\,,
\end{equation}
where we again changed the order of summation over \(n\) and \(m\). Then we apply a slightly modified formula (23) from \cite{milgram2017sums}
\begin{multline}
    \sum\limits_{m=0}^{n}\frac{\psi(m+1)}{m+\frac{1}{2}}=-\sum\limits_{m=0}^{n}\frac{\psi\left(m+\frac{1}{2}\right)}{m+1}+\psi(n+2)\psi\left(n+\frac{3}{2}\right)+2\psi(n+2)-2\psi\left(n+\frac{3}{2}\right)- \\
    -\psi(1)\psi\left(\frac{1}{2}\right)-2\psi(1)+2\psi\left(\frac{1}{2}\right)\,,
\end{multline}
which leads to the following identity
\begin{align}\label{transformation_identity1}
    & \sum\limits_{l=0}^{+\infty}\frac{\psi(l+1)\psi'\left(l+\frac{1}{2}\right)}{l+\frac{1}{2}}=-\sum\limits_{l=0}^{+\infty}\frac{\psi\left(l+\frac{1}{2}\right)\psi'\left(l+\frac{1}{2}\right)}{l+1}+\frac{49}{180}\pi^4+8\gamma_E \log 2+ \\
    & +16(-1+\log 2)\log 2-\frac{2}{3}\log^4 2+\frac{1}{3}\pi^2(-3\gamma_E+2(1+(-6+\log 2)\log 2))-16\Li_4\!\left(\frac{1}{2}\right)- \notag \\
    & -\frac{7}{2}(-4+3\gamma_E+6\log 2)\zeta_3\,. \notag
\end{align}

Our strategy is to derive a second, independent equation for the sums above and solve them as a system of linear equations. To achieve this we will use the formula similar to (C.50) from \cite{milgram2017sums} but with the different sign factor
\begin{equation}\label{psi_double_sum2}
    \sum\limits_{m=1}^{+\infty}\sum\limits_{l=1}^{+\infty}(-1)^{m}\frac{\psi(l+m)}{lm(l+m)}=-\frac{17}{1440}\pi^4+\frac{5}{8}\gamma_E \zeta_3\,.
\end{equation}
Our next step is the following: we split the summation over \(m\) in \eqref{psi_double_sum2} into the summation over odd and even \(m\), namely
\begin{equation}\label{even_odd_sums}
\sum\limits_{m=1}^{+\infty}\sum\limits_{l=1}^{+\infty}(-1)^{l}\frac{\psi(l+m)}{lm(l+m)} = \sum\limits_{j=1}^{+\infty}\left(\frac{1}{2j-1} \sum\limits_{l=1}^{+\infty} (-1)^{l} \frac{\psi(l+2j-1)}{l(l+2j-1)}+\frac{1}{2j} \sum\limits_{l=1}^{+\infty} (-1)^{l} \frac{\psi(l+2j)}{l(l+2j)}\right)\,.
\end{equation}
Thanks to the formulae (C.47) and (C.48) from \cite{milgram2017sums} we can perform the sums over \(l\) in the right hand sides of \eqref{even_odd_sums}. After conducting some intermediate summations over \(j\) with the usage of the sums from \cite{milgram2017sums}, we obtain the following equation
\begin{align}\label{sums_eq2}
    & \sum\limits_{l=0}^{+\infty}\left(\frac{\psi\left(l+\frac{1}{2}\right)\psi'\left(l+\frac{1}{2}\right)}{l+1}+\frac{\psi(l+1)\psi'(l+1)}{l+\frac{3}{2}}\right)=\frac{11}{360}\pi^4+\frac{1}{3}\pi^4(2+\log 2 (-12+ \\
    & +5\log 2))+\frac{1}{2}\gamma_E(16+\pi^2(-2+2\log 2)-21\zeta_3)+\frac{1}{3}\left(-48+24\log^2 2+\log^4 2+\right. \notag \\
    & \left.+24\Li_4\!\left(\frac{1}{2}\right)-21(-3+2\log 2)\zeta_3\right)\,. \notag
\end{align}
The second sum on the left hand side of \eqref{sums_eq2} contains exactly one of the sums to be found, whereas the denominator of the second one is \(l+3/2\) instead of \(l+1/2\). This can be overcome by using the properties of polygamma functions and we are able to express
\begin{align}\label{transformation_identity2}
    & \sum\limits_{l=0}^{+\infty}\frac{\psi(l+1)\psi'(l+1)}{l+\frac{3}{2}}=\sum\limits_{l=0}^{+\infty}\frac{\psi(l+1)\psi'(l+1)}{l+\frac{1}{2}}-16+8\gamma_E+16\log 2- \\
    & -8\gamma_E \log 2-8\log^2 2+7\zeta_3\,. \notag
\end{align}

Substitution of the identities \eqref{transformation_identity1} and \eqref{transformation_identity2} into \eqref{sums_eq1} and \eqref{sums_eq2} respectively gives us the following system of equations
\begin{equation}\label{system_for_sums}
\begin{cases}
    \sum\limits_{l=0}^{+\infty}\left(-\frac{\psi\left(l+\frac{1}{2}\right)\psi'\left(l+\frac{1}{2}\right)}{l+1}+\frac{\psi(l+1)\psi'(l+1)}{l+\frac{1}{2}}\right)=-\frac{19}{360}\pi^4+8\Li_4\!\left(\frac{1}{2}\right)+ \\
    +\frac{7}{2}(\gamma_E-4+4\log 2)\zeta_3+\frac{1}{3}\log^4 2-16\log ^2 2+16\log 2- \\
    -8\gamma_E \log 2-\frac{1}{3}\pi^2(2+\gamma_E(3\log 2-3)+\log 2 (7\log 2-12))\,, \\
    \sum\limits_{l=0}^{+\infty}\left(\frac{\psi\left(l+\frac{1}{2}\right)\psi'\left(l+\frac{1}{2}\right)}{l+1}+\frac{\psi(l+1)\psi'(l+1)}{l+\frac{1}{2}}\right)=\frac{11}{360}\pi^4+8\Li_4\!\left(\frac{1}{2}\right)- \\
    -\frac{7}{2}(3\gamma_E-4+4\log 2\zeta_3+\frac{1}{3}\log^4 2+16\log^2 2+8\gamma_E \log 2-16\log 2+ \\
    +\frac{1}{3}\pi^2(2+3\gamma_E(\log 2-1)+\log 2 (5\log 2-12))\,.
\end{cases}
\end{equation}
Solving the system of equations \eqref{system_for_sums}, we find the answer
\begin{align}\label{sums_solution}
    & \sum\limits_{l=0}^{+\infty}\frac{\psi(l+1)\psi'(l+1)}{l+\frac{1}{2}}=-\frac{\pi^4}{90}-\frac{\pi^2 \log^2 2}{3}+\frac{\log^4 2}{3}+8\Li_4\!\left(\frac{1}{2}\right)-\frac{7}{2}\gamma_E \zeta_3\,, \\
    & \sum\limits_{l=0}^{+\infty}\frac{\psi\left(l+\frac{1}{2}\right)\psi'\left(l+\frac{1}{2}\right)}{l+1}=\frac{\pi^4}{24}+2\pi^2 \log^2 2+\pi^2 \gamma_E \log 2-14\log 2 \zeta_3-7\gamma_E \zeta_3-4\pi^2 \log 2- \notag \\
    & -\pi^2 \gamma_E+14\zeta_3+\frac{2\pi^2}{3}+16\log^2 2+8\gamma_E \log 2-16\log 2\,. \notag
\end{align}
To sum up, substituting the result \eqref{sums_solution} into \eqref{3rd_order_sums} we derive the coefficient in front of \(\gamma^1\) in the small \(\gamma\) expansion of \eqref{I_sums}
\begin{equation}
    -\frac{1}{1024\pi^4}\left(\pi^2+\pi^2 \log 2-\frac{21}{2}\zeta_3+\frac{\pi^4}{40}+\frac{\log^4 2}{2}+12\Li_4\!\left(\frac{1}{2}\right)\right)\,.
\end{equation}

\section{Computation of double sums in the BFKL limit of Checkerboard CFT}
\label{BFKLdoublesums}

The particular case of double sums from \cite{Derkachev:2022lay} for \(d=2\) and \(\alpha_1=\alpha_3=a_1\) and \(\alpha_2=\alpha_4=a_2\) is decribed by the expressions
\begin{align}\label{I_sums_d=2}
    I_1(a_1,a_2,\delta) &=A_0^2(a_1)A_0(a_2)A_0(2-a_{12})A_0(a_2+\delta)A_0(2-a_{12}-\delta) \times \\
    &\times \left[_3F_2(\delta,1-a_1,-1+a_{12}+\delta,2-a_{12},a_2+\delta;1)\right]^2\,, \notag \\    I_2(a_1,a_2,\delta) &= A_0^2(a_2)A_0(\delta) A_0(-1+a_{12}+\delta) A_0(2-a_2-\delta) A_0(3-a_1-2a_2-\delta) \times \notag \\
    &\times \left[_3F_2(a_1,2-a_{12}-\delta,1-a_2,2-a_2-\delta,3-a_1-2a_2-\delta;1)\right]^2\,, \notag \\
    I_3(a_1,a_2,\delta) &= A_0(a_1) A_0(\delta) A_0(3-2a_{12}-\delta) A_0(2-a_{12}-\delta) A_0(-1+a_1+2a_2+\delta) \notag \\
    &\times A_0(a_{12}) \left[_3F_2(a_2,-2+2a_{12}+\delta,-1+a_{12}+\delta,a_{12},-1+a_1+2a_2+\delta;1)\right]^2\, , \notag
\end{align}
where $a_{12} = a_1 + a_2$. We are to analyse the sum \((I_1+I_2+I_3)(a_1,a_2,\delta)\) in two cases (we exchanged $a_1$ and $a_2$ as \((I_1+I_2+I_3)(a_1,a_2,\delta)\) does not depend on it)
\begin{equation}\label{asdelta_case1}
    a_1=1+u\,, a_2=-1-u+\Delta_+\,, \; \delta=2-\Delta_+-\frac{\Delta}{2}\, ,
\end{equation}
and
\begin{equation}\label{asdelta_case2}
    a_1=1+u\,, a_2=1-u-\Delta_+\,, \; \delta=\Delta_+-\frac{\Delta}{2}\,.
\end{equation}
Let us first focus on the first case, and prove that the limit
\begin{equation}
    \lim\limits_{\Delta_+ \rightarrow 0}(I_1+I_2+I_3)\left(-1-u+\Delta_+,1+u,2-\Delta_+-\frac{\Delta}{2}\right)
\end{equation}
is finite for arbitrary \(u\). The singularities in \(\Delta_+\) come from the prefactor of \(I_1\) and from the hypergeometric function in \(I_3\). Namely, we obtain for \(I_1\)
\begin{multline}\label{I1_exp_1}
    I_1\left(-1-u+\Delta_+,1+u,2-\Delta_+-\frac{\Delta}{2}\right)=\frac{(1+u)^2}{\Delta_+}\frac{\Gamma(-u)}{\Gamma(1+u)}\frac{\Gamma\left(u+\frac{\Delta}{2}\right)}{\Gamma\left(1-u-\frac{\Delta}{2}\right)} \\
    \times \frac{\Gamma\left(1-\frac{\Delta}{2}\right)}{\Gamma\left(\frac{\Delta}{2}\right)}\left(_3F_2\left(-u,1-\frac{\Delta}{2},2-\frac{\Delta}{2};2,1-u-\frac{\Delta}{2};1\right)\right)^2+\mathcal{O}(\Delta_+^0)\,.
\end{multline}
To get the small \(\Delta_+\) expansion of \(I_3\) we need the expansion of the hypergeometric function in it, which is
\begin{multline}
    \,_3F_2\left(1-\frac{\Delta}{2},-1-u+\Delta_+,-\frac{\Delta}{2}+\Delta_+;\Delta_+,-u-\frac{\Delta}{2}+\Delta_+;1\right) \\
    = \frac{\Delta}{2\Delta_+}\left(\frac{\Delta}{2}-1\right)\frac{u+1}{u+\frac{\Delta}{2}}\,_3F_2\left(-u,1-\frac{\Delta}{2},2-\frac{\Delta}{2};2,1-u-\frac{\Delta}{2};1\right)+\mathcal{O}(\Delta_+^0)\, .
\end{multline}
Substituting into the formula for \(I_3\) gives us
\begin{multline}\label{I3_exp_1}
    I_3\left(-1-u+\Delta_+,1+u,2-\Delta_+-\frac{\Delta}{2}\right)=-\frac{(1+u)^2}{\Delta_+}\frac{\Gamma(-u)}{\Gamma(1+u)}\frac{\Gamma\left(u+\frac{\Delta}{2}\right)}{\Gamma\left(1-u-\frac{\Delta}{2}\right)} \\
    \times \frac{\Gamma\left(1-\frac{\Delta}{2}\right)}{\Gamma\left(\frac{\Delta}{2}\right)}\left(_3F_2\left(-u,1-\frac{\Delta}{2},2-\frac{\Delta}{2};2,1-u-\frac{\Delta}{2};1\right)\right)^2+\mathcal{O}(\Delta_+^0)\,.
\end{multline}
Hence the singularities in \eqref{I1_exp_1} and \eqref{I3_exp_1} cancel each other and we see that
\begin{equation}\label{limit1}
    (I_1+I_2+I_3)\left(-1-u+\Delta_+,1+u,2-\Delta_+-\frac{\Delta}{2}\right) = \mathcal{O}(\Delta_+^0)\,.
\end{equation}

A similar analysis applies to the second choice of parameters $(a_1,a_2,\delta)$, see \eqref{asdelta_case2}. The singularity in \(I_1\) now comes from the hypergeometric function, whereas the singularity in \(I_3\) comes from the prefactor.
Following the same steps as for \eqref{limit1}, one can check that
\begin{equation}\label{limit2}
    \lim\limits_{\Delta_+ \rightarrow 0}(I_1+I_2+I_3)\left(1-u-\Delta_+,1+u,\Delta_+-\frac{\Delta}{2}\right) = \mathcal{O}(\Delta_+^0)\,.
\end{equation}

Taking into account that the limits \eqref{limit1} and \eqref{limit2} are finite for any \(u\), we can now address the question of the small \(u\) expansion of these quantities. This procedure yields
\begin{align}
    & (I_1+I_2+I_3)\left(-1-u,1+u,2-\frac{\Delta}{2}\right)= \\
    & =\frac{1}{u^2}+\frac{2}{u}\left(\psi\left(\frac{\Delta}{2}\right)+\psi\left(1-\frac{\Delta}{2}\right)-2\psi(1)\right)+\mathcal{O}(u^0)\,, \notag \\
    & (I_1+I_2+I_3)\left(1-u,1+u,-\frac{\Delta}{2}\right)= \\
    & =\frac{1}{u^2}+\frac{2}{u}\left(\psi\left(\frac{\Delta}{2}\right)+\psi\left(1-\frac{\Delta}{2}\right)-2\psi(1)\right)+\mathcal{O}(u^0)\,. \notag
\end{align}

\section{SoV Representation of Feynman Diagrams}\label{app:SoV}

\subsection{2D Rectangular Fishnets}
\label{app:SoV2D}

We present in this appendix some of the details of the derivation of \eqref{rectangular BD even} and \eqref{rectangular BD odd} in two dimensions. For brevity, we will state several results without presenting their proofs, which can be easily adapted from those of \cite{Derkachov:2001yn,Derkachov:2014gya,Derkachov:2018rot}. Once the 2D case is understood, it is relatively straightforward to treat the 4D case in a similar fashion, following \cite{Derkachov:2019tzo,Derkachov:2020zvv}.

We work in a slightly more general setting than what was presented in section~\ref{sec:BD} where the four fields have a spin. This means that for each field $i$, we now have two complex numbers $\Delta_i$ and $\bar{\Delta}_i$ whose difference is an integer: they encode the scaling dimension $\Delta_i+\bar{\Delta}_i\in\mathbb{C}$ and the spin $\Delta_i-\bar{\Delta}_i\in\mathbb{Z}$ of the field. The scalar propagators are replaced by the spinning ones according to
\begin{equation}
    |z|^{-2\Delta_i}\longrightarrow [z]^{-\Delta_i} = |z|^{-\Delta_i-\bar{\Delta}_i}\, \ee^{\ii(\bar{\Delta}_i-\Delta_i)\theta} \quad \text{where} \quad z = |z|\ee^{\ii\theta}\, .
\end{equation}
Thus, the kernel of the graph-building operator is now
\begin{multline}
    \bra{y_1,\dots,y_{N}} \Lambda_{N}\ket{z_1,\dots,z_{N}} = \prod_{k=0}^{\left\lfloor\frac{N-1}{2}\right\rfloor} [y_{2k+1} - z_{2k+1}]^{-2\Delta_1} [y_{2k,2k+1}]^{-2\Delta_2}\\
    \times \prod_{k=1}^{\left\lfloor\frac{N}{2}\right\rfloor} [y_{2k} - z_{2k}]^{-2\Delta_3} [y_{2k-1,2k}]^{-2\Delta_4}\, ,
\end{multline}
where $y_0 = 0$. The simpler case $\Delta_1 = \Delta_3 = \tilde{\Delta}_2 = \tilde{\Delta}_4$ was treated in \cite{Derkachov:2001yn}. The main subtlety in our setting is that we can only diagonalise the product $\Lambda'_{N}\Lambda_{N}$, see equations \eqref{right eigenvectors} and \eqref{left eigenvectors} below, but we are still able to find vectors that transform nicely under the action of $\Lambda_{N}$ alone, cf. \eqref{pseudo-eigenvalue Lambda} for instance. 

For future convenience, we have to introduce some notation: for $\gamma\in\mathbb{C}$, we set $\tilde\gamma = 1 - \gamma$. For $\gamma$ and $\bar\gamma$ two complex numbers whose difference is an integer, we also define
\begin{equation}
    A(\gamma) = \frac{\Gamma(1-\bar\gamma)}{\Gamma(\gamma)}\, .
\end{equation}
This function satisfies $A(\gamma)A(\tilde\gamma) = (-1)^{\gamma-\bar\gamma} = [-1]^\gamma$. Then, for $F$ a function (or an operator) depending on $(\Delta_1,\Delta_2,\Delta_3,\Delta_4)$, we define 
\begin{equation}
    F' = F\big|_{(\Delta_1,\Delta_2)\leftrightarrow (\Delta_3,\Delta_4)}\, , \quad\text{and}\quad \overset{\,\circ}{F} = F\big|_{(\Delta_1,\Delta_3)\leftrightarrow (\tilde{\Delta}_2,\tilde{\Delta}_4)}\, .
\end{equation}
For $(\bfu,\bar{\bfu}) = (\frac{m}{2} + \ii u,-\frac{m}{2} + \ii u)$ with $m$ integer\footnote{To be precise, $m\in\mathbb{Z} + \frac{\sigma}{4}$, where $\sigma\in\{0,1,2,3\}$ is (fixed and) such that $a - \bar a \in \mathbb{Z}$ and $b - \bar b \in \mathbb{Z}$. For simplicity, we shall assume that the parameters $\Delta_1, \Delta_2, \Delta_3,$ and $\Delta_4$ are such that $\sigma = 0$.} and $u$ real, we define the following two functions:
\begin{align}
    a(\bfu) &= \frac{\tilde{\Delta}_2}{2} + \frac{\Delta_1-\Delta_3}{4} - \bfu = \frac{\Delta_1}{2} + \frac{\Delta_4 - \Delta_2}{4} - \bfu\, ,\\
    b(\bfu) &= \frac{\tilde{\Delta}_2}{2} + \frac{\Delta_3-\Delta_1}{4} + \bfu = \frac{\Delta_3}{2} + \frac{\Delta_4 - \Delta_2}{4} + \bfu\, .
\end{align}
These functions satisfy the following properties:
\begin{equation}
    a+b+\Delta_2 = a'+b'+\Delta_4 = a+b'+\tilde{\Delta}_1 = a'+b+\tilde{\Delta}_3 = 1\, ,
\end{equation}
and
\begin{equation}
    (\overset{\circ}{a},\overset{\circ}{b}) = (a,b')\, .
\end{equation}

The eigenvectors of $\Lambda_{N}\Lambda'_{N}$ and $\Lambda'_{N}\Lambda_{N}$ are built iteratively using some ``layer'' operators $\Pi_N(\bfu)$ (as well as the variants $\Pi'_N(\bfu)$, $\overset{\circ}{\Pi}_N(\bfu)$, $\overset{\circ\ \ }{\Pi'_N}(\bfu)$). These are defined for $N\geqslant 2$ by
\begin{multline}
    \bra{y_1,\dots,y_{N}}\Pi_{N}(\bfu)\ket{z_1,\dots,z_{N-1}}\\
    = [y_{01}]^{-\tilde{b}'(\bfu)} \prod_{k=1}^{\left\lfloor\frac{N}{2}\right\rfloor} \frac{A(\tilde{a}(\bfu)) A(\tilde{b}(\bfu))}{[y_{2k-1,2k}]^{\Delta_2} [z_{2k-1} - y_{2k-1}]^{b(\bfu)} [y_{2k} - z_{2k-1}]^{a(\bfu)}}\\
    \times \prod_{k=1}^{\left\lfloor\frac{N-1}{2}\right\rfloor} \frac{A(\tilde{a}'(\bfu)) A(\tilde{b}'(\bfu))}{[y_{2k,2k+1}]^{\Delta_4} [z_{2k} - y_{2k}]^{b'(\bfu)} [y_{2k+1} - z_{2k}]^{a'(\bfu)}}\, ,
\end{multline}
where the normalisation ensures that
\begin{equation}\label{pseudo-eigenvalue Lambda}
    \Pi_{N+1}(\bfu_1) \overset{\circ}{\Pi}_{N}(\bfu_2) = \Pi_{N+1}(\bfu_2) \overset{\circ}{\Pi}_{N}(\bfu_1)\, .
\end{equation}

Moreover, it holds that
\begin{equation}
    \Lambda_N\Pi_N(\bfu) = \lambda_N(\bfu) \Pi'_{N}(\bfu) \overset{\circ}{\Lambda}_{N-1}
\end{equation}
with
\begin{equation}
    \lambda_N(\bfu) = \left(\frac{A(\Delta_3)}{A(\tilde{\Delta}_2)}\right)^{\left\lfloor\frac{N}{2}\right\rfloor} \left(\frac{A(\Delta_1)}{A(\tilde{\Delta}_4)}\right)^{\left\lfloor\frac{N-1}{2}\right\rfloor} \frac{A(\Delta_1)}{A(b'(\bfu))} \times \begin{cases}
    [-1]^{\Delta_1+\Delta_3} A(\tilde{a}'(\bfu)) & \text{if $N$ is even}\\
    A(\tilde{a}(\bfu)) & \text{if $N$ is odd}
    \end{cases}\, .
\end{equation}

We also need to construct left eigenvectors. To that end, we introduce the following conjugate ``layer'' operators for $N\geqslant 2$:
\begin{multline}
    \bra{y_1,\dots,y_{N-1}} \Xi_{N}(\bfu) \ket{z_1,\dots,z_{N}}\\
    = [z_{01}]^{-b(\bfu)} \prod_{k=1}^{\left\lfloor\frac{N}{2}\right\rfloor} \frac{[-1]^{\Delta_4} A(a'(\bfu)) A(b'(\bfu))}{[z_{2k-1,2k}]^{-\Delta_4} [y_{2k-1} - z_{2k-1}]^{\tilde{b}'(\bfu)} [z_{2k} - y_{2k-1}]^{\tilde{a}'(\bfu)}}\\
    \times \prod_{k=1}^{\left\lfloor\frac{N-1}{2}\right\rfloor} \frac{[-1]^{\Delta_2} A(a(\bfu)) A(b(\bfu))}{[z_{2k,2k+1}]^{-\Delta_2} [y_{2k} - z_{2k}]^{\tilde{b}(\bfu)} [z_{2k+1} - y_{2k}]^{\tilde{a}(\bfu)}}\, .
\end{multline}

We have
\begin{equation}
    \Xi_{N}(\bfu_1) \overset{\circ}{\Xi}_{N+1}(\bfu_2) = \Xi_{N}(\bfu_2) \overset{\circ}{\Xi}_{N+1}(\bfu_1)\, ,
\end{equation}
and
\begin{equation}
    \Xi_N(\bfu) \Lambda_N = \lambda_N(\bfu) \overset{\circ}{\Lambda}_{N-1} \Xi'_N(\bfu)\, .
\end{equation}
Moreover, assuming $\bfu\neq \bfv$ and $N\geqslant 3$, it holds that
\begin{equation}
    \Xi_N(\bfu) \Pi'_N(\bfv) = \frac{1}{(\bfu-\bfv)(\bar \bfv - \bar \bfu)} \overset{\circ\ }{\Pi'}_{N-1}(\bfv) \overset{\circ}{\Xi}_{N-1}(\bfu)\, .
\end{equation}

Let $\bra{\bfu}$ and $\ket{\bfu}$ be the left and right eigenvectors for $N=1$, they are
\begin{equation}
    \bra{\bfu}\ket{y} = \frac{1}{[y_{01}]^{b'(\bfu)}}\quad \text{and} \quad \bra{y}\ket{\bfu} = \frac{1}{[y_{01}]^{\tilde{b}'(\bfu)}}\, .
\end{equation}
We have
\begin{equation}
    \bra{\bfu_1} \Xi_2(\bfu_{2}) = \bra{\bfu_2} \Xi_2(\bfu_{1})\, ,\quad \Pi_2(\bfu_{2}) \overset{\circ\ }{\ket{\bfu_1}} = \Pi_2(\bfu_{1}) \overset{\circ\ }{\ket{\bfu_2}}\, ,
\end{equation}
and, assuming $\bfu\neq \bfv$,
\begin{equation}
    \Xi_2(\bfu) \Pi'_2(\bfv) = \frac{1}{(\bfu-\bfv)(\bar \bfv - \bar \bfu)} \ket{\bfv} \bra{\bfu}\, .
\end{equation}
For $N\geqslant 2$, we define
\begin{equation}
    \bra{\bfu_1,\dots,\bfu_N}' = \begin{cases}
    \bra{\bfu_1} \Xi_2(\bfu_{2}) \cdots \overset{\circ}{\Xi}_{N-1}(\bfu_{N-1}) \Xi_N(\bfu_N) & \text{if $N$ is even}\\
    \overset{\circ}{\bra{\bfu_1}} \overset{\circ}{\Xi}_2(\bfu_{2}) \cdots \overset{\circ}{\Xi}_{N-1}(\bfu_{N-1}) \Xi_N(\bfu_N) & \text{if $N$ is odd}
    \end{cases}\, ,
\end{equation}
and
\begin{equation}
    \ket{\bfu_1,\dots,\bfu_N} = \begin{cases}
    \Pi_N(\bfu_N) \overset{\circ}{\Pi}_{N-1}(\bfu_{N-1}) \cdots \Pi_2(\bfu_{2}) \overset{\circ\ }{\ket{\bfu_1}} & \text{if $N$ is even}\\
    \Pi_N(\bfu_N) \overset{\circ}{\Pi}_{N-1}(\bfu_{N-1}) \cdots \overset{\circ}{\Pi}_2(\bfu_{2}) \ket{\bfu_1} & \text{if $N$ is odd}
    \end{cases}\, .
\end{equation}
Using the properties of the ``layer'' operators stated above, it is straightforward to show that these functions are completely symmetric under permutation of the parameters $\bfu_1,\dots,\bfu_N$. And that they are eigenvectors of $\Lambda'_N\Lambda_N$:
\begin{equation}\label{right eigenvectors}
    \Lambda'_N\Lambda_N \ket{\bfu_1,\dots,\bfu_N} = \prod_{k=1}^N \rho(\bfu_k) \ket{\bfu_1,\dots,\bfu_N}
\end{equation}
and
\begin{equation}\label{left eigenvectors}
    \bra{\bfu_1,\dots,\bfu_N} \Lambda'_N\Lambda_N = \prod_{k=1}^N \rho(\bfu_k) \bra{\bfu_1,\dots,\bfu_N}\, ,
\end{equation}
where
\begin{equation}
    \rho = A(\Delta_1) A(\Delta_3) \frac{A(\tilde a) A(\tilde{a}')}{A(b) A(b')}\, .
\end{equation}

They also satisfy the following orthogonality relations:
\begin{equation}
    \bra{\bfu_1,\dots,\bfu_N}\ket{\bfv_1,\dots,\bfv_N} = \mu^{(2)}_N(\bfu_1,\dots,\bfu_N)^{-1} \sum_{\sigma\in\mathfrak{S}_N} \prod_{k=1}^N \delta(\bfu_k - \bfv_{\sigma(k)})\, ,
\end{equation}
where $\bfu_k = \frac{m_k}{2} + \ii u_k$, $\bfv_k = \frac{n_k}{2} + \ii v_k$, 
\begin{equation}
    \delta(\bfu - \bfv) = \delta_{m,n}\delta(u-v)\, ,
\end{equation}
and the measure is
\begin{equation}
    \mu^{(2)}_N(\bfu_1,\dots,\bfu_N) = (2\pi)^{-N} \prod_{1\leqslant i < j \leqslant N} \left[(u_i-u_j)^2 + \frac{(m_i - m_j)^2}{4}\right]\, .
\end{equation}

From this, we conjecture the resolution of the identity
\begin{equation}
    \frac{1}{N!}\sum_{m_1,\dots,m_N = -\infty}^{+\infty} \int_{\mathbb{R}^N} \mu^{(2)}_N(\bfu_1,\dots,\bfu_N) \ket{\bfu_1,\dots,\bfu_N} \bra{\bfu_1,\dots,\bfu_N} \dd u_1 \dots \dd u_N = \operatorname{Id}\, .
\end{equation}
The analogous resolution of the identity for the usual $SL(2,\mathbb{C})$ spin chain (relevant for the Basso--Dixon diagrams computed in \cite{Derkachov:2018rot}) was proved in \cite{Manashov:2023ehg}. This was done using techniques that were first applied in \cite{Derkachov:2021wja} to the $SL(2,\mathbb{R})$ chain. We expect that these techniques can be adapted to the present situation, but we have not done it and we will content ourselves with assuming that the statement holds, as was originally done in \cite{Derkachov:2001yn,Derkachov:2014gya,Derkachov:2018rot}.

Finally, we need the reduction of the eigenvectors when all the external points coincide:
\begin{multline}
    \bra{\bfu_1,\dots,\bfu_N}\ket{z,\dots,z} = A(\tilde{\Delta}_2)^{\frac{1}{2}\left\lfloor\frac{N}{2}\right\rfloor \left\lfloor\frac{N+2}{2}\right\rfloor} \left(A(\tilde{\Delta}_4) A(\Delta_3)\right)^{\frac{1}{2}\left\lfloor\frac{N-1}{2}\right\rfloor \left\lfloor\frac{N+1}{2}\right\rfloor}\\
    \times A(\Delta_1)^{\frac{1}{2}\left\lfloor\frac{N-2}{2}\right\rfloor \left\lfloor\frac{N}{2}\right\rfloor} [z_0-z]^{-b'_N - b_{N-1} - b'_{N-2} - b_{N-3} - \dots}\, ,
\end{multline}
and
\begin{multline}
    \bra{y,\dots,y} \prod_{k=0}^{\left\lfloor\frac{N-1}{2}\right\rfloor}[y_{2k,2k+1}]^{\Delta_4} \prod_{k=1}^{\left\lfloor\frac{N}{2}\right\rfloor} [y_{2k-1,2k}]^{\Delta_2}\ket{\bfu_1,\dots,\bfu_N}= \\
    =[z_0-y]^{-a'_N - a_{N-1} - a'_{N-2} - a_{N-3} - \dots} \times \\
    \times A(\tilde{\Delta}_2)^{\frac{1}{2}\left\lfloor\frac{N}{2}\right\rfloor \left\lfloor\frac{N+2}{2}\right\rfloor} \left(A(\tilde{\Delta}_4) A(\Delta_3)\right)^{\frac{1}{2} \left\lfloor\frac{N-1}{2}\right\rfloor \left\lfloor\frac{N+1}{2}\right\rfloor} A(\Delta_1)^{\frac{1}{2}\left\lfloor\frac{N-2}{2}\right\rfloor \left\lfloor\frac{N}{2}\right\rfloor}\\
    \times [-1]^{N\left\lfloor\frac{N-1}{2}\right\rfloor(\Delta_1+\Delta_4)} \prod_{k=1}^N \left(\frac{A(\tilde{a}(\bfu_k))}{A(b(\bfu_k))}\right)^{\left\lfloor\frac{N}{2}\right\rfloor} \left(\frac{A(\tilde{a}'(\bfu_k))}{A(b'(\bfu_k))}\right)^{\left\lfloor\frac{N-1}{2}\right\rfloor}\, .
\end{multline}

Combining the previous results, one obtains
\begin{align}
    & I_{N,2L-1}(y,z)= \\
    & =A(\tilde{\Delta}_2)^{\left\lfloor\frac{N}{2}\right\rfloor \left\lfloor\frac{N+2}{2}\right\rfloor} \left(A(\tilde{\Delta}_4) A(\Delta_3)\right)^{\left\lfloor\frac{N-1}{2}\right\rfloor \left\lfloor\frac{N+1}{2}\right\rfloor} A(\Delta_1)^{\left\lfloor\frac{N-2}{2}\right\rfloor \left\lfloor\frac{N}{2}\right\rfloor} \frac{[z_0-y]^{Y_N}[z_0-z]^{Z_N}}{N!} \times \notag \\
    & \times  [-1]^{N\left\lfloor\frac{N-1}{2}\right\rfloor(\Delta_1+\Delta_4)} \sum_{m_1,\dots,m_N = -\infty}^{+\infty} \int_{\mathbb{R}^N} \mu_N(\bfu_1,\dots,\bfu_N) \prod_{k=1}^N f_{N,2L-1}(\eta;\bfu_k) \dd u_1 \dots \dd u_N\,, \notag
\end{align}
where $\eta = (y-z_0)/(z-z_0)$,
\begin{equation}
    Y_N = - \left\lfloor\frac{N+1}{2}\right\rfloor \left(\frac{\Delta_3}{2} + \frac{\Delta_2 - \Delta_4}{4}\right) - \left\lfloor\frac{N}{2}\right\rfloor \left(\frac{\Delta_1}{2} + \frac{\Delta_4 - \Delta_2}{4}\right)\, ,
\end{equation}
\begin{equation}
    Z_N = - \left\lfloor\frac{N+1}{2}\right\rfloor \left(\frac{\Delta_1}{2} + \frac{\Delta_2 - \Delta_4}{4}\right) - \left\lfloor\frac{N}{2}\right\rfloor \left(\frac{\Delta_3}{2} + \frac{\Delta_4 - \Delta_2}{4}\right)\, ,
\end{equation}
and 
\begin{equation}
    f_{N,2L-1}(\eta;\bfu) = \left[\eta\right]^{\bfu} \rho({\bfu})^L \left(\frac{A(\tilde{a}(\bfu))}{A(b(\bfu))}\right)^{\left\lfloor\frac{N}{2}\right\rfloor} \left(\frac{A(\tilde{a}'(\bfu))}{A(b'(\bfu))}\right)^{\left\lfloor\frac{N-1}{2}\right\rfloor}\, .
\end{equation}
This can also be expressed as a determinant:
\begin{align}\label{determinant SoV 2D}
    & I_{N,2L-1}(y,z)= \\
    & =A(\tilde{\Delta}_2)^{\left\lfloor\frac{N}{2}\right\rfloor \left\lfloor\frac{N+2}{2}\right\rfloor} \left(A(\tilde{\Delta}_4) A(\Delta_3)\right)^{\left\lfloor\frac{N-1}{2}\right\rfloor \left\lfloor\frac{N+1}{2}\right\rfloor} A(\Delta_1)^{\left\lfloor\frac{N-2}{2}\right\rfloor \left\lfloor\frac{N}{2}\right\rfloor} [z_0-y]^{Y_N}[z_0-z]^{Z_N} \times \notag \\
    & \times [-1]^{N\left\lfloor\frac{N-1}{2}\right\rfloor(\Delta_1+\Delta_4)} \det_{1\leqslant i,j\leqslant N}\left(\sum_{m=-\infty}^{+\infty} \int \left(\frac{m}{2}+\ii u\right)^{i-1} \left(\frac{m}{2}-\ii u\right)^{j-1} f_{N,2L-1}(\eta;\bfu)\frac{\dd u}{2\pi} \right)\,. \notag
\end{align}
We presented the formula for $I_{N,2L}$, but the previous results also imply
\begin{align}
    & I_{N,2L}(y,z)= \\
    & =A(\Delta_3)^{\left\lfloor\frac{N^2+3}{4}\right\rfloor} \left(A(\Delta_1) A(\tilde{\Delta}_2) A(\tilde{\Delta}_4)\right)^{\left\lfloor\frac{N^2}{4}\right\rfloor} \frac{[z_0-y]^{Y_N}[z_0-z]^{Z'_N}}{N!} [-1]^{\left\lfloor\frac{N}{2}\right\rfloor(\Delta_1+\Delta_3)} \times \notag \\
    & \times  [-1]^{N\left\lfloor\frac{N-1}{2}\right\rfloor(\Delta_1+\Delta_4)} \sum_{m_1,\dots,m_N = -\infty}^{+\infty} \int_{\mathbb{R}^N} \mu_N(\bfu_1,\dots,\bfu_N) F_{N,2L}(\eta;\bfu_1,\dots,\bfu_N) \dd u_1 \dots \dd u_N\,, \notag
\end{align}
where
\begin{equation}
    F_{N,2L}(\eta;\bfu_1,\dots,\bfu_N) = \frac{\prod_{k=1}^N f_{N,2L-1}(\eta;\bfu_k) \times \begin{cases}
    A(\tilde{a}(\bfu_k)) & \text{if $N$ is even}\\
    A(\tilde{a}'(\bfu_k)) & \text{if $N$ is odd}
    \end{cases}}{A(b(\bfu_N)) A(b'(\bfu_{N-1})) A(b(\bfu_{N-2})) \cdots}\, .
\end{equation}

\subsection{Remark on the 4D Formula}


It is unclear whether \eqref{rectangular BD even} can be further simplified. In 2D, one can write it as a determinant, see \eqref{determinant SoV 2D}. In 4D, we focus, for clarity, on the particular case $\Delta_2 = \Delta_4 = 1$ where
\begin{equation}
    f_{N,2L}(r;\bfu) = \frac{r^{2\ii u}}{\left((u-\ii\gamma)^2+\frac{(m+1)^2}{4}\right)^{L+\left\lfloor\frac{N}{2}\right\rfloor} \left((u+\ii\gamma)^2+\frac{(m+1)^2}{4}\right)^{L+\left\lfloor\frac{N-1}{2}\right\rfloor}}\, ,
\end{equation}
with $\gamma = \frac{1-\Delta_1}{2} = \frac{\Delta_3-1}{2}$. For convergence of the various integrals, we have to assume that $|\Re(\gamma)|<1/2$. Following the procedure of \cite{Basso:2021omx}, we can then recast the integral in the following form:
\begin{multline}
    I_{N,2L}(y,z) = A_0(\Delta_3)^{\left\lfloor\frac{N-1}{2}\right\rfloor \left\lfloor\frac{N+1}{2}\right\rfloor - \left\lfloor\frac{N-2}{2}\right\rfloor \left\lfloor\frac{N}{2}\right\rfloor} \frac{y^{2 Y_N} z^{2 Z_N}}{(-4\ii)^N N!}\\
    \times \int_{\mathbb{R}^N} \frac{\det(A)}{\prod_{k=1}^{N} \sh\left(\frac{x_k-\ii\theta}{2}\right) \sh\left(\frac{x_k+\ii\theta}{2}\right)} \dd x_1 \dots \dd x_N\, ,
\end{multline}
where $A$ is a $2N\times 2N$ matrix whose elements are
\begin{equation}
    A_{i,2k-1} = A_i^+(x_k)\, ,\quad A_{i,2k} = A_i^-(x_k)\, ,
\end{equation}
with
\begin{equation}
    A_i^{\pm}(x) = \int_{\mathbb{R}\pm\ii\epsilon} \frac{\xi^{i-1} \ee^{\ii \xi(\sigma\pm x)}}{(\xi-\ii\gamma)^{L+\left\lfloor\frac{N}{2}\right\rfloor} (\xi+\ii\gamma)^{L+\left\lfloor\frac{N-1}{2}\right\rfloor}} \frac{\dd\xi}{2\ii\pi}
\end{equation}
for any $\epsilon > |\!\Re(\gamma)|$, and we introduced $\sigma = \ln r$. Clearly,
\begin{equation}
   A_i^{\pm}(x) \propto \theta(\mp\sigma -x)
\end{equation}
so that the integrals over the variables $x_1,\dots,x_N$ are actually restricted to $x_k<-|\sigma|$. However, we cannot go much further in the computation because $\det(A)$ seems difficult to evaluate, even though the functions $A_i^{\pm}$ are easy to compute.

\subsection{Diamond Correlators}
\label{app:diamond SoV}

We consider a graph made of $N\times L$ R-matrices; they correspond to the diagrams of figure~\ref{class_II_diag} where, using a conformal transformation, we send $x_1$ to infinity and $x_3$ to $0$. We call them $B_{N,L}$ and they are related to $G^{(II)}_{N,L}$ through
\begin{equation}
    B_{N,L}(y,z) = \lim_{x^2_1\to +\infty} \left[x_1^{2L(\Delta_2+\Delta_3)} G^{(II)}_{N,L}(x_1,z,0,y)\right]
\end{equation}
and
\begin{equation}
    G^{(II)}_{N,L}(x_1,x_2,x_3,x_4) = \frac{B_{N,L}\left(\frac{x_{41}}{x_{41}^2} - \frac{x_{31}}{x_{31}^2},\frac{x_{21}}{x_{21}^2} - \frac{x_{31}}{x_{31}^2}\right)}{x_{21}^{2N(\Delta_3+\Delta_4)} x_{31}^{2L(\Delta_2+\Delta_3)} x_{41}^{2N(\Delta_1+\Delta_2)}}\, .
\end{equation}
The computation is very close to that of \cite{Derkachov:2018rot}, the main difference being that the graph-building operator is now made of R-matrices and not just ``half'' of them, i.e. the triangles. The eigenvectors are unchanged but the reduction is not the same: now both left and right eigenvectors are directly evaluated at coinciding points. In dimension $d=2$ or $d=4$, the result is
\begin{multline}\label{BNL d}
    B_{N,L}(y,z) =  \frac{A_0(\Delta_1)^{NL} A_0(\Delta_4)^{NL} |y|^{-N(\Delta_1+\Delta_2)} |z|^{-N(\Delta_3+\Delta_4)}}{N!\left[\prod_{k=1}^{\frac{d}{2}}(\Delta_1+\Delta_2-k)(\Delta_3+\Delta_4-k)\right]^{\frac{N(N-1)}{2}}}\\
    \times \sum_{m_1,\dots,m_N = -\infty}^{+\infty} \prod_{k=1}^N P^{(d)}_{m_k}(\theta) \int_{\mathbb{R}^N} \mu_N^{(d)}(\bfu_1,\dots,\bfu_N) \prod_{k=1}^N r^{2\ii u_k}  g^L_{m_k}(\bfu_k) \dd u_k\, ,
\end{multline}
where
\begin{equation}
    g_m(\bfu) = A_m\left(\frac{d}{2}+\frac{\Delta_2-\Delta_1}{2}+\ii u\right) A_m\left(\frac{d}{2}+\frac{\Delta_3-\Delta_4}{2}-\ii u\right)\, .
\end{equation}

\section{Particular Cases of \texorpdfstring{$B_{N,L}$}{BNL} in 4 Dimensions}
\label{app:diamond SoV check}

Some of the integrals $B_{N,L}$ can be trivially computed using the star-triangle relation, as explained in section~\ref{sec:diamonds_II}. We verify here on the two simplest examples that the SoV representation does reproduce the correct result. We will also compute the simplest non-trivial integral.

We will be computing the integrals over the separated variables using the residue theorem. Assuming, for instance, that
\begin{equation}
    \Re(\Delta_1) > \Re(\Delta_2)\, ,\quad \Re(\Delta_4) > \Re(\Delta_3)\, ,\quad \text{and} \quad r > 1
\end{equation}
means that we can deform the contours in the upper half-plane, and that the relevant poles only come from one of the Gamma functions.

\subsection{\texorpdfstring{$L=1,N=1$}{L=1,N=1}}

The simplest integral is straightforward to compute. The sum over residues gives
\begin{align*}
    & B_{1,1}(y,z)= \\
    & = \frac{A_0(\Delta_1) A_0(\Delta_4)}{y^{2(2-\Delta_3)} z^{2\Delta_3}} \sum_{l,n=0}^{+\infty} r^{-l-2n} C^{(1)}_l(\cos\theta) (l+1) \frac{(-1)^n \Gamma(\Delta_1+\Delta_4-2 + l + n)}{n!\, \Gamma(l+n+2) \Gamma(4-\Delta_1-\Delta_4 - n)}\\
    & = \frac{A_0(\Delta_1) A_0(\Delta_4) A_0(4-\Delta_1-\Delta_4)}{y^{2(2-\Delta_3)} z^{2\Delta_3}} \sum_{m=0}^{+\infty} r^{-m} \sum_{n=0}^{\lfloor \frac{m}{2}\rfloor} (m+1-2n)\\
    & \qquad\qquad\qquad\qquad\qquad\qquad\times  \frac{ (\Delta_1+\Delta_4-2)_{m-n} (\Delta_1+\Delta_4-3)_{n}}{n!\, \Gamma(m-n+2)} C^{(1)}_{m-2n}(\cos\theta)\\
    & = \frac{A_0(\Delta_1) A_0(\Delta_4) A_0(4-\Delta_1-\Delta_4)}{y^{2(2-\Delta_3)} z^{2\Delta_3}} \sum_{m=0}^{+\infty} r^{-m} C_m^{(\Delta_1+\Delta_4 - 2)}(\cos\theta)\\
    & = \frac{A_0(\Delta_1) A_0(\Delta_4) A_0(4-\Delta_1-\Delta_4)}{y^{2\Delta_2} z^{2\Delta_3} (y-z)^{2(\Delta_1+\Delta_4 - 2)}}\, ,
\end{align*}
where $C^{(\lambda)}_l$ are the Gegenbauer polynomials. This result coincides with a simple application of the chain relation. 

\subsection{\texorpdfstring{$L=1,N=2$}{L=1,N=2}}


Let us try to evaluate $B_{1,2}$ from the SoV representation \eqref{BNL d}. We first compute the integrals using the residue theorem:
\begin{multline}\label{B12 residues}
    B_{1,2}(y,z)\propto \sum_{l_1,l_2,n_1,n_2 = 0}^{+\infty} (n_2 - n_1)(l_1 - l_2 + n_1 - n_2)(l_1 + 1 + n_1 - n_2)\\
    \times (l_2 + 1 + n_2 - n_1) \tilde{\lambda}_{l_1,n_1}(r,\theta) \tilde{\lambda}_{l_2,n_2}(r,\theta)\, ,
\end{multline}
where
\begin{equation}
    \tilde{\lambda}_{l,n}(r,\theta) = (\ee^{\ii(l+1)\theta} - \ee^{-\ii(l+1)\theta}) r^{-l-2n} \frac{(-1)^{n} (l+1) \Gamma\left(\kappa + l + n\right)}{n!\,  \Gamma\left(l + n + 2\right) \Gamma\left(2 - \kappa -  n\right)}\, ,
\end{equation}
where $\kappa = \Delta_1+\Delta_4 - 2 = 2 - \Delta_3-\Delta_2$. Collecting the terms with the same powers of $r$ and $\ee^{\ii\theta}$, one may write
\begin{equation}
    B_{1,2}(y,z)\propto \sum_{M,P = 0}^{+\infty} c_{M,P} r^{-M} \left(\ee^{\ii P \theta} + \ee^{-\ii P \theta}\right)\, .
\end{equation}
First, it is easy to see that $c_{M,P} = 0$ unless $M-P\in 2\mathbb{N}$, since there is simply no such term in the expansion \eqref{B12 residues}. However, because a simple application of the star-triangle relation shows that $B_{1,2}(y,z) = 0$ when $y\neq z$, we actually expect all the coefficients $c_{M,P}$ to vanish. This seems non-trivial from the representation as a sum over residues.

One may easily verify that $c_{P,P} = 0$ and that
\begin{align}
    c_{P+2,P} &= \frac{-2}{\Gamma(2-\kappa) \Gamma(-\kappa)} \Bigg[\frac{\Gamma(\kappa) \Gamma(\kappa+P)}{\Gamma(P-1)}
    + \sum_{l=0}^{P-2} \frac{(P-2l) (l-1) \Gamma(\kappa+P-l) \Gamma(\kappa+l)}{(P-l) \Gamma(l+1) \Gamma(P-1-l)} \Bigg]\\
    &= \frac{-2}{\Gamma(2-\kappa) \Gamma(-\kappa)} \sum_{l=2}^{P-2} \frac{(P-2l) (l-1) \Gamma(\kappa+P-l) \Gamma(\kappa+l)}{(P-l) \Gamma(l+1) \Gamma(P-1-l)} = 0\, ,
\end{align}
for instance. However, some of the other coefficients are more involved: one has
\begin{multline}
    c_{2M,0} = \sum_{l=0}^M \sum_{n=0}^l \frac{(-1)^l (M+1-l)^2 (l-2n)^2 (M +1 - 2n) (M + 1 + 2l - 2n)}{n!\, (l-n)!\, \Gamma(M - l + n + 2) \Gamma(M - n + 2)}\\
    \times \frac{\Gamma(\kappa + M + n - l) \Gamma(\kappa + M - n)}{\Gamma(2 - \kappa - n)  \Gamma(2 - \kappa + n - l)}\, .
\end{multline}
We checked, using Mathematica, that $c_{2M,0} = 0$ for $M\leqslant 20$.

\subsection{\texorpdfstring{$L=2,N=1$}{L=2,N=1}}

This integral is non-trivial. On the one hand, using the star-triangle relation, we can express it as
\begin{equation}
    B_{2,1}(y,z) = \frac{\left[A_0(\Delta_1) A_0(\Delta_4) A_0(2-\kappa)\right]^2}{y^{2\Delta_2} z^{2\Delta_3}} \int \frac{\pi^{-2}\dd^4 w}{w^{2(2-\kappa)} (w-y)^{2\kappa} (w-z)^{2\kappa}}\, ,
\end{equation}
with $\kappa = \Delta_1+\Delta_4 - 2 = 2 - \Delta_3-\Delta_2$.

On the other hand, from the SoV representation, we obtain
\begin{multline}
    B_{2,1}(y,z) = \frac{\left[A_0(\Delta_1) A_0(\Delta_4) \right]^2}{y^{2(2-\Delta_3)} z^{2\Delta_3}} \sum_{l,n = 0}^{+\infty} r^{-l-2n} C^{(1)}_l(\cos\theta) \frac{(l+1) \Gamma^2(\kappa+l+n)}{\left[n!\, \Gamma(l+n+2) \Gamma(2-\kappa-n)\right]^2}\\
    \times \left[\ln r + \psi(n+1) + \psi(l+n+2) - \psi(2-\kappa-n) - \psi(\kappa+l+n)\right]\, .
\end{multline}

\bibliographystyle{JHEP}
\bibliography{Checkerboard}

\end{document}